\documentclass[journal]{IEEEtran}
\usepackage[english]{babel}
\usepackage{soul,xcolor}
\usepackage{blindtext}
\usepackage{pstricks}
\usepackage{amsmath,amssymb,amscd,latexsym}
\usepackage{breqn}
\usepackage{comment,psfrag,subfigure,enumerate}
\usepackage{cancel}
\usepackage{cite}
\usepackage{url}
\usepackage{tensor}
\usepackage{graphicx}
\usepackage{pstool}
\usepackage{float,color}
\usepackage{rotating}
\usepackage{multirow}
\usepackage{multicol}
\usepackage{setspace}
\usepackage{hyperref}
\usepackage{graphicx} 
\usepackage{epstopdf} 
\usepackage{subfigure}
\usepackage{times}
\usepackage{tikz}
\usepackage{verbatim}
\usepackage{pst-optexp}
\usepackage[framemethod=TikZ]{mdframed}
\usepackage{pgfplotstable,enumerate}
\usetikzlibrary[pgfplots.groupplots]
\usepgflibrary{decorations.pathmorphing} 
\usetikzlibrary{arrows,shapes}
\usetikzlibrary{arrows,patterns}

\setstcolor{red}

\newcommand{\eqlab}[2]{\begin{align} \label{#1} #2 \end{align}}

\newcommand{\eqrf}[1]{\eqref{#1}}
\newtheorem{theorem}{Theorem}

\newcommand{\Rmnum}[1]{\expandafter\@slowromancap\romannumeral #1@}
\newcommand{\HR}[1]{{\color{black}#1}}

\usetikzlibrary{calc}
\mdfdefinestyle{MyFrame}{%
    usetwoside=false,
    linecolor=black,
    outerlinewidth=0.5pt,
    roundcorner=10pt,
    innertopmargin=\baselineskip,
    innerbottommargin=\baselineskip,
    innerrightmargin=10pt,
    leftmargin=10pt,
    rightmargin=10pt,
    skipabove = -10pt,
    skipbelow = 0pt,
    innerleftmargin=10pt,
    userdefinedwidth = 1.0\textwidth
    backgroundcolor=gray!1!white}
    
    \mdfdefinestyle{MyFrame1}{%
    usetwoside=false,
    linecolor=black,
    outerlinewidth=0.5pt,
    roundcorner=10pt,
    innertopmargin=\baselineskip,
    innerbottommargin=\baselineskip,
    innerrightmargin=10pt,
    leftmargin=10pt,
    rightmargin=10pt,
    skipabove = 10pt,
    skipbelow = 0pt,
    innerleftmargin=10pt,
    userdefinedwidth = 1.0\columnwidth
    backgroundcolor=gray!1!white}

\newcolumntype{f}{>{$}l<{$}}
\newcolumntype{n}{l}
\newcolumntype{N}{>{\scriptsize}l}
\newcolumntype{v}[1]{>{\raggedright\hspace{0pt}}p{#1}}
\newcolumntype{V}[1]{>{\scriptsize\raggedright\hspace{0pt}}p{#1}}
\newcolumntype{B}[1]{>{\boldmath\DC@{.}{,}{#1}}l<{\DC@end}}
\newcolumntype{d}[1]{>{\DC@{.}{,}{#1}}l<{\DC@end}}
\newcolumntype{i}[1]{>{\DC@{.}{,}{#1}\mathnormal\bgroup}l<{\egroup\DC@end}}
\newcolumntype{s}[1]{>{\DC@{.}{,}{#1}\mathsf\bgroup}l<{\egroup\DC@end}}
\newcolumntype{R}[1]{%
  >{\begin{turn}{90}\begin{minipage}{#1}\scriptsize\raggedright\hspace{0pt}}l%
  <{\end{minipage}\end{turn}}%
}
\newcolumntype{x}{>{\scriptsize\raggedright\hspace{0pt}}X}
\makeatother

\usetikzlibrary{intersections,arrows,decorations.pathmorphing,snakes}

\pgfdeclaredecoration{fiber}{initial}
{
    \state{initial}[width=1.5\pgfdecorationsegmentamplitude, next state=final]{
    \pgfpathmoveto{\pgfpointorigin}
    \pgflineto{\pgfpoint{\pgfdecoratedinputsegmentlength/2}{0pt}}
    \pgfpathcircle{\pgfpoint{\pgfdecoratedinputsegmentlength/2}{.5\pgfdecorationsegmentamplitude}}{.5\pgfdecorationsegmentamplitude}
    \pgfpathcircle{\pgfpoint{\pgfdecoratedinputsegmentlength/2-.25\pgfdecorationsegmentamplitude}{.5\pgfdecorationsegmentamplitude}}{.5\pgfdecorationsegmentamplitude}
    \pgfpathcircle{\pgfpoint{\pgfdecoratedinputsegmentlength/2+.25\pgfdecorationsegmentamplitude}{.5\pgfdecorationsegmentamplitude}}{.5\pgfdecorationsegmentamplitude}
    \pgfpathmoveto{\pgfpoint{\pgfdecoratedinputsegmentlength/2}{0pt}}
    }
   \state{final}{
        \pgfpathlineto{\pgfpointdecoratedpathlast}
   }
}

\newtheorem{ProofLemma}{Proof of Lemma}

\newtheorem{corollary}{Corollary}
\newtheorem{lemma}{Lemma}
\newcommand*{\QEDA}{\small\hfill\ensuremath{ \blacksquare}}%
\begin{document}

\title{A\HR{n Improved} Model of Nonlinear Fiber Propagation in the Presence of Kerr Nonlinearity and Stimulated Raman Scattering}
\author{Hami~Rabbani, Gabriele Liga, {\it{Member}}, {\it{IEEE}}, Vin\'{i}cius~Oliari, Lotfollah~Beygi, Erik Agrell,  {\it{Fellow}}, {\it{IEEE}}, Magnus Karlsson, {\it{Senior Member}}, {\it{IEEE}}, {\it{Fellow}}, {\it{OSA}}, Alex~Alvarado, {\it{Senior Member}}, {\it{IEEE}}
\thanks{H.~Rabbani, G. Liga, V.~Oliari and A. Alvarado are with the Information and Communication Theory Lab, Signal Processing Systems Group, Department of Electrical Engineering, Eindhoven University of Technology, Eindhoven 5600 MB, The Netherlands. E-mails: \{h.rabbani,g.liga,v.oliari.couto.dias,a.alvarado\}@tue.nl}
\thanks{L.~Beygi is with the EE Dept. of K. N. Toosi University of Technology. E-mail: beygi@kntu.ac.ir}
\thanks{E.~Agrell is with the Dept. of Electrical Engineering, Chalmers University of Technology, Sweden. E-mail: agrell@chalmers.se}
\thanks{M.~Karlsson is with the Dept. of Microtechnology and Nanoscience, Photonics Laboratory, Chalmers University of Technology, Sweden. E-mail: magnus.karlsson@chalmers.se}
}

\markboth{Preprint, \today}{}%
\maketitle

\newpage

\begin{abstract}
Ultra-wideband fiber optical transmission \HR{suffers from both Kerr nonlinearity and stimulated Raman scattering (SRS).} Mathematical models that address the interplay between Kerr nonlinearity and SRS exist. 
These models are based on the Gaussian-noise (GN) and enhanced Gaussian noise (EGN) models. 
\HR{In this paper, we propose a modulation-format-dependent model, accounting for an enhanced link function, which is valid for non-identical spans where the fiber span loss and SRS gain/loss are not necessarily compensated for by the amplifier at the end of each span. A new signal power profile is also introduced, comprising the frequency-dependent fiber attenuation. The proposed analytical model takes all terms of nonlinear interference (NLI), including self-channel interference, cross-channel interference, and multi-channel interference, into account. It is also shown that the proposed model has the power to predict the NLI power with greater precision than the previous models can.
}
Split-step Fourier simulations indicate that when both SRS and arbitrary modulation formats are considered, previous models may inaccurately predict the NLI power. This difference could be up to 4.3 dB for a 10.011 THz system with 1001 channels at 10 Gbaud. 
\end{abstract}



\section{Introduction}

\IEEEPARstart{T}{he} tremendous growth in the demand for high data rates is gradually leading to a capacity crunch of optical networks operating transmission in the C-band \cite{chralyvy2009}. To cope with this capacity shortage, transmission in the C+L band and beyond is currently seen by the optical communication community as one of the most promising solutions (see e.g.,~\cite{CaiECOCpdp2018, IonescuOFC2019, RenaudierOFC2019}). The most dominant factor currently restricting the capacity of optical fiber transmission systems is the Kerr nonlinearity \cite{Essiambre_2010}, which leads to signal distortion and decreased transmission quality. Although wideband optical transmission provides a clear path to a linear scaling of the system throughput, stronger nonlinear interference is also incurred as the number of channels is increased. Moreover, due to the large optical bandwidth, these systems are significantly affected by the stimulated Raman scattering (SRS) effect, which causes the power profile of the transmitted signal to change as a function of the channel location in the optical spectrum \cite{agrawal_linear_2006}.   

Finding efficient ways to estimate the transmission performance of optical transmission systems in the presence of Kerr and SRS effects is then of key importance for modern optical links. \HR{T}he split-step Fourier method to solve the nonlinear Schr{\"o}dinger equation (NLSE) \HR{is} not a viable option due to the high computational complexity caused by the wide transmission bandwidth considered. On the other hand, many approximated analytical models for nonlinear fiber propagation are currently available in the literature\cite{ CarenaGaussian2012, Mecozzi2012, Pontus_JLT_modeling_2013, dar2013properties, Curri2013, carena2014egn}. All of these models aim to accurately predict the nonlinear interference (NLI) power caused by the Kerr effect, in order to quantify the system transmission performance. This remarkable modelling effort enables NLI power prediction in a wide variety of system scenarios such as multiple optical channels, flexible channel symbol rates and frequency spacing, different modulation formats, different amplification schemes, etc.
Among the models, the Gaussian noise (GN) model \cite[Ch.~13]{kurtzke1995kapazitatsgrenzen},\cite{splett1993ultimate, CarenaGaussian2012, Poggiolini2012, Poggio2014}, and the enhanced Gaussian noise (EGN) model \cite{carena2014egn,poggiolini2015simple} have risen to popularity due to their wide scope of application and availability of relatively accurate closed-form expressions. All of the above models, however, neglect the Kerr-SRS interplay.   

The importance of the effect of SRS on the NLI power has only recently been recognised and modelled in \cite{Roberts2017, semrau2018gaussian, cantono2018modelling, cantono2018interplay, cantono2017introducing}. The SRS models proposed so far mainly extend the scope of the GN model to wideband transmission scenarios. Also, correction terms to include modulation format dependency of the NLI in the Kerr-SRS context have been derived \HR{in \cite{Semrau.2019.MD.jlt,Lasagniecoc,serena2020sims}.}  

In this paper, we propose a\HR{n} analytical model which accurately captures the effect on NLI of the main features of interest for modern wideband optical communication systems. These include: \HR{an enhanced link function which is valid for heterogeneous fiber spans in the presence of SRS, new power profile accounting for the frequency-dependent attenuation}, flexible modulation formats across different wavelength-division multiplexing (WDM) channels,
and finally varying symbol rate. In what follows, we briefly review some of the main models available in the literature. We then explain our contributions.

\subsection{Main NLI Models in the Absence of SRS}
Since the early 2010s a large amount of analytical models based on perturbation methods have been proposed to estimate the effect of the fiber Kerr nonlinearity on the transmission performance. The GN model was derived based on the assumption that the field at the input of the fiber can be modelled as a Gaussian process \cite{Carena2012,Poggiolini2012}. Similar derivations to GN model were also presented in \cite{Pontus_JLT_modeling_2013,serena2013alternative}. One drawback of all the aforementioned GN-based models is that they often significantly overestimate the NLI power due to the assumption that the transmitted signal, after transmission, statistically
behaves as stationary Gaussian noise \HR{\cite[Sec.~II-A]{Poggio2014}}.

The first modulation-format dependent model was introduced in \cite{A.Mecozzi2012, dar2013properties}, using a time-domain perturbational approach. This model only considers cross-phase modulation (XPM) as a dominant nonlinear effect. The advantages of such a model in accurately capturing the effect of the modulation format on the NLI were highlighted in \cite{dar2013properties,dar2014accumulation}. 

Following a similar approach as in \cite{dar2013properties}, the authors of \cite{carena2014egn} derived a new perturbation model (in the frequency domain) dropping the assumption of Gaussianity of the transmitted signal. This model was labelled enhanced Gaussian noise (EGN) model. The EGN model resulted in a number of additional correction terms compared to the GN model formulation, which fully captured the modulation format dependency of the NLI. Moreover, the frequency-domain approach in \cite{carena2014egn} allows the model to fully account for all the different contributions of the NLI in a WDM spectrum, including: the self-channel interference (SCI), and unlike \cite{dar2013properties}, all cross-channel interference (XCI) and multi-channel interference (MCI) terms. In \cite{poggiolini2016analytical} the time domain, GN, and EGN models were compared in subcarrier-multiplexed systems via simulation results, and it was found that both the GN and time-domain model in \cite{A.Mecozzi2012,dar2013properties} failed to accurately predict the NLI falling over the channel of interest (COI), whilst the EGN model was able to capture both the modulation format and the symbol rate dependency of the NLI. \HR{A time-domain extended EGN model including an auto-correlation function of the received NLI was presented in \cite{Serena2015}. The variance of nonlinear signal-signal and signal-noise distortions was derived in \cite{Ghazisaeidi2017}. A statistical model of NLI was proposed in \cite{golani2016modeling}, which is accurately able to assess the bit error rate of optical systems. The GN and EGN models were reviewed in \cite[Ch.~7]{zhou2016enabling}, where a link function for heterogeneous spans was introduced irrespective of the SRS effects.} In Table~\ref{Recent_works}, we show a summary of the GN-like channel models with applicability to a bandwidth regime where SRS can be safely neglected (e.g.~C-band transmission).

\begin{table*}
\begin{center}

\caption{\small{GN-like channel models proposed until 2019. Properties of the model are shown: Modulation Dependent (MD); Frequency Domain (FD), Time Domain (TD); Considered NLI terms such as SCI, XPM, XCI, and MCI; Valid for Gaussian (G) or non-Gaussian (NG) signals; Accounting for SRS.}}
\label{Recent_works}
\renewcommand{\arraystretch}{0.7}
\parbox{1\textwidth}{
\centering
\begin{tabular}{llllllll}
\hline

\hline

\small{\color{black}Year}&
\small{\color{black}Ref.}&
\multicolumn{1}{V{1.5em}}{\color{black}MD?}&
\multicolumn{1}{V{2.5em}}{\color{black}FD/TD}&
\multicolumn{1}{V{5.5em}}{\color{black}SCI, XCI, MCI}&
\multicolumn{1}{V{2.5em}}{\color{black}Signal}&
\multicolumn{1}{V{3.5em}}{\color{black}Accounting for SRS?}&
\multicolumn{1}{V{5.5em}}{\color{black}Remarks}
\\
\hline
\scriptsize{1993}&\scriptsize{\cite{splett1993ultimate}} &\scriptsize{No}&\scriptsize{FD}&\scriptsize{SCI, XCI, MCI}&\scriptsize{G}&\scriptsize{No}&\scriptsize{First GN-type model}\\
\hline
\scriptsize{1995}&\scriptsize{\cite[Ch.~13]{kurtzke1995kapazitatsgrenzen}} &\scriptsize{No}&\scriptsize{FD}&\scriptsize{SCI, XCI, MCI}&\scriptsize{G}&\scriptsize{No}&\scriptsize{First GN-type model}\\
\hline
\scriptsize{}&\scriptsize{\cite{Carena2012}} &\scriptsize{No}&\scriptsize{FD}&\scriptsize{SCI, XCI, MCI}&\scriptsize{G}&\scriptsize{No}&\scriptsize{GN model}\\
\scriptsize{\color{black}2012}&\scriptsize{\cite{A.Mecozzi2012}} &\scriptsize{Yes}&\scriptsize{TD}&\scriptsize{XPM}&\scriptsize{NG}&\scriptsize{No}&\multicolumn{1}{V{15em}}{\scriptsize{Alternative perturbation model for the NLI assuming large accumulated dispersion}}\\
\scriptsize{}&\scriptsize{\cite{Beygi_2012}} &\scriptsize{Yes}&\scriptsize{TD}&\scriptsize{SCI}&\scriptsize{NG}&\scriptsize{No}&\multicolumn{1}{V{15em}}{\scriptsize{Interpreting channel as a white Gaussian channel for high enough symbol rate}}\\
\hline
\scriptsize{}&\scriptsize{\cite{Pontus_JLT_modeling_2013}} &\scriptsize{No}&\scriptsize{FD}&\scriptsize{SCI, XCI, MCI}&\scriptsize{G}&\scriptsize{No}&\multicolumn{1}{V{15em}}{\scriptsize{Inclusion of higher order dispersion and a more formal derivation}}\\
\scriptsize{\color{black}2013}&\scriptsize{\cite{serena2013alternative}} &\scriptsize{No}&\scriptsize{FD}&\scriptsize{SCI, XCI, MCI}&\scriptsize{G}&\scriptsize{No}&\scriptsize{Alternative GN model}\\
\scriptsize{}&\scriptsize{\cite{bononi2012alternative}} &\scriptsize{No}&\scriptsize{FD}&\scriptsize{SCI, XPM}&\scriptsize{G}&\scriptsize{No}&\scriptsize{Alternative GN model}\\
\scriptsize{}&\scriptsize{\cite{dar2013properties}} &\scriptsize{Yes}&\scriptsize{TD}&\scriptsize{XPM}&\scriptsize{NG}&\scriptsize{No}&\scriptsize{Adding a correction term to XPM by comparing \cite{Carena2012,A.Mecozzi2012}}\\
\hline
\scriptsize{\color{black}2014}&\scriptsize{\cite{Pontus2014}} &\scriptsize{Yes}&\scriptsize{FD}&\scriptsize{SCI, XPM}&\scriptsize{G}&\scriptsize{No}&\scriptsize{Valid for flex-grid WDM systems}\\
\scriptsize{}&\scriptsize{\cite{carena2014egn}} &\scriptsize{Yes}&\scriptsize{FD}&\scriptsize{SCI, XCI, MCI}&\scriptsize{NG}&\scriptsize{No}&\scriptsize{EGN model. Adding correction terms to SCI, XCI, and MCI}\\
\hline
\scriptsize{2015}&\scriptsize{\cite{poggiolini2015simple}} &\scriptsize{Yes}&\scriptsize{FD}&\scriptsize{XPM}&\scriptsize{NG}&\scriptsize{No}&\scriptsize{A simple approximate closed-form for XPM}\\
\scriptsize{}&\scriptsize{\cite{Serena2015}} &\scriptsize{Yes}&\scriptsize{TD}&\scriptsize{XPM}&\scriptsize{NG}&\scriptsize{Yes}&\scriptsize{Time-domain version of the EGN model}\\
%
\hline
\scriptsize{}&\scriptsize{\cite{poggiolini2016analytical}} &\scriptsize{Yes}&\scriptsize{TD}&\scriptsize{SCI, XCI, MCI}&\scriptsize{NG}&\scriptsize{No}&\multicolumn{1}{V{15em}}{\scriptsize{Comparing time domain, GN and EGN models in sub-carrier multiplexed systems}}\\
\scriptsize{\color{black}2016}&\scriptsize{\cite{rademacher2016nonlinear}} &\scriptsize{No}&\scriptsize{FD}&\scriptsize{SCI, XCI, MCI}&\scriptsize{G}&\scriptsize{No}&\multicolumn{1}{V{15em}}{\scriptsize{GN model for multimode fiber}}\\
\scriptsize{}&\scriptsize{\cite{Antonelli2016JLT}} &\scriptsize{Yes}&\scriptsize{TD}&\scriptsize{XPM}&\scriptsize{NG}&\scriptsize{No}&\multicolumn{1}{V{15em}}{\scriptsize{Modulation dependent model for multimode fiber}}\\
\scriptsize{}&\scriptsize{\cite{golani2016modeling}} &\scriptsize{Yes}&\scriptsize{TD}&\scriptsize{SCI, XCI, MCI}&\scriptsize{NG}&\scriptsize{No}&\multicolumn{1}{V{15em}}{\scriptsize{Analyzing the bit error rate performance using the second order statistical properties of NLI}}\\
\scriptsize{}&\scriptsize{\cite[Ch.~7]{zhou2016enabling}} &\scriptsize{Yes}&\scriptsize{FD}&\scriptsize{SCI, XCI, MCI}&\scriptsize{NG}&\scriptsize{No}&\multicolumn{1}{V{15em}}{\scriptsize{Proposing a link function for non-identical spans, regardless of the SRS effects}}\\
\hline
\scriptsize{}&\scriptsize{\cite{antonelli2017nonlinear}} &\scriptsize{Yes}&\scriptsize{TD}&\scriptsize{SCI, XCI, MCI}&\scriptsize{NG}&\scriptsize{No}&\scriptsize{A comprehensive model for multimode fiber}\\
\scriptsize{\color{black}2017}&\scriptsize{\cite{Ghazisaeidi2017}} &\scriptsize{Yes}&\scriptsize{TD}&\scriptsize{SCI, XCI, MCI}&\scriptsize{NG}&\scriptsize{No}&\multicolumn{1}{V{15em}}{\scriptsize{Deriving the variance of nonlinear signal-signal and signal-noise interactions}}\\
\scriptsize{}&\scriptsize{\cite{Roberts2017,roberts2018corrections}} &\scriptsize{No}&\scriptsize{FD}&\scriptsize{SCI, XCI, MCI}&\scriptsize{G}&\scriptsize{Yes}&\multicolumn{1}{V{15em}}{\scriptsize{Deriving the first link function in the presence of SRS for identical spans where the span loss and SRS gain/loss are perfectly compensated for by the amplifier at the end of each span}}\\
\hline
\scriptsize{}&\scriptsize{\cite{cantono2018interplay}} &\scriptsize{No}&\scriptsize{FD}&\scriptsize{SCI, XCI, MCI}&\scriptsize{G}&\scriptsize{Yes}&\multicolumn{1}{V{15em}}{\HR{\scriptsize{Deriving the link function for identical spans where the span loss and SRS gain/loss are perfectly compensated for by the amplifier at the end of each span}}}\\
\scriptsize{\color{black}2018}&\scriptsize{\cite{semrau2018gaussian}} &\scriptsize{No}&\scriptsize{FD}&\scriptsize{SCI, XCI, MCI}&\scriptsize{G}&\scriptsize{Yes}&\multicolumn{1}{V{15em}}{\HR{\scriptsize{Deriving the link function for identical spans where the span loss and SRS gain/loss are perfectly compensated for by the amplifier at the end of each span}}}\\
\scriptsize{}&\scriptsize{\cite{Semrauecoc2018}} &\scriptsize{No}&\scriptsize{FD}&\scriptsize{SCI, XCI, MCI}&\scriptsize{G}&\scriptsize{Yes}&\multicolumn{1}{V{15em}}{\scriptsize{Deriving the link function for non-identical spans where the span loss and SRS gain/loss are perfectly compensated for by the 
amplifier at the end of each span}}\\
\hline
\scriptsize{}&\scriptsize{\cite{Semrau2019jlt}} &\scriptsize{No}&\scriptsize{FD}&\scriptsize{SCI, XPM}&\scriptsize{G}&\scriptsize{Yes}&\multicolumn{1}{V{15em}}{\scriptsize{Proposing the simple closed-form approximations for the SPM and XPM terms  using the link function given in \cite{roberts2018corrections,semrau2018gaussian,cantono2018interplay}}}\\
\scriptsize{\color{black}2019}&\scriptsize{\cite{Semrau.2019.MD.jlt,Semrau_Correction}} &\scriptsize{Yes}&\scriptsize{FD}&\scriptsize{SCI, XPM}&\multicolumn{1}{V{4em}}{\scriptsize{SCI (G), XPM (NG)}}&\scriptsize{Yes}&\multicolumn{1}{V{15em}}{\scriptsize{Proposing the simple closed-form approximations for the SPM and XPM terms  using the link function given in \cite{roberts2018corrections,semrau2018gaussian,cantono2018interplay}}}\\
\hline
\scriptsize{2020}&\scriptsize{\cite{serena2020sims}} &\scriptsize{Yes}&\scriptsize{FD}&\scriptsize{SCI, XCI, MCI}&\scriptsize{NG}&\scriptsize{Yes}&\scriptsize{Using a simple closed form for the link function as in \cite{Semrau2019jlt}}\\
\hline
\scriptsize{\bf{This Work}}&\scriptsize{} &\scriptsize{Yes}&\scriptsize{FD}&\scriptsize{SCI, XCI, MCI}&\scriptsize{NG}&\scriptsize{Yes}&\multicolumn{1}{V{15em}}{\scriptsize{Deriving an enhanced link function for heterogeneous fiber spans where the span loss and SRS gain/loss are not fully compensated for by the amplifier at the end of each span, Frequency-dependent attenuation coefficient is also included in the link function
}
}\\
\hline
\end{tabular}
}
\end{center}
\end{table*}

\subsection{GN and EGN Models with SRS}

All of the works discussed in the previous section are based on the assumption that all frequency components attenuate in the same manner. This assumption is no longer satisfied for ultra-wideband transmission systems due to the SRS effect. In this scenario, the power evolution of signal substantially depends on the SRS loss/gain that each frequency component experiences during propagation along a link. In order to include the SRS effect, the conventional NLSE equation that governs pulse propagation in the presence of Kerr nonlinearity needs to be modified to include the Raman term \cite[eq.~(3)]{franccois1991nonlinear}. 

Channel models in the presence of SRS, which stems from the mathematical description in \cite[eq.~(3)]{franccois1991nonlinear}, are also available in the literature  \cite{semrau2017achievable,Roberts2017,roberts2018corrections,semrau2018gaussian,cantono2018interplay}. Such models generalized the approach followed in the GN model derivation to include the effect of SRS. A closed-form expression was presented in \cite{Semrau2017JLT} to compute the NLI power for first- and second-order backward-pumped Raman amplified links. The achievable information rate (AIR) degradation in coherent ultra-wideband systems was studied in \cite{semrau2017achievable}, using a modified GN model in order to simultaneously take into account both SRS and Kerr nonlinearity such that the approximated NLI coefficient \cite{Poggio2014} for each channel was obtained by defining an effective attenuation coefficient. An effective attenuation coefficient for each channel matches the actual effective length of the corresponding channel in the presence of SRS. In \cite{Roberts2017}, the signal power profile was obtained based on the linearity assumption of the attenuation profile in frequency. 
Another derivation of GN model in the presence of SRS was presented in \cite{semrau2018gaussian}, which is capable of taking into account an arbitrary frequency-dependent signal power profile. The model derived in \cite{semrau2018gaussian} is valid for Gaussian-modulated signals such as probabilistically-shaped high-order modulation signals. Very recently, \cite{Semrau.2019.MD.jlt, Semrau_Correction} proposed an approximate GN model for SCI and XPM. The authors of \cite{Semrau.2019.MD.jlt} added a modulation format correction term to XPM, while SCI was computed under a Gaussian assumption. \HR{A link function for non-identical spans was presented in \cite{Semrauecoc2018} no matter the SRS gain/loss at the end of each span. A modulation-format-dependent model was proposed in \cite{serena2020sims}, accounting for all the NLI terms such as SCI, XCI, and MCI}.
A summary of the channel models is given in Table~\ref{Recent_works}. 

\subsection{Contributions of the Proposed Model}

%
\HR{The main contribution of this work is,  on the one hand, the combination of an enhanced link function for multiple different spans with arbitrary amplifier gains and a new signal power profile supplemented by a frequency-dependent attenuation, and on the other hand, the study of all the NLI terms with a new formulation. 
Unlike the previous models addressing ultra-wideband transmission \cite[Ch.~7]{zhou2016enabling}, \cite{Roberts2017,cantono2018interplay,semrau2018gaussian,Semrauecoc2018, serena2020sims}, an enhanced link function\footnote{\HR{During the peer-review process of this paper, \cite{mukherjeespringer} has emerged, in which an improved link function is presented in \cite[Ch.~9, eq.~(46)]{mukherjeespringer}.}} is derived for multiple different spans such that the amplifier gains do not compensate for the span loss and SRS gain/loss. The detailed derivation of this link function in the presence of SRS is given in Appendix~B,
which has not been done before. The link function is able to coherently evaluate the NLI. 
This enhanced link function is in contrast to \cite[Ch.~7, eq.~(A.8)]{zhou2016enabling}, which was derived without regard for the SRS effects.
We provide a new formulation in {\it Theorem}~\ref{main.result} ahead, which streamlines the NLI calculation, not least in partially loaded link where channels can have different bandwidths, by allowing for SCI, XCI, and MCI alike.
We show that
previously published SRS models \cite{cantono2018modelling,semrau2018gaussian,Semrau.2019.MD.jlt, serena2020sims,Lasagniecoc} have not the power to estimate the NLI precisely. 
}

The rest of this paper is organized as follows: in Sec.~\ref{preliminaries} we describe the system model and the SRS phenomenon. The main result of this work is presented in Sec.~\ref{NLIN power}. 
Numerical results are presented in Section.~\ref{numerical results}, where our results are benchmarked against previous models accounting for SRS. Finally, 
Sec.~\ref{conclusion.sec} concludes this paper. 

\subsection{Notation Convention}\label{notation.sec}
We have three delta functions in this paper: $\delta(f)$ is used for the continuous domain, i.e., $\int_{-\infty}^{\infty}\text{d}f\delta (f)=1$, and $\delta_{i,j}$ is used for the discrete domain (Kronecker delta), i.e.,
\begin{equation}\label{delta}
\begin{split}
\delta_{i,j}=\left\{
\begin{array}{rl}
1, & \text{if $i=j$}\\
0, & \text{otherwise}.\\
\end{array}\right.
\end{split}
\end{equation}
In this paper we also use $\bar{\delta}_{i,j}=1-\delta_{i,j}$.
The Fourier transform of $s(t)$ is defined as 
\begin{equation}
S(f)=\int_{-\infty}^{\infty} \text{d}t s(t)\text{exp}(-\imath2\pi ft).
\end{equation}
The imaginary unit is denoted by $\imath$.

Throughout this paper we use $(\cdot)_{\text{x}}$ and $(\cdot)_{\text{y}}$ to represent variables associated to polarizations $\text{x}$ and $\text{y}$, resp. We also use the notation $(\cdot)_{\text{x}/\text{y}}$ to show that a certain expression is valid for both $\text{x}$ and $\text{y}$ polarizations. Expectations are denoted by $\mathbb{E}\{\cdot\}$, and two-dimensional complex (time- and frequency-domain) functions are denoted using boldface symbols.

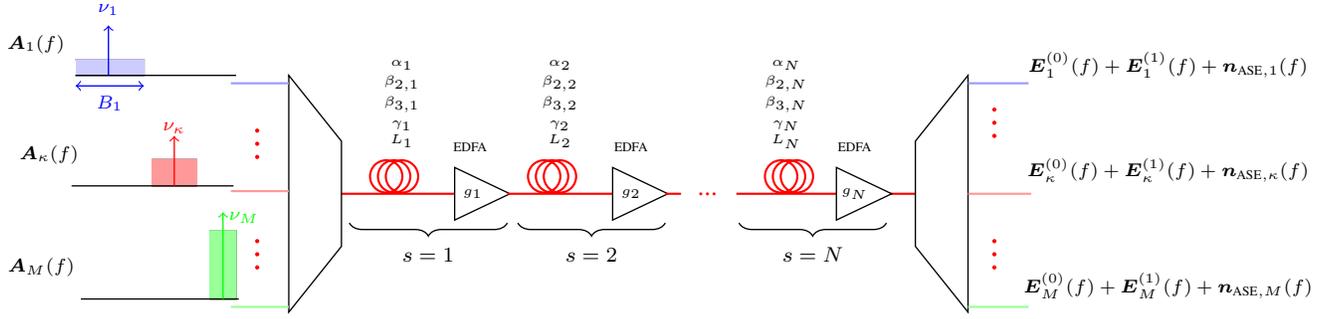
\begin{figure*}
\begin{center}
 \begin{tikzpicture}[scale=0.7, line width=0.5pt]
\node at ( 4.9+1.4,1.3){\tiny{EDFA}};
\node at ( 8.5+0.85,1.3){\tiny{EDFA}};
\node at ( 16.3-2.7,1.3){\tiny{EDFA}};
\node at (21-1-2.75+3-0.7,0.85){\scriptsize{${\boldsymbol{E}}^{(0)}_{\kappa}(f)+\boldsymbol{E}_{\kappa}^{(1)}(f)+{\boldsymbol{n}}_{\text{\tiny{ASE}},\kappa}(f)$}};
\node at (21-1-2.75+3-0.7,0.85+2){\scriptsize{${\boldsymbol{E}}^{(0)}_{1}(f)+\boldsymbol{E}_{1}^{(1)}(f)+{\boldsymbol{n}}_{\text{\tiny{ASE}},1}(f)$}};
\node at (21-1-2.75+3-0.7,0.85-2.2){\scriptsize{${\boldsymbol{E}}^{(0)}_M(f)+\boldsymbol{E}_{M}^{(1)}(f)+{\boldsymbol{n}}_{\text{\tiny{ASE}},M}(f)$}};
\draw [red, line width=0.8pt](0+2+1.85,0.4) -- (4.5+1.5,0.4);
\node at ( 2-0.05+1+1.5+0.55,0.73) [inner sep=0pt,minimum size=4.3mm,draw=red,line width=0.8pt,circle] (2){};
\node at ( 2+0.3+1+1+0.55,0.73) [inner sep=0pt,minimum size=4.3mm,draw=red,line width=0.8pt,circle] (2){};
\node at ( 2+0.05+0.6+1+0.5+0.55,0.73) [inner sep=0pt,minimum size=4.3mm,draw=red,line width=0.8pt,circle] (2){};
\draw (4.5+1.5,-0.1) -- (4.5+1.5,0.9) -- (5.3+1.75,0.4) -- cycle;
\draw [red, line width=0.8pt] (5.5+1.5,0.4) -- (9-0.8+1.5,0.4);
\node at ( 2-0.05+1+1.5+0.55+3,0.73) [inner sep=0pt,minimum size=4.3mm,draw=red,line width=0.8pt,circle] (2){};
\node at ( 2+0.3+1+1+0.55+3,0.73) [inner sep=0pt,minimum size=4.3mm,draw=red,line width=0.8pt,circle] (2){};
\node at ( 2+0.05+0.6+1+0.5+0.55+3,0.73) [inner sep=0pt,minimum size=4.3mm,draw=red,line width=0.8pt,circle] (2){};
\draw [fill=white]  (9-1+1,-0.1) -- (9-1+1,0.9) -- (9.8-1+1.25,0.4) -- cycle;
\draw [red, line width=0.8pt](9.8-0.8+1,0.4) -- (9.8-1+1.5,0.4) node at (9.8-1+2,0.4) {...} ;
\draw [red, line width=0.8pt] (13+0.6-2.25,0.4) -- (17-2.25,0.4);
\node at ( 2-0.05+1+1.5+0.55+7.5,0.73) [inner sep=0pt,minimum size=4.3mm,draw=red,line width=0.8pt,circle] (2){};
\node at ( 2+0.3+1+1+0.55+7.5,0.73) [inner sep=0pt,minimum size=4.3mm,draw=red,line width=0.8pt,circle] (2){};
\node at ( 2+0.05+0.6+1+0.5+0.55+7.5,0.73) [inner sep=0pt,minimum size=4.3mm,draw=red,line width=0.8pt,circle] (2){};
\draw [fill=white]  (17-1-2.75,-0.1) -- (17-1-2.75,0.9) -- (17.8-1-2.5,0.4) -- cycle;
\draw(3.85,-1+0.4)--(2.85,-1.5-0.75+0.4)--(2.85,1.5+0.75+0.4)--(3.85,1+0.4)--cycle;
\draw  (18.5-1-2.75,-1+0.4)--(18.5-1-2.75+1,-1.5-0.75+0.4)--(18.5-1-2.75+1,1.5+0.75+0.4)--(18.5-1-2.75,1+0.4)--cycle;
\draw  [red!40, line width=0.8pt](18.5-1-2.75+1,0.4) -- (18.5-1-2.75+2+0.2,0.4);
\draw [postaction={decorate,draw=blue!20,fill=blue!20}]
(-0.6-0.6,0+2.65) -- (0.1,0+2.65)--(0.1,0.3+2.65)--(-0.6-0.6,0.3+2.65)   node at (-1.75-0.2,2.75+0.5) {\scriptsize{${\boldsymbol{A}}_{1}(f)$}};
\draw [black] (-0.6-0.6,0+2.65) -- (-0.6+3-0.55,0+2.65) ;
\draw [->,blue] (-0.85/2-0.35+0.2,0+2.65) -- (-0.85/2-0.35+0.2,0+2.65+0.95) node at (-0.85/2-0.35+0.2,0+2.65+1.25) {\scriptsize{$\nu_1$}} ;
\draw [<->,blue] (-0.6-0.6,0+2.45) -- (0-0.35+0.45,0+2.45) node [sloped,midway,below] {\scriptsize{$B_{1}$}} ;
\draw [postaction={decorate,draw=red!40,fill=red!40}]
(-0.2-1.15+1.1+0.5,0.55) -- (-0.2-1.15+0.85+1.1+0.5,0.55) --(-0.2-1.15+0.85+1.1+0.5,1.05)--(-0.2-1.15+1.1+0.5,1.05) node at (-1.5-1.5+1+0.3,0.3+0.35+0.5) {\scriptsize{${\boldsymbol{A}}_\kappa(f)$}};
\draw [black] (-0.975-0.25-1.15+1.1,0.55) -- (-0.6+3-0.55-1.15+1.1,0.55) ;
\draw [->,red] (-0.2-1.15+0.85/2+1.1+0.5,0+2.25-2.25+0.55) -- (-0.2-1.15+0.85/2+1.1+0.5,0+2.25-2.25+0.55+0.95) node at (-0.2-1.15+0.85/2+1.1+0.5,0+2+1.2-2.25+0.55+0.15) {\scriptsize{$\nu_{\kappa}$}} ;
\draw [postaction={decorate,draw=green!40,fill=green!40}]
(2.35-1,0-1.7+0.1) -- (2.6-0.75,0-1.7+0.1)-- (2.6-0.75,-1+0.7)-- (2.35-1,-1+0.7) node at (-1.85,-1.7+0.2+0.5) {\scriptsize{${\boldsymbol{A}}_M(f)$}};
\draw [black] (-0.6-0.5,0-1.7+0.1) -- (-0.6+3-0.5,0-1.7+0.1) ;
\draw [->,green] (1.725+0.3-0.85/2,0+2.25-2.25-1.7+0.1) -- (1.725+0.3-0.85/2,0+2.25-2.25-1.7+0.1+0.95+0.7) node at (1.725+0.3-0.85/2+0.4,0+2.25+1.25-2.25-1.7+0.4) {\scriptsize{$\nu_{M}$}} ;

\fill[red] (-1-0.5+3.75,1.5-2) circle (0.4mm);
\fill[red] (-1-0.5+3.75,1.25-2) circle (0.4mm);
\fill[red] (-1-0.5+3.75,1-2) circle (0.4mm);

\fill[red] (-1-0.5+3.75+14,1.5-2) circle (0.4mm);
\fill[red] (-1-0.5+3.75+14,1.25-2) circle (0.4mm);
\fill[red] (-1-0.5+3.75+14,1-2) circle (0.4mm);

\fill[red] (-1-0.5+3.75+14,2) circle (0.4mm);
\fill[red] (-1-0.5+3.75+14,1.75) circle (0.4mm);
\fill[red] (-1-0.5+3.75+14,1.5) circle (0.4mm);

\fill[red] (-1-0.5+3.75,1.5+0.1) circle (0.4mm);
\fill[red] (-1-0.5+3.75,1.25+0.1) circle (0.4mm);
\fill[red] (-1-0.5+3.75,1+0.1) circle (0.4mm);



\draw[blue!40, line width=0.8pt](2.85,-1.75+4.25) --(1.75,-1.75+4.25);
\draw[red!40, line width=0.8pt](2.85,0.45) --(1.75,0.45);
\draw[green!40, line
width=0.8pt](2.85,-1.75) --(1.75,-1.75);
\draw[green!40, line
width=0.8pt](2.9+14,-1.75) --(1.75+14,-1.75);
\draw[blue!40, line width=0.8pt](2.9+14,-1.75+4.25) --(1.75+14,-1.75+4.25);
\draw [decorate,decoration={brace,amplitude=5pt,mirror,raise=4ex}]
  (4,0.75) -- (7,0.75) node[midway,yshift=-3em]{\footnotesize$s=1$};
  \draw [decorate,decoration={brace,amplitude=5pt,mirror,raise=4ex}]
  (4+3.2,0.75) -- (6.9+3.1,0.75) node[midway,yshift=-3em]{\footnotesize$s=2$};
    \draw [decorate,decoration={brace,amplitude=5pt,mirror,raise=4ex}]
  (4+3.2+4.2,0.75) -- (6.9+3.1+4.2,0.75) node[midway,yshift=-3em]{\footnotesize$s=N$};
  \node at  (5,2.75+0.1) {\tiny$\alpha_1$};
  \node at  (5,2.4+0.1) {\tiny$\beta_{2,1}$};
  \node at  (5,2+0.1) {\tiny$\beta_{3,1}$};
  \node at  (5,1.6+0.1) {\tiny$\gamma_1$};
  \node at  (5,1.3+0.1) {\tiny$L_1$};
    \node at  (5+3,2.75+0.1) {\tiny$\alpha_2$};
  \node at  (5+3,2.4+0.1) {\tiny$\beta_{2,2}$};
  \node at  (5+3,2+0.1) {\tiny$\beta_{3,2}$};
  \node at  (5+3,1.6+0.1) {\tiny$\gamma_2$};
  \node at  (5+3,1.3+0.1) {\tiny$L_2$};
   \node at  (5+7.3,2.75+0.1) {\tiny$\alpha_N$};
  \node at  (5+7.3,2.4+0.1) {\tiny$\beta_{2,N}$};
  \node at  (5+7.3,2+0.1) {\tiny$\beta_{3,N}$};
  \node at  (5+7.3,1.6+0.1) {\tiny$\gamma_N$};
  \node at  (5+7.3,1.3+0.1) {\tiny$L_N$};
  \node at  (6.35,0.4) {\tiny$g_1$};
  \node at  (3+6.35,0.4) {\tiny$g_2$};
  \node at  (6+7.6,0.4) {\tiny$g_N$};
\end{tikzpicture}
\caption{Fiber-optic system studied in this work: an ideal WDM multiplexing is used to transmit $M$ optical channels, followed by an optical link comprising $N$ different spans with span lengths $L_{s}$ and EDFAs with gains \HR{$g_{s}(f)$} at the end of each span. At the receiver, ideal demultiplexing and down-conversion are performed. The output for channel $\kappa$ includes a linear component ${\boldsymbol{E}}^{(0)}_\kappa(f)+{\boldsymbol{n}}_{\text{\tiny{ASE}},\kappa}(f)$ and a nonlinear component ${\boldsymbol{E}}^{(1)}_\kappa(f)$, where ${\boldsymbol{n}}_{\text{\tiny{ASE}},\kappa}(f)$
 is the ASE noise, originated from the EDFAs along the fiber. \HR{The links are dispersion uncompensated.}}
\label{Multi_span}
\end{center}
\end{figure*}

\section{Preliminaries}\label{preliminaries}
\subsection{System Model}\label{key_results}

We consider multi-channel optical transmission of independent and identically distributed (i.i.d.) random complex symbol sequences
$(b_{\text{x},\kappa,1},b_{\text{x},\kappa,2},\ldots)$ and $(b_{\text{y},\kappa,1},b_{\text{y},\kappa,2},\ldots)$, 
selected from arbitrary dual-polarization (DP) constellations, where $\kappa=1,2,\ldots,M-1,M$ is the channel index. We further assume that the transmitted symbols on polarization x and y are independent of each other. We also assume that different channels across the spectrum can use different modulation formats, and that all formats have zero mean. \HR{Additionally, channels can have different bandwidths across the spectrum.}

The low-pass equivalent of the DP transmitted signal is denoted by ${\boldsymbol{a}}_\kappa(t)=({a}_{\text{x},\kappa}(t),{a}_{\text{y},\kappa}(t))$, which is assumed to be periodic with an arbitrarily large signal period $T_0$ \HR{\cite[eq.~(30)]{Carena2012}, \cite[eq.~(3)]{carena2014egn}}, i.e.,
\eqlab{ak}{
{{\boldsymbol{a}}}_{\kappa}(t)=\text{e}^{\imath 2\pi \HR{\nu_{\kappa}} t}\sum_{n=-\infty}^{\infty}{{\boldsymbol{p}}}_{\kappa}(t-n T_0), 
}
where \HR{$\nu_{\kappa}$} is the center frequency of channel ${\kappa}$ with bandwidth of \HR{$B_{\kappa}$} and $T_0=WT_s$. The signal ${{\boldsymbol{p}}}_{\kappa}(t)=(p_{\text{x},\kappa}(t), p_{\text{y},\kappa}(t))$ consists of $W$ symbols,
 where \HR{\cite[eq.~(31)]{Carena2012}} 
\eqlab{PeriodicSignal_x}
{
p_{\text{x}/\text{y},\kappa}(t)=\sum_{w=1}^{{\scriptsize W}}{b_{\text{x}/\text{y},\kappa,w}}s_{\kappa}(t-wT_s),
}
in which $s_{\kappa}(t)$ is the pulse used by channel $\kappa$. As discussed in \cite[Sec.~II-B]{CarenaGaussian2012}, the assumption of a periodic signal results in no loss of generality, as an aperiodic signal can be seen as the limit of a periodic signal for its period tending to infinity. 

The Fourier transform of the signal $\boldsymbol{{a}}_{\kappa}(t)$ in \eqref{ak}, denoted by ${\boldsymbol{{A}}}_{\kappa}(f)=(A_{\text{x},\kappa}(f),A_{\text{y},\kappa}(f))$, can be expressed as \HR{\cite[eqs.~(35) and (36)]{Carena2012}}
\eqlab{FouierSeries}
{
{\boldsymbol{{A}}}_{\kappa}(f)=\sqrt{f_0}\sum_{n=-\infty}^{\infty}{\boldsymbol\xi}_{{\kappa},n}\delta(f-nf_0-\HR{\nu_\kappa}),
}
where $f_0 = 1/T_0$ and ${\boldsymbol\xi}_{{\kappa},n}=({\xi}_{{\text{x},\kappa},n},{\xi}_{\text{x},{\kappa},n})$ in which
\eqlab{FourierSerisCoef1}
{
&{\xi}_{{\text{x/y},\kappa},n}=\sqrt{f_0}S_{\kappa}(n f_0)\sum_{w=1}^{W}{{b}}_{\text{x/y},{\kappa},w}\text{e}^{-\imath\frac{2\pi}{W}nw}
}
are the Fourier series coefficients of ${{{a}}}_{\text{x},/\text{y},\kappa}(t)$, and $S_{\kappa}(f)$ is the Fourier transform of $s_{\kappa}(t)$, which has \HR{a near rectangular} shape\footnote{
\HR{For large roll-off factors the excess terms must be taken into consideration \cite[Appendix~B, eqs. (39)-(43)]{golani2016modeling}}.} with amplitude \HR{$1/{B_\kappa}$} and support \HR{${B_\kappa}$} around frequency $f=0$.

The power transmitted over channel $\kappa$ is given by \HR{\cite[eq.~(4)]{carena2014egn}}
\eqlab{Power_channel}{
P_\kappa=\mathbb{E}\{|{b}_{\text{x},\kappa}|^2+|{b}_{\text{y},\kappa}|^2\}=\mathbb{E}\{|{b}_{\text{x},\kappa}|^2\}+\mathbb{E}\{|{b}_{\text{y},\kappa}|^2\},
}
and due to the assumption of identically distributed symbols $b_{x/y,\kappa,w}$,
\eqlab{PowerX/Y}{
\mathbb{E}\{|b_{x,\kappa}|^2\}=\mathbb{E}\{|b_{y,\kappa}|^2\}=\frac{\mathbb{E}\{|b_{\kappa}|^2\}}{2}=\frac{P_\kappa}{2}.
}


In this work, we consider the optical system depicted in Fig.~\ref{Multi_span}. The system includes a wide-band transmitter, where the entire WDM bandwidth is populated with $M$ channels. The fiber-optic link consists of $N$ spans, where each span can have different \HR{frequency-dependent} attenuation coefficients ($\alpha_1(f),\ldots,\alpha_N(f)$), different span lengths ($L_1,\ldots,L_N$), different group velocity dispersion coefficients ($\beta_{2,1},\ldots,\beta_{2,N}$), different third-order dispersion terms ($\beta_{3,1},\ldots,\beta_{3,N}$), and different nonlinear coefficients ($\gamma_1,\ldots,\gamma_N$). Optical \HR{erbium-doped
fiber amplifiers (EDFA)
at the end of each span are assumed to also have different frequency-dependent gains ($g_1(f),\ldots,g_N(f)$).
}
At the receiver, each channel is assumed to be ideally demultiplexed, i.e., filtered and down-converted around the zero frequency. The spectrum of the demultiplexed signal includes a linear ($\boldsymbol{E}^{0}_{\kappa}(f)$) and a nonlinear  ($\boldsymbol{E}^{1}_{\kappa}(f)$) component, both defined in the next section, and an amplified spontaneous emission (ASE) noise component $\boldsymbol{n}_{\text{ASE},\kappa}(f)$ added by the EDFA amplifiers and impinging on channel $\kappa$. As discussed in the next sections, the main goal of the model is to derive the power spectral density (PSD) of $\boldsymbol{E}^{1}_{\kappa}(f)$ in order to compute the NLI power.


\subsection{Nonlinear Propagation}

The propagation of DP signals in an optical fiber is governed by the Manakov equation \cite[Ch.~2]{Book_agrawal}, \cite[eq.~(5)]{cantono2018interplay}, \cite[eq.~(13)]{semrau2018gaussian} which in the frequency domain can be written as
\eqlab{eq1}
{
\frac{\partial}{\partial z} {\boldsymbol{E}}(z,f)={\Gamma}(z,f){\boldsymbol{E}}(z,f)+{\boldsymbol{Q}}(z,f),
}
where
\begin{equation}\label{gamma}
{\Gamma}(z,f)=\frac{g(z,f)}{2}+\imath 2\pi^{2}\beta_2 f^{2}+\imath\frac{4}{3}\pi^3 \beta_3f^3, 
\end{equation}
and
\eqlab{Kerr-term1}
{
{\boldsymbol{Q}}(z,f)=&\imath \gamma\frac{8}{9}\Big[E_{\text{x}}(z,f)*E_{\text{x}}^{\small{*}}(z,-f)\nonumber\\&+E_{\text{y}}(z,f)*E_{\text{y}}^{\scriptsize{*}}(z,-f)\Big]*{\boldsymbol{E}}(z,f).
}
is the ``Kerr term''. 

In \eqref{eq1}, ${\boldsymbol{E}} = \big(E_{\text{x}}, E_{\text{y}}\big)$ is the spectrum of the electrical field of the propagating DP signal. We model the effect of SRS through the generic frequency- and distance-dependent power gain coefficient $g(z,f)$
(see \cite[Appendix]{cantono2018interplay} and \cite[Appendix~A]{semrau2018gaussian}).
 The term ${\boldsymbol{Q}}$ in \eqref{Kerr-term1} is the  DP Kerr-term vector ${\boldsymbol{Q}}= \big(Q_{\small{\text{x}}},Q_{\small{\text{y}}}\big)$, where $*$ stands for convolution and $(\cdot)^{\small{*}}$ denotes the complex conjugate.

An analytical approximation to (\ref{eq1}) can be written as \HR{\cite[eqs.~(46) and (49)]{poggiolini2012detailed}}
\eqlab{MAN_sol}
{
{\boldsymbol{E}}(z,f)\approx{\boldsymbol{E}}^{(0)}(z,f)+{\boldsymbol{E}}^{(1)}(z,f).
}
In \eqref{MAN_sol}, ${\boldsymbol{E}}^{(0)}=\big(E_{\text{x}}^{(0)}, E_{\text{y}}^{(0)}\big)$ is almost the linear solution in the absence of Kerr nonlinearity, which is given by \HR{\cite[eq.~(50)]{poggiolini2012detailed}}
\eqlab{MAN_LIN_sol}
{
{\boldsymbol{E}}^{(0)}(z,f)=\text{e}^{\tilde{\Gamma}(z,f)}{\boldsymbol{E}}(0,f),
}
where 
\eqlab{gamma.tilde.12}{
\tilde{\Gamma}(z,f)=\int_{0}^{z}\text{d}z'{\Gamma}(z',f),
}
and ${\boldsymbol E}(0,f)$ is the spectrum of the electrical field of the DP signal at the input of the fiber-optic link, which can be expressed as
\begin{align}\label{E_input}
{\boldsymbol E}(0,f) & =\sum_{\kappa=1}^{M}{\boldsymbol{A}}_{\kappa}(f)\\
\label{E_input.final}
&=\sum_{\kappa=1}^{M}\sqrt{f_0}\sum_{n=-\infty}^{\infty}{\boldsymbol\xi}_{{\kappa},n}\delta(f-nf_0-\HR{\nu_\kappa}),
\end{align}
where \eqref{E_input.final} follows from \eqref{FouierSeries}.  

To compute the nonlinear solution (perturbative term) $\boldsymbol{E}^{(1)}$ in (\ref{MAN_sol}), we use the well-known perturbation approach (similar to \cite{Pontus_JLT_modeling_2013,bononi2012alternative,Poggiolini2012, carena2014egn}) which gives
\eqlab{F_kerr}
{
{\boldsymbol{Q}}(z,f)=&\imath \gamma\frac{8}{9}\Big[E_{\text{x}}^{(0)}(z,f)*E_{\text{x}}^{(0)\scriptsize{*}}(z,-f)\nonumber\\&+E_{\text{y}}^{(0)}(z,f)*E_{\text{y}}^{(0)\scriptsize{*}}(z,-f)\Big]*{\boldsymbol{E}}^{(0)}(z,f).
}
We then insert (\ref{MAN_sol}) into (\ref{eq1}), and  use (\ref{MAN_LIN_sol}) and (\ref{F_kerr}) to obtain \HR{\cite[eq.~(51)]{poggiolini2012detailed}}
\eqlab{MAN_NLI_sol}
{
{\boldsymbol{E}}^{(1)}(z,f)=\text{e}^{\tilde{\Gamma}(z,f)}\int_{0}^{z}\text{d}z'{\boldsymbol{Q}}(z',f)\text{e}^{-\tilde{\Gamma}(z',f)}.}

\HR{In is noticeable that the model in this paper is not totally consistent with the perturbation approach since  $\boldsymbol{E}^{(0)}$ is not strictly the linear solution of the Manakov equation \eqref{eq1}, as it accounts for the nonlinear Raman part.
However, we follow the approaches proposed in \cite{Roberts2017, cantono2018interplay, semrau2018gaussian} which are based on the perturbation approach.}
The investigation of this problem goes beyond the scope of
this paper and is left to future research.

\section{Stimulated Raman Scattering}\label{srs}
 In optical WDM systems, low wavelength channels act as low power pump channels and provide gain for high wavelength channels, an effect known as SRS. Raman optical amplifiers are built based on this phenomenon. The frequency dependent attenuation coefficient and the coupling between short and long wavelengths which stems from the SRS process result in each frequency component having different power evolutions.  To evaluate the power profile of channel $\kappa$ in a WDM system, the set of coupled ordinary differential equations \cite[eq.~(1)]{Zirngibl
}, \cite[eq.~(1)]{Roberts2017}, \cite[eq.~(6)] 
{semrau2018gaussian},
\begin{equation}\label{SRS_gl}
\begin{split}
\frac{\partial P_{\kappa}}{\partial z}&=-\sum_{i=1}^{\kappa-1}\frac{\nu_i}{\nu_\kappa}g_r(\Delta f)P_\kappa P_i+\sum_{i=\kappa+1}^{M}g_r(\Delta f)P_\kappa P_i\\&
-\HR{\alpha}(\HR{\nu_\kappa})P_\kappa,
 \end{split}
\end{equation}
must be solved for $\kappa = 1,\ldots,M$, where $\Delta f=|\HR{\nu_i-\nu_\kappa}|$, $g_r(\Delta f)$ is the Raman gain spectrum \HR{normalized to the effective core area $A_{\text{eff}}$} (see \cite[Fig.~2]{stolen1973raman} and \cite[Fig.~1]{semrau2017achievable}), \HR{and $\alpha(f)$ is the optical power attenuation}. The first term in the right hand side of (\ref{SRS_gl}) accounts for depletion of channel $\kappa$ by channels whose central frequencies are smaller than $\nu_{\kappa}$, while the second term accounts for depletion of channels with central frequencies above $\nu_{\kappa}$. The factor ${\nu_i}/{\nu_\kappa}$ in the first term of \eqref{SRS_gl} accounts for the energy difference between channels $i$ and $\kappa$. Here, following \cite{Zirngibl}
 we assume this ratio is equal to one, i.e., ${\nu_i}/{\nu_\kappa}\approx 1$.
  \HR{\begin{theorem}
Under the linear assumption of the Raman gain
spectrum and negligible photon conversion loss (${\nu_i}/{\nu_\kappa}\approx 1$), the $M$ coupled equations in \eqref{SRS_gl} can be written as a single integro-differential equation in the continuous domain, which gives \cite[eq.~(2)]{Zirngibl}
\begin{align}\label{integro.differential.equations}
   \frac{\text{d}G(z,f)}{\text{d}z}=-\alpha(f)G(z,f)-C_rG(z,f)\int\text{d}\omega (f-\omega)G(z,\omega),
\end{align}
where $G(z,f)$ is the signal spectrum at distance $z$. The solution to \eqref{integro.differential.equations} can be written as
\begin{align}\label{rho.3}
    G(z,f)=\rho(z,f)G(0,f),
\end{align}
where the signal power profile is expressed as
 \begin{align}\label{general.link.function}
     &\rho(z,f)=\\&\nonumber
     \frac{\int\text{d}\omega'G(0,\omega')\text{e}^{-\alpha(\omega')z}\cdot \text{e}^{-\alpha(f)z-C_rf\int\text{d}\omega G(0,\omega)L_{\text{eff}}(z,\omega)}}{\int\text{d}\omega'G(0,\omega')\text{e}^{-\alpha(\omega')z-C_r\omega'\int\text{d}\omega'' G(0,\omega'')L_{\text{eff}}(z,\omega'')}},
 \end{align}
in which the effective length is
  \eqlab{L_eff.gnrl}{L_{\text{eff}}(z,f)=(1-\text{e}^{-\alpha(f) z})/\alpha(f),
 }
 and $G(0,f)=\sum_{\kappa=1}^{M}P_\kappa S_\kappa(f-\nu_\kappa)$ is the signal spectrum at the input of fiber, and $C_r$ is the slope of the Raman gain spectrum.
 \end{theorem}
 \begin{IEEEproof}
 See Appendix~A
 \end{IEEEproof}
 
  Considering \eqref{MAN_LIN_sol} and \eqref{gamma}, we can express \eqref{general.link.function} as\footnote{By excluding the dispersion terms from \eqref{gamma}, and combining \eqref{MAN_LIN_sol} and \eqref{gamma.tilde.12}, we have $\rho(z,f)=|\boldsymbol{E}^{(0)}(z,f)|^2/P_{\text{tot}}$.} \cite[eq.~(26)]{cantono2018interplay}, \cite[Appendix~A]{semrau2018gaussian}
\begin{equation}\label{rho.1}
\text{e}^{\int_{0}^{z}\text{d}z' g(z',f)}=\rho(z,f).
\end{equation}

 }

\HR{Assuming the negligible attenuation variation, \eqref{general.link.function} reduces to}
\cite[eq.~(9)]{Zirngibl}, \cite[eq.~(8)]{semrau2018gaussian}
\HR{\eqlab{rho}{
\rho(z,f)=\frac{P_{\text{tot}} \text{e}^{-\alpha z-P_{\text{tot}}C_rL_{\text{eff}}(z)f}}{\int \text{d}\omega G(0,\omega)\text{e}^{-P_{\text{tot}}C_rL_{\text{eff}}(z)\omega}},}
}
where
$P_{\text{tot}}$
 is the total launch power within the entire WDM spectrum, and 
 \eqlab{L_eff}{L_{\text{eff}}(z)=(1-\text{e}^{-\alpha z})/\alpha
 }
 is the effective length of each fiber span. Eq.~\eqref{rho} describes the normalized signal power profile of each frequency component.

\begin{table*}[t]
    \footnotesize
    \centering
    \caption{Integral expressions for the terms used in Theorem~\ref{main.result}. The values of $\Upsilon$, $\mu$ and $\varphi$ are given in Tables~\ref{terms.different} and \ref{terms.identical} 
    for nonidentical and identical spans, resp.}
    \label{DEFGH}
    \begin{tabular}{|c|l|}
    \hline
    
    \hline
    Term & Integral Expression \\
    \hline
    
    \hline
$\displaystyle D_\kappa(\kappa_1,\kappa_2,\HR{\kappa_3})$ &  $\displaystyle \frac{16}{27} \HR{B_{\kappa_1}B_{\kappa_2}B_{\kappa_3}}\int_{-\HR{B_{\kappa}/2}}^{\HR {B_{\kappa}/2}}\text{d}f\int_{\HR{-B_{\kappa_1}/2}}^{\HR{B_{\kappa_1}/2}}\text{d}f_1\int_{\HR{-B_{\kappa_2}/2}}^{\HR{B_{\kappa_2}/2}}\text{d}f_2 |{S}_{\HR{\kappa_1}}(f_1)|^{2} |{S}_{\HR{\kappa_2}}({f}_2)|^{2}|{S}_{\HR{\kappa_3}}(f_1+{f}_2-{f}-\HR{\nu_\kappa+\Omega})|^{2}$\\
&\hspace{24em}$\cdot
|\Upsilon(f_1+\HR{\nu_{\kappa_1}},f_2+\HR{\nu_{\kappa_2}},f+\HR{\nu_{\kappa}})|^{2}$
\\
\hline
$\displaystyle E_\kappa(\kappa_1,\kappa_2,\HR{\kappa_3})$ &  $\displaystyle \frac{80}{81}\HR{B_{\kappa_1}B_{\kappa_2}}\int_{\HR{-B_{\kappa}/2}}^{\HR{B_{\kappa}/2}}\text{d}f\int_{\HR{-B_{\kappa_1}/2}}^{\HR{B_{\kappa_1}/2}}\text{d}f_1\int_{\HR{-B_{\kappa_2}/2}}^{\HR{B_{\kappa_2}/2}}\text{d}f_2\int_{\HR{-B_{\kappa_1}/2}}^{\HR{B_{\kappa_1}/2}}\text{d}f'_1 
|S_{\HR{\kappa_2}}(f_2)|^{2}S_{\HR{\kappa_1}}(f_1)S_{\HR{\kappa_1}}^{*}(f'_1) S_{\HR{\kappa_3}}^{*}(f_1+f_2-f-\HR{\nu_\kappa+\Omega})$\\
&\hspace{12em} $\cdot S_{\HR{\kappa_3}}(f'_1+f_2-f-\HR{\nu_\kappa+\Omega})\Upsilon(f_1+\HR{\nu_{\kappa_1}},f_2+\HR{\nu_{\kappa_2}},f+\HR{\nu_{\kappa}})\Upsilon^{*}(f'_1+\HR{\nu_{\kappa_1}},f_2+\HR{\nu_{\kappa_2}},f+\HR{\nu_{\kappa}})$
\\
\hline
$\displaystyle F_\kappa(\kappa_1,\kappa_2,\HR{\kappa_3})$ &
$\displaystyle \frac{16}{81}\HR{B_{\kappa_1}B_{\kappa_3}}\int_{\HR{-B_\kappa/2}}^{\HR{B_\kappa/2}}\text{d}f\int_{\HR{-B_{\kappa_1}/2}}^{\HR{B_{\kappa_1}/2}}\text{d}f_1\int_{\HR{-B_{\kappa_2}/2}}^{\HR{B_{\kappa_2}/2}}\text{d}f_2\int_{\HR{-B_{\kappa_1}/2}}^{\HR{B_{\kappa_1}/2}}\text{d}f'_1 S_{\HR{\kappa_1}}(f_1)S_{\HR{\kappa_2}}(f_2)S_{\HR{\kappa_1}}^{*}(f'_1)  |S_{\HR{\kappa_3}}(f_1+f_2-f-\HR{\nu_\kappa+\Omega})|^{2}$\\
&\hspace{12em} $\cdot
S_{\HR{\kappa_2}}^{*}(f_1+f_2-f_1')\Upsilon(f_1+\HR{\nu_{\kappa_1}},f_2+\HR{\nu_{\kappa_2}},f+\HR{\nu_{\kappa}}) \Upsilon^{*}(f_1'+\HR{\nu_{\kappa_1}},f_1+f_2-f_1'+\HR{\nu_{\kappa_2}},f+\HR{\nu_{\kappa}})$
\\
\hline
$\displaystyle G_\kappa(\kappa_1,\kappa_2,\HR{\kappa_3})$ & 
$\displaystyle \frac{16}{81}\HR{B_{\kappa_1}} \int_{\HR{-B_{\kappa}/2}}^{\HR{B_{\kappa}/2}}\text{d}f\int_{\HR{-B_{\kappa_1}/2}}^{\HR{B_{\kappa_1}/2}}\text{d}f_1\int_{\HR{-B_{\kappa_2}/2}}^{\HR{B_{\kappa_2}/2}}\text{d}f_2\int_{\HR{-B_{\kappa_1}/2}}^{\HR{B_{\kappa_1}/2}}\text{d}f'_1\int_{\HR{-B_{\kappa_2}/2}}^{\HR{B_{\kappa_2}/2}}\text{d}f'_2 S_{\HR{\kappa_1}}(f_1)S_{\HR{\kappa_2}}(f_2)S_{\HR{\kappa_1}}^{*}(f'_1) S_{\HR{\kappa_3}}^{*}(f_1+f_2-f-\HR{\nu_\kappa+\Omega}) $\\
&\hspace{10em} $\cdot S_{\HR{\kappa_2}}^{*}(f'_2) S_{\kappa_3}(f_1'+f'_2-f-\HR{\nu_\kappa+\Omega})\Upsilon(f_1+\HR{\nu_{\kappa_1}},f_2+\HR{\nu_{\kappa_2}},f+\HR{\nu_{\kappa}})\Upsilon^{*}(f_1'+\HR{\nu_{\kappa_1}},f_2'+\HR{\nu_{\kappa_2}},f+\HR{\nu_{\kappa}})$
\\
\hline

\hline
    \end{tabular}
\end{table*}

\section{Key Result: Nonlinear Noise power}\label{NLIN power}

The NLI power on the COI caused by $\boldsymbol{E}_{\kappa}^{(1)}$ is given by
\eqlab{NLI_Power_Channel_Kappa}
{
\sigma^{2}_{\text{NLI},\kappa}=\int_{\HR{\nu_{\kappa}-{B_{\kappa}}/{2}}}^{\HR{\nu_{\kappa}+{B_{\kappa}}/{2}}}\text{d}f G_{\text{NLI},\kappa}(f),
}
where $G_{\text{NLI},\kappa}(f)$ is the PSD of the dual-polarization (DP) nonlinear electrical field of channel $\kappa$ at the input of the receiver. This PSD is
\eqlab{G_x.noBias_Def}
{
&G_{\text{NLI},\kappa}(f)=G_{\text{NLI},\text{x},\kappa}(f)+G_{\text{NLI},\text{y},\kappa}(f)=2G_{\text{NLI},\text{x},\kappa}(f),
}
where we used the fact that the NLI PSD is equal on both polarizations.


The following theorem is the main result of the paper, which gives an analytical expression for the NLI power in \eqrf{NLI_Power_Channel_Kappa}.
\begin{theorem}[Nonidentical Spans]\label{main.result}
The NLI power on channel $\kappa$ in \eqrf{NLI_Power_Channel_Kappa} is given by
\eqlab{EGN_fixed}
{
\sigma_{\tiny{\text{NLI},{\kappa}}}^{2}
=\!\!\!\!\!\!\!\sum_{\HR{\kappa_1,\kappa_2,\kappa_3\in\mathcal{T}_\kappa}}\!\!\!\!\!\!P_{\kappa_1}P_{\kappa_2}P_{\HR{\kappa_3}}
\Big(&\!D_\kappa\!+\!\delta_{\kappa_1,\HR{\kappa_3}}\Phi_{\kappa_1}E_\kappa
\!+\!\delta_{\kappa_1,\kappa_2}\Phi_{\kappa_1}F_\kappa\nonumber\\&+\delta_{\kappa_1,\kappa_2}\delta_{\kappa_2,\HR{\kappa_3}}\Psi_{\kappa_1}G_\kappa\Big),}
where
\begin{align}\label{tk}
\mathcal{T}_{\kappa} =&\HR{\{(\kappa_1,\kappa_2,\kappa_3) \in \{1,\ldots,M\}^3:} \nonumber\\& \HR{-\frac{|\hat{B}-B_{\kappa}|}{2}\leq\Omega-\nu_{\kappa}\leq\frac{|\hat{B}-B_{\kappa}|}{2}\}},
\end{align}
\HR{where $\hat{B}=B_{\kappa_1}+B_{\kappa_2}+B_{\kappa_3}$ and $\Omega=\nu_{\kappa_1}+\nu_{\kappa_2}-\nu_{\kappa_3}$}, and
\begin{align}
\label{excess.kurtosis}\Phi_i &\triangleq\frac{\mathbb{E}\{|{{b}}_{i}|^{4}\}}{\mathbb{E}^{2}\{|{{b}}_{i}|^{2}\}}-2\\\label{Psi}
\Psi_i & \triangleq\frac{\mathbb{E}\{|{{b}}_{i}|^{6}\}}{\mathbb{E}^{3}\{|{{b}}_{i}|^{2}\}}-9\frac{\mathbb{E}\{|{{b}}_{i}|^{4}\}}{\mathbb{E}^{2}\{|{{b}}_{i}|^{2}\}}+12
\end{align}
and the terms $D_\kappa$, $E_\kappa$, $F_\kappa$, and $G_\kappa$ are given in Table~\ref{DEFGH}, where the terms $\Upsilon$, $\mu_s$, $\varphi_s$, and $\rho_s$ are given by the expressions in Table~\ref{terms.different}. 
\end{theorem}
\begin{IEEEproof}
See Appendix~B
\end{IEEEproof}

\begin{table}[t]
    \renewcommand{\arraystretch}{2.0}
    \footnotesize
    \centering
    \caption{Expressions for the terms $\Upsilon$, $\mu_s$ and $\varphi_s$ used in Theorem~\ref{main.result} for different spans, each with fiber parameters $\alpha_s$, $\gamma_s$, $L_s$, $\beta_{2,s}$, and $\beta_{3,s}$. The EDFA at the end of each span has gain $g_s(f)$, not necessarily \HR{compensating for the span attenuation and SRS gain/loss \cite{hashemi2020joint}. The link is dispersion-uncompensated.}}
    \label{terms.different}
    \begin{tabular}{@{}c@{}|@{~}l@{}}
    \hline
    
    \hline
    Term & Expression \\
    \hline
    
    \hline
    $\Upsilon(f_1,f_2,f)$ & 
    $\displaystyle\sum_{s=1}^{N}\gamma_{s}\mu_s(f_1,f_2,f)$\\
    &$\cdot{\text{e}^{\imath 4\pi^{2}(f_1-f)(f_2-f)\sum_{s'=1}^{s-1}(\beta_{2,s'}L_{s'}+\pi(f_1+f_2)\beta_{3,s'}L_{s'})}}$\\&$\cdot\prod_{s'=1}^{s-1} \sqrt{\HR{g_{s'}(f_1)}\rho_{s'}(L_{s'},f_1)}$\\&$\cdot\sqrt{\HR{g_{s'}(f_1-f+f_2)}\rho_{s'}(L_{s'},f_1-f+f_2)}$\\
    &$ \cdot \sqrt{\HR{g_{s'}(f_2)}\rho_{s'}(L_{s'},f_2)}
    \prod_{s'=s}^{N}\sqrt{\HR{g_{s'}(f)}\rho_{s'}(L_{s'},f)}$
    \\
    \hline
    $\mu_s(f_1,f_2,f)$ & 
    $\displaystyle\int_{0}^{L_{s}}\text{d}z'\rho_{s}(z',f_1+f_2-f)\text{e}^{\imath\varphi_s(f_1,f_2,f,z')}$
    \\
    \hline
    $\varphi_s(f_1,f_2,f,z')$ &
    $\displaystyle 4\pi^{2}(f_1-f)(f_2-f)\Big[\beta_{2,s}+\pi\beta_{3,s}(f_1+f_2)\Big]z'$
    \\
    \hline
    $\rho_s(z',f)$ &
    \HR{$\displaystyle \frac{ \text{e}^{-\alpha_s(f)z-C_rf\int\text{d}\omega G(0,\omega)L_{\text{eff}}(z,\omega)}}{\int\text{d}f'G(0,\omega')\text{e}^{-\alpha_s(\omega')z-C_r\omega'\int\text{d}\omega'' G(0,\omega'')L_{\text{eff}}(z,\omega'')}}$}\\&\HR{$\cdot \int\text{d}\omega'G(0,\omega')\text{e}^{-\alpha_s(\omega')z}$}\\
    \hline
    
    \hline
    \end{tabular}
\end{table}

Theorem~\ref{main.result} together with Tables~\ref{DEFGH} and \ref{terms.different} give an expression for the NLI power coherently accumulated along a fiber-optic link with different spans and \HR{no dispersion compensation}, where the loss of each span and \HR{SRS gain/loss} are not necessarily compensated for by the gain of the amplifier at the end of span. \HR{The term $D_\kappa$ in Table~\ref{DEFGH} corresponds to the GN model terms, whereas $E_\kappa$, $F_\kappa$, and $G_\kappa$ are the non-Gausianity correction terms. 

{\it Theorem}~\ref{main.result} has the ability to compute all the NLI terms for partially loaded spectrum where channels can have different bandwidths. In particular,\eqref{EGN_fixed} shows that the three different frequencies $f_1$, $f_2$, $f_3=f_1+f_2-f$ located in channels $\kappa_1$, $\kappa_2$, $\kappa_3$, resp., interact with each other and create an interfering frequency $f$ in channel $\kappa$. The SCI term (the pink lozenge-shaped island in \cite[Fig.~7]{carena2014egn}) is obtained from \eqref{EGN_fixed} when $\kappa_1=\kappa_2=\kappa_3=\kappa$, meaning that $f_1$, $f_2$, $f_3$, and $f$ are in a single channel, namely $\kappa$. The XCI terms are produced when the interfering frequencies $f_1$, $f_2$, $f_3$, and $f$ are located in two different channels. Eq.~\eqref{EGN_fixed} hence has the ability to separate the XCI terms: $\kappa_1=\kappa_3\neq\kappa_2=\kappa$ gives X1 or XPM in \cite[Fig.~7]{carena2014egn},  $\kappa_1=\kappa_3=\kappa\neq\kappa_2$ gives X2 in \cite[Fig.~7]{carena2014egn}, $\kappa_1=\kappa_2=\kappa\neq\kappa_3$ gives X3 in \cite[Fig.~7]{carena2014egn}, and $\kappa_1=\kappa_2=\kappa_3\neq\kappa$ gives X4 in \cite[Fig.~7]{carena2014egn}.
The MCI terms are generated when the interfering frequencies are in three or four different channels. These terms can also be readily obtained from \eqref{EGN_fixed} when $\kappa_1=\kappa_3\neq\kappa_2\neq\kappa$ (corresponding to M1 and M2 in \cite[Fig.~7]{carena2014egn}), $\kappa_1=\kappa_2\neq\kappa_3\neq\kappa$ (corresponding to M3 in \cite[Fig.~7]{carena2014egn}), or $\kappa_1\neq\kappa_3\neq\kappa_2\neq\kappa$ (M0 in \cite[Fig.~7]{carena2014egn}).

The term $\Upsilon(\cdot)$ in Table~\ref{terms.different} is an enhanced link function, which is able to coherently capture the impact of both Kerr nonlinearity and stimulated Raman scattering at the end of the link. The term $\mu_s(\cdot)$ captures the nonlinear disturbance imposed on signal at span $s$. The term $\varphi_s(\cdot)$ is the phase matching coefficient \cite[eq.~(8)]{serena2020sims}, multiplied by $z'$. $\rho_s(\cdot)$ is a new signal power profile defined in \eqref{general.link.function}.}
We note that \eqref{excess.kurtosis} and \eqref{Psi} are defined as in\cite[eq.~(6)]{carena2014egn}. The expressions ${\mathbb{E}\{|{{b}}_{i}|^{4}\}}/{\mathbb{E}^{2}\{|{{b}}_{i}|^{2}\}}$ and ${\mathbb{E}\{|{{b}}_{i}|^{6}\}}/{\mathbb{E}^{3}\{|{{b}}_{i}|^{2}\}}$ are referred to in \cite{dar2014accumulation} as the second and third order modulation factors, resp. The authors of \cite{Semrau.2019.MD.jlt} called the expression in \eqref{excess.kurtosis} the excess kurtosis of the modulation format.

The next corollary shows how Theorem~\ref{main.result} particularizes to the case of multiple \emph{identical} spans where span loss, \HR{SRS gain/loss are fully compensated for by the EDFA at the end of span. 
}
\begin{corollary}[Identical Spans]\label{main.result.identical}
For systems with multiple identical spans, where amplifiers perfectly compensate for the span loss \HR{and SRS gain/loss}, the NLI power is given by \eqref{EGN_fixed} and Table~\ref{DEFGH}, where the terms $\Upsilon$, $\mu$, $\varphi$, and $\rho$ are given by the expressions in Table~\ref{terms.identical}. 
\end{corollary}
\begin{IEEEproof}
See Appendix~C
\end{IEEEproof}


\begin{table}[t]
    \renewcommand{\arraystretch}{2.0}
    \footnotesize
    \centering
   \caption{Expressions for the terms $\Upsilon$, $\mu$ and $\varphi$ used in Corollary~\ref{main.result.identical} for multiple identical spans where spans have the same parameters: $\alpha_s=\alpha$, $L_s=L$, $\gamma_s=\gamma$, $\beta_{2,s}=\beta_2$, and $\beta_{3,s}=\beta_3$ $\forall s\in\{1,2,\ldots,N\}$. We also assume that all the EDFAs should have the same gain and ideally compensate the span loss.}
    \label{terms.identical}
    \begin{tabular}{@{}c@{~}|@{~}l@{}}
    \hline
    
    \hline
    Term & Expression \\
    \hline
    
    \hline
    $\Upsilon(f_1,f_2,f)$ & 
    $\mu(f_1,f_2,f)\frac{1-\text{e}^{4\pi^{2}(f_1-f)(f_2-f)NL[\beta_2+\pi\beta_3(f_1+f_2)]}}{1-\text{e}^{4\pi^{2}(f_1-f)(f_2-f)L[\beta_2+\pi\beta_3(f_1+f_2)]}}
    $
    \\
    \hline
    $\mu(f_1,f_2,f)$ & 
    $\displaystyle \int_{0}^{L}\text{d}z'\rho(z',f_1+f_2-f) \text{e}^{\imath\varphi(f_1,f_2,f,z')} $
    \\
    \hline
    $\varphi(f_1,f_2,f,z')$ &
    $\displaystyle 4\pi^{2}(f_1-f)(f_2-f)\Big[\beta_{2}+\pi\beta_{3}(f_1+f_2)\Big]z'$
    \\
 \hline
    $\rho(z',f)$ &
    \HR{$\displaystyle \frac{ \text{e}^{-\alpha(f)z-C_rf\int\text{d}\nu G(0,\nu)L_{\text{eff}}(z,\nu)}}{\int\text{d}\nu'G(0,\nu')\text{e}^{-\alpha(\nu')z-C_r\nu'\int\text{d}\nu'' G(0,\nu'')L_{\text{eff}}(z,\nu'')}}$}\\&\HR{$\cdot \int\text{d}\nu'G(0,\nu')\text{e}^{-\alpha(\nu')z}$}\\
    \hline
    \end{tabular}
\end{table}


\section{Numerical Results}\label{numerical results}

In this section, our model is numerically validated. A comparison with the previously published models on SRS is also presented. \HR{We assume that the attenuation variation across the spectrum is negligible and thus we use \eqref{rho}, even though {\it{Theorem}}~\ref{main.result} holds for an arbitrary loss profile.} The numerical study was conducted for \HR{two  transmission scenarios, including a single $L=100$~km fiber span and a link with three non-identical spans.} Two different scenarios were considered: i) $B_{\text{tot}}=1$ THz  with an artificially increased Raman gain slope of 1.12 [1/W/km/THz], and ii) $B_{\text{tot}}=10$~THz with a more typical Raman gain slope of 0.028 [1/W/km/THz]. In both cases, the product $C_r P_{\text{tot}} B_{\text{tot}}$ was fixed to 0.089 [1/km], resulting in a power profile gap of 
$\Delta \rho(L)\approx 8.2$ dB between the outermost channels for both scenarios, consistently with the approach employed in \cite[Sec.~III]{semrau2018gaussian}. The system parameters for the two cases are shown in Table~\ref{system_parameters_table}. 

Our model's analytical expressions for \begin{align}\label{eta.1}
\eta_\kappa \triangleq \frac{\sigma_{\text{NLI},\kappa}^2}{P^3},
\end{align}
assuming $P_\kappa = P$ for all $\kappa$, were first evaluated using Monte-Carlo (MC) numerical integration. We numerically evaluated the expression in {\it{Theorem}}~\ref{main.result}.
Split-step Fourier method (SSFM) simulations using the Manakov equation \eqref{eq1}--\eqref{Kerr-term1} were then performed to provide an arbitrarily accurate reference for the \emph{true} $\eta_{\kappa}$ values. The function $g(z,f)$ in \eqref{gamma} is equal to $\text{d}\ln{(\rho(z,f))}/\text{d}z$ according to \eqref{rho.1}, where $\rho(z,f)$ is obtained from \eqref{rho}. The group velocity dispersion and third-order dispersion given in \eqref{gamma} are expressed as
$\beta_2=-D\lambda^2/(2\pi c)$
and
$ \beta_3=(2D+\lambda S)\lambda^3/(2\pi c)^2$,
resp., where $\lambda$ is the operating wavelength, $c$ is the light velocity, and the dispersion coefficient $D$ and dispersion slope $S$ are given in Table~\ref{system_parameters_table}.
\HR{In a single span scenario,} the SSFM results were compared to our model in {\it{Theorem}}~\ref{main.result}, \HR{and the \HR{previous models on SRS} \cite{semrau2018gaussian,cantono2018interplay,Lasagniecoc, serena2020sims} \cite[eq.~(3)]{Semrau_Correction}}\footnote{Erratum \cite{Semrau_Correction} replaced \cite[eq.~(16)]{Semrau.2019.MD.jlt} with the correct expression given in \cite[eq.~(3)]{Semrau_Correction}.}, \HR{under the assumption that the SRS effect is fully compensated at the end of span}. Three modulation formats were investigated: PM-QPSK, PM-16QAM and PM-64QAM. Moreover, the $\eta_{\kappa}$ performance of a polarization-multiplexed, two-dimensional Gaussian-distributed constellation (PM-2D-Gauss) was also studied as a reference case. The results for PM-2D-Gauss were obtained using  {\it{Theorem}}~\ref{main.result}, and \HR{\cite{semrau2018gaussian}, \cite{cantono2018interplay}, \cite{Lasagniecoc, serena2020sims}, \cite[eq.~(3)]{Semrau_Correction}} for Gaussian modulation. We note that, when a Gaussian constellation is selected, our model's expression in {\it{Theorem}}~\ref{main.result} coincides, as expected, to the interchannel SRS-GN model presented in \cite{semrau2018gaussian,cantono2018interplay}.

\begin{table}[t]
\small
\begin{center}
\caption{Values of $\Phi$ and $\Psi$}
\label{tab2}
\begin{tabular}{lll}
\hline

\hline
\small {Format} &\small {$\Phi$} & \small{$\Psi$}\\
\hline 

\hline 
\small {PM-QPSK} &\small {-1} & \small{4}\\
\small {PM-16QAM} &\small {-0.68} & \small{2.08}\\
\small {PM-64QAM} &\small {-0.619} & \small{1.7972}\\
\small {Gaussian} &\small {0} & \small{0}\\
\hline
\end{tabular}
\end{center}
\end{table}

\begin{table}
\small
\begin{center}
    \caption{System Parameters for numerical verification}
    \label{system_parameters_table}
    \begin{tabular}{l|cc}
\hline 

\hline 
    \multirow{2}{*}{\bf{Parameters}} & \multicolumn{2}{c}{\textbf{Values}}\\
    & (a) & (b)\\
\hline 

\hline 
    \multicolumn{1}{l|}{Loss ($\alpha$) [dB/km]}&0.2 & 0.2  \\ \hline
    
     \multicolumn{1}{l|}{Dispersion ($D$) [ps/nm/km]}&  17& 17\\ \hline
     
      \multicolumn{1}{l|}{Dispersion slope ($S$) [ps/nm\textsuperscript{2}/km]}&  0& 0.067\\ \hline
       
         \multicolumn{1}{l|}{Nonlinear coefficient ($\gamma$) [1/W/km]}&  1.2& 1.2\\ \hline
         
          \multicolumn{1}{l|}{Raman gain slope ($C_r$) [1/W/km/THz]}&  1.12& 0.028\\ \hline
           
          \multicolumn{1}{l|}{Span length ($L$) [km]}& 100& 100 \\ \hline  
          
          \multicolumn{1}{l|}{Total launch power ($P_{\text{tot}}$) [dBm]}& 19 & 25 \\ \hline  
          
           
          \multicolumn{1}{l|}{}&  & 10 \\
          
          \multicolumn{1}{l|}{Symbol rate ($R$) [Gbaud]}&  10&  40
          \\ 
          \multicolumn{1}{l|}{}&  & 100
          \\ \hline  
      \multicolumn{1}{l|}{Roll-off factor $[\%]$}&  0.01&0.01\\ \hline
    \multicolumn{1}{l|}{Channel spacing ($R$) [GHz]}&  10.001& $1.0001\cdot R$  \\ \hline
    \multicolumn{1}{l|}{}&   &1001\\ 
    \multicolumn{1}{l|}{Number of channels ($2M+1$)}&  101 & 251\\
    \multicolumn{1}{l|}{}&   &101\\
    
    \hline

  \multicolumn{1}{l|}{} &  &10.011\\ 
    \multicolumn{1}{l|}{Optical bandwidth ($B_{\text{tot}}$) [THz]} & 1.01 &10.041\\ 
      \multicolumn{1}{l|}{} &  &10.101\\ 
  \hline

    \end{tabular}
    \center
    \end{center}
 \end{table}

    
     
       
         
           
           
    



\begin{figure*}[!t]
\pgfplotsset{compat=1.3}
\begin{center}
\begin{tikzpicture}[scale=1]
\begin{groupplot}[group style={
group name=my plots, group size=3 by 2,horizontal sep=18pt,vertical sep=50pt},
width=6.8cm,height=6cm]

\nextgroupplot
[
legend columns=4,
ylabel={$\eta_\kappa \ [\text{dB}(W^{-2})]$},
grid=major,
ymin=28.5,ymax=42.5,
xmin=-500,xmax=500,
xlabel={Channel Frequency [GHz]}, 
legend style={at={(0.00,-0.4)},anchor=south ,font=\footnotesize},
title=PM-2D-Gauss (101$\times$10 Gbaud),
ytick={28,30,...,43},
]

\addplot [black, line width=0.5mm, dash dot] table[x index=0,y index=1]
{Integration-DAN-SER/gauss-SerIntegral-1THz.txt};

\addplot [black, line width=0.5mm,dashed,x filter/.code={\pgfmathparse{(\pgfmathresult-51)*10}\pgfmathresult}]table[x index=0,y index=1] {Dat/gauss-1thz-closedform.txt};

\addplot [black, line width=0.5mm, loosely dotted] table[x index=0,y index=1]
{Integration-DAN-SER/gauss-DanIntegral-1THz.txt};

\addplot [line width=0.5mm, solid,x filter/.code={\pgfmathparse{(\pgfmathresult-51)*10}\pgfmathresult}]table[x index=0,y index=1] {Dat/NLI_GN_101CH_10GHZ_25DBM.dat};

\node at (axis cs:430,41.5){\textbf{(a)}};

\node at (axis cs:-400,29.25) {\small{1 THz}};

\draw[->,>=stealth,thick] (axis cs:200,36.56) -- (axis cs:200,39.77);
\node at (axis cs:120,37.85) {\footnotesize{$3.2$ dB}};

\draw (axis cs:0,40.639) -- (axis cs:500,40.639);
\node [inner sep=1.3,circle,fill=black] at (axis cs:0,40.639){};

\nextgroupplot
[
legend columns=4,
grid=major,
ymin=28.5,ymax=42.5,
xmin=-500,xmax=500,
xlabel={Channel Frequency [GHz]}, 
legend style={at={(0.00,-0.4)},anchor=south ,font=\footnotesize},
title=PM-16QAM (101$\times$10 Gbaud),
ytick={28,30,...,43},
yticklabels={,,}
]

\draw[->,>=stealth,thick] (axis cs:-400,37.87) -- (axis cs:-400,40.0946684976828);
\node at (axis cs:-310,39) {\footnotesize{$2.2$ dB}};

\draw (axis cs:-500,40.639) -- (axis cs:500,40.639);

\addplot [line width=0.5mm, blue,x filter/.code={\pgfmathparse{(\pgfmathresult-51)*10}\pgfmathresult}]table[x index=0,y index=1] {Dat/NLI_16QAM_101CH_10GHZ_25DBM.dat};

\addplot [blue, line width=0.5mm, dash dot] table[x index=0,y index=1]
{Integration-DAN-SER/16qam-SerIntegral-1THz.txt};

\addplot [blue, line width=0.5mm, loosely dotted] table[x index=0,y index=1]
{Integration-DAN-SER/16qam-DanIntegral-1THz.txt};

\addplot [blue,only marks,solid, mark=*,each nth point={3},mark repeat=3,  mark size = 1pt, line width = 0.8, mark options={fill=white,solid,scale=1.5},x filter/.code={\pgfmathparse{(\pgfmathresult-51)*10}\pgfmathresult}]table[x index=0,y index=1] {Dat/ssfm_16qam_srs.dat};

\addplot [blue, line width=0.5mm,dashed,x filter/.code={\pgfmathparse{(\pgfmathresult-51)*10}\pgfmathresult}]table[x index=0,y index=1] {Dat/NLIN_Dan_16qam_closedform.dat};

\draw (axis cs:-200,39.354) -- (axis cs:500,39.354);
\node [inner sep=1.3,circle,fill=black] at (axis cs:-200,39.354){};

\node at (axis cs:430,41.5){\textbf{(b)}};

\node at (axis cs:-400,29.25) {\small{1 THz}};

\nextgroupplot
[
legend columns=6,
grid=major,
ymin=28.5,ymax=42.5,
xmin=-500,xmax=500,
xlabel={Channel Frequency [GHz]}, 
legend style={at={(-0.725,-0.45)},anchor=south ,font=\footnotesize},
title=PM-QPSK (101$\times$10 Gbaud),
ytick={28,30,...,43},
yticklabels={,,}
]
\addlegendimage{empty legend}

\addlegendimage{black, line width=0.5mm}
\addlegendimage{black, line width=0.5mm,dashed}
\addlegendimage{black,only marks,solid, mark=*,mark size = 1pt, line width = 0.8, mark options={fill=white,solid,scale=1.5}}

\addlegendimage{black, line width=0.5mm,dash dot}
\addlegendimage{black, line width=0.5mm,loosely dotted}

\addlegendimage{empty legend}
\addlegendimage{area legend,draw opacity=0, fill=red}
\addlegendimage{area legend,draw opacity=0, fill=blue}
\addlegendimage{area legend,draw opacity=0, fill=black}

\addlegendentry{Line style: }

\addlegendentry{ {\it{Theorem}}~\ref{main.result}}
\addlegendentry{Closed Form \cite{Semrau_Correction}}
\addlegendentry{SSFM Simulations}
\addlegendentry{Integration \cite{serena2020sims,Lasagniecoc} }
\addlegendentry{Integration \cite[int. model]{Semrau.2019.MD.jlt}}

\addlegendentry{Color coding: \ \ \   }
\addlegendentry{ PM-QPSK}
\addlegendentry{ PM-16QAM}
\addlegendentry{PM-2D-Gauss \cite{cantono2018interplay,semrau2018gaussian}}

\addplot [red, line width=0.5mm, solid,x filter/.code={\pgfmathparse{(\pgfmathresult-51)*10}\pgfmathresult}]table[x index=0,y index=1] {Dat/NLI_101CH_10GHZ_25DBM.dat};










\addplot [red, line width=0.5mm, dash dot] table[x index=0,y index=1]
{Integration-DAN-SER/qpsk-SerIntegral-1THz.txt};

\addplot [red, line width=0.5mm, loosely dotted] table[x index=0,y index=1]
{Integration-DAN-SER/qpsk-DanIntegral-1THz.txt};

\draw[->,>=stealth,thick] (axis cs:-200, 37.624) -- (axis cs:-200, 39.354);
\node at (axis cs:-110,38.5) {\footnotesize{$1.7$ dB}};

\draw (axis cs:-200,39.354) -- (axis cs:-500,39.354);
\node [inner sep=1.3,circle,fill=black] at (axis cs:-200,39.354){};

\addplot [red, line width=0.5mm,dashed,x filter/.code={\pgfmathparse{(\pgfmathresult-51)*10}\pgfmathresult}]table[x index=0,y index=1] {Dat/NLIN_DAN_QPSK_CLOSEDFORM.dat};





\addplot [red,only marks,solid, mark=*,each nth point={3},mark repeat=3, mark size = 1pt, line width = 0.8, mark options={fill=white,solid,scale=1.5},x filter/.code={\pgfmathparse{(\pgfmathresult-51)*10}\pgfmathresult}]table[x index=0,y index=1] {Dat/ssfm_qpsk_srs.dat};


\draw (axis cs:0,40.639) -- (axis cs:-500,40.639);
\node [inner sep=1.3,circle,fill=black] at (axis cs:0,40.639){};

\draw[->,>=stealth,thick] (axis cs:0,36.74) -- (axis cs:0,40.639);
\node at (axis cs:90,39.15) {\footnotesize{$3.9$ dB}};

\draw[->,>=stealth,thick] (axis cs:300,31.053500) -- (axis cs:300,35.34);
\node at (axis cs:390,33) {\footnotesize{$4.3$ dB}};

\draw[->,>=stealth,thick] (axis cs:-400,34.58) -- (axis cs:-400,38.38);
\node at (axis cs:-320,36) {\footnotesize{$3.8$ dB}};

\node at (axis cs:-400,29.25) {\small{1 THz}};
\node at (axis cs:430,41.5){\textbf{(c)}};

\end{groupplot}

\end{tikzpicture}
\\
\vspace*{-2pt}
\end{center}
\setlength{\belowcaptionskip}{-12pt}
\caption{The parameter $\eta_\kappa$ as a function of channel number $\kappa$ for 1 THz transmission and after a single $L=100$ km fiber span. Results are shown for transmission with (a), and without (b) SRS, and with a spectrally flat input power profile for PM-QPSK, PM-16QAM, and PM-2D-Gauss constellations. 
SSFM simulations are represented by circles whilst MC integrations of {\it{Theorem}}~\ref{main.result} are represented by solid lines. Dashed lines represent the closed-form expression in \cite[eq.~(3)]{Semrau_Correction}.} 
\label{nlc_fig}
\end{figure*}
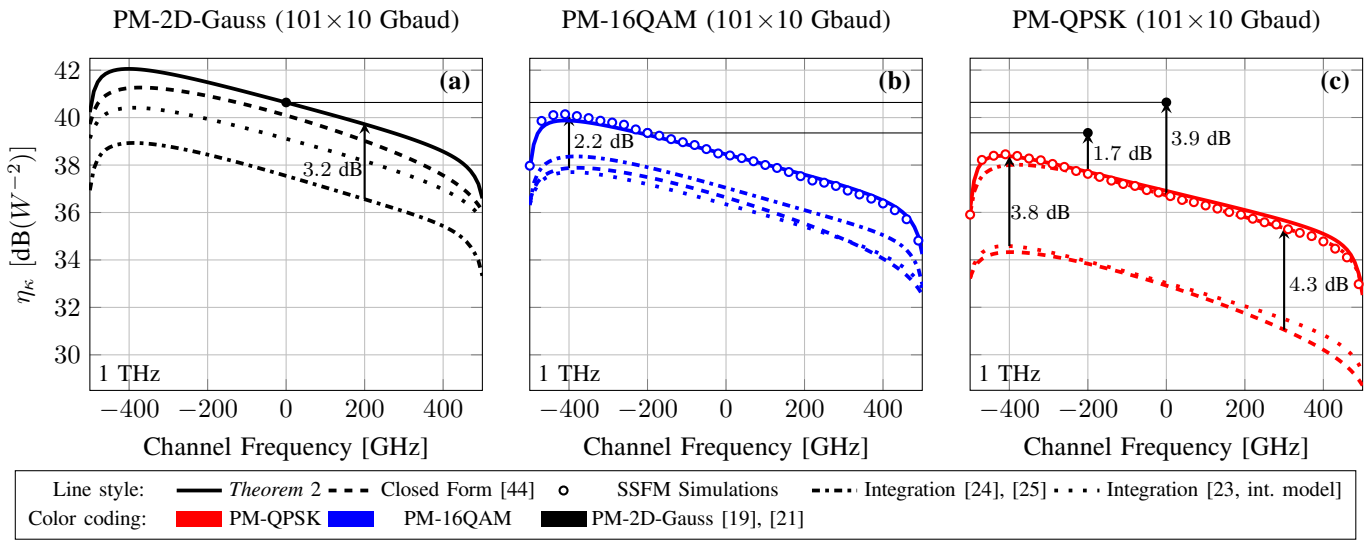

\subsection{Monte-Carlo Numerical Integration}

An analytical expression for $\eta_\kappa$ in the transmission scenario of interest is given by
\eqlab{EGN_fixed_eta}
{
\eta_{\kappa}
= \!\!\!\!\!\!\!\sum_{\kappa_1,\kappa_2,\kappa_3\in\mathcal{T}_\kappa}\!\!\!\!\!\!P_{\kappa_1}P_{\kappa_2}P_{\kappa_3}
\Big(&\!D_\kappa\!+\!\delta_{\kappa_1,\kappa_3}\Phi_{\kappa_1}E_\kappa
\!+\!\delta_{\kappa_1,\kappa_2}\Phi_{\kappa_1}F_\kappa\nonumber\\&+\delta_{\kappa_1,\kappa_2}\delta_{\kappa_2,\kappa_3}\Psi_{\kappa_1}G_\kappa\Big),}
which is obtained by combining \eqref{eta.1} and {\it{Theorem}}~\ref{main.result}. The terms $D_\kappa,E_\kappa,F_\kappa$, and $G_\kappa$, given in \eqref{EGN_fixed_eta}, are expressed in Table~\ref{DEFGH}, and the terms $\Upsilon$, $\mu$, $\varphi$, and $\rho$ are given by the expressions in Table~\ref{terms.different}.
The terms in Table~\ref{DEFGH} are obtained via MC integration, since their integrals are usually not solvable in closed-form. For the integration, we used uniform sampling \HR{as in \cite{dar2014accumulation}}, and increased the number of integration points until convergence was achieved within a $0.05$ dB accuracy. Nonuniform MC numerical integrations are also possible when solving this type of problem \cite{serena2020sims}. As the GN term $D$ represents the most time-consuming integration to perform, it was transformed into hyperbolic coordinates for faster convergence, as shown in \cite{Poggiolini2012}. The remaining EGN terms $E,F$, and $G$ were obtained keeping the  integration in Cartesian coordinates. \HR{Unlike \cite{Semrau.2019.MD.jlt, serena2020sims,Lasagniecoc}, we compute $\mu_s(\cdot)$, given in Table~\ref{terms.different}, by numerical integration over $z$ rather than by a closed-form expression as proposed in \cite[eq.~(14)]{Semrau2019jlt}.}

\subsection{SSFM Numerical Simulations}\label{subsec:SSFMnumsims}
In order to validate the $\eta_{\kappa}$ model estimates, ASE-noise-free ultra-wideband SSFM numerical simulations were performed. In the absence of other noise sources, $\eta_{\kappa}$ can be estimated via the received signal-to-noise ratio (SNR) for each channel $\kappa$, denoted by $\text{SNR}_\kappa$ via the relationship 
\begin{equation}\label{eq:snr}
\eta_\kappa\approx \dfrac{1}{\text{SNR}_\kappa P_{}^2}.
\end{equation} The approximate equality in \eqref{eq:snr} is due to the fact that the SSFM-based SNR estimates also contain perturbation terms higher than the first one. The SNR for a constellation with $N_s$ symbols was computed as
\begin{equation}\label{eq:snr2}
    \text{SNR} = \dfrac{\sum_{i=1}^{{N_s}}|\overline{y}_i|^2}{\sum_{i=1}^{{N_s}}\mathbb{E}\{| Y - \overline{y}_i|^2 \ | \ X = x_i \}},
\end{equation}
where $X$ and $Y$ are the random variables representing the transmitted and received symbols, resp., $x_i$ is the $i$-th constellation point, and $\overline{y}_i= \mathbb{E}\{Y | X = x_i\}$.

As discussed in \cite[Sec.~III-A]{serena2020sims}, the number of transmitted symbols is important to avoid both undesired cyclic effects in the received signal and to guarantee a good accuracy for the $\eta_{\kappa}$ estimation. Transmitting sequences of $2^{16}$ symbols on each channel and discarding the first 500 and last 500 symbols from the transmitted and received sequences was found sufficient to achieve convergence of the $\eta_{\kappa}$ values within 0.1 dB.

The SSFM simulations were performed using an adaptive-step approach, where the maximum instantaneous nonlinear phase rotation per step was fixed \cite{Sinkin2003}. The maximum nonlinear phase rotation per step was fixed, in our simulations to $10^{-5}$ [rad]. This value proved to be low enough to obtain estimates of the NLI power with an accuracy within 0.1 dB for both 1 THz and 10 THz transmission scenarios.


 The received signal was ideally compensated for chromatic dispersion. After dispersion compensation, matched filtering and symbol-rate sampling was applied. The resulting received symbols were used to calculate the SNR. Since the system is symmetric with respect to x and y polarizations, both the polarization channels were used for the SNR estimation.

\subsection{1 THz Results}\label{subsec:1THz_results}



 In the 1 THz scenario, $101$ WDM channels with symbol rate $R=10$ Gbaud are transmitted, each spaced by $10.001$ GHz, as shown in Table~\ref{system_parameters_table}. Each channel was shaped by an ideal root-raised-cosine with $0.01\%$ roll-off factor. The launch power was set to $-1$ dBm per channel, yielding a total launch power of $19$ dBm. 


Fig.~\ref{nlc_fig} shows $\eta_\kappa$ in $\text{dB}(W^{-2})=10\log_{10}(\eta_{\kappa}\cdot 1W^2)$ as a function of channel number 
for systems without (a) and with (b) SRS for PM-QPSK (red), PM-16QAM (blue), and PM-2D-Gauss (black) modulation formats. The results presented in solid line were obtained using our model in {\it{Theorem}}~\ref{main.result}, \HR{which are compared to  \cite{semrau2018gaussian,cantono2018interplay, Semrau.2019.MD.jlt,Lasagniecoc, serena2020sims}, \cite[eq.~(3)]{Semrau_Correction}.} SSFM simulation results (circles) are presented for PM-QPSK and PM-16QAM.

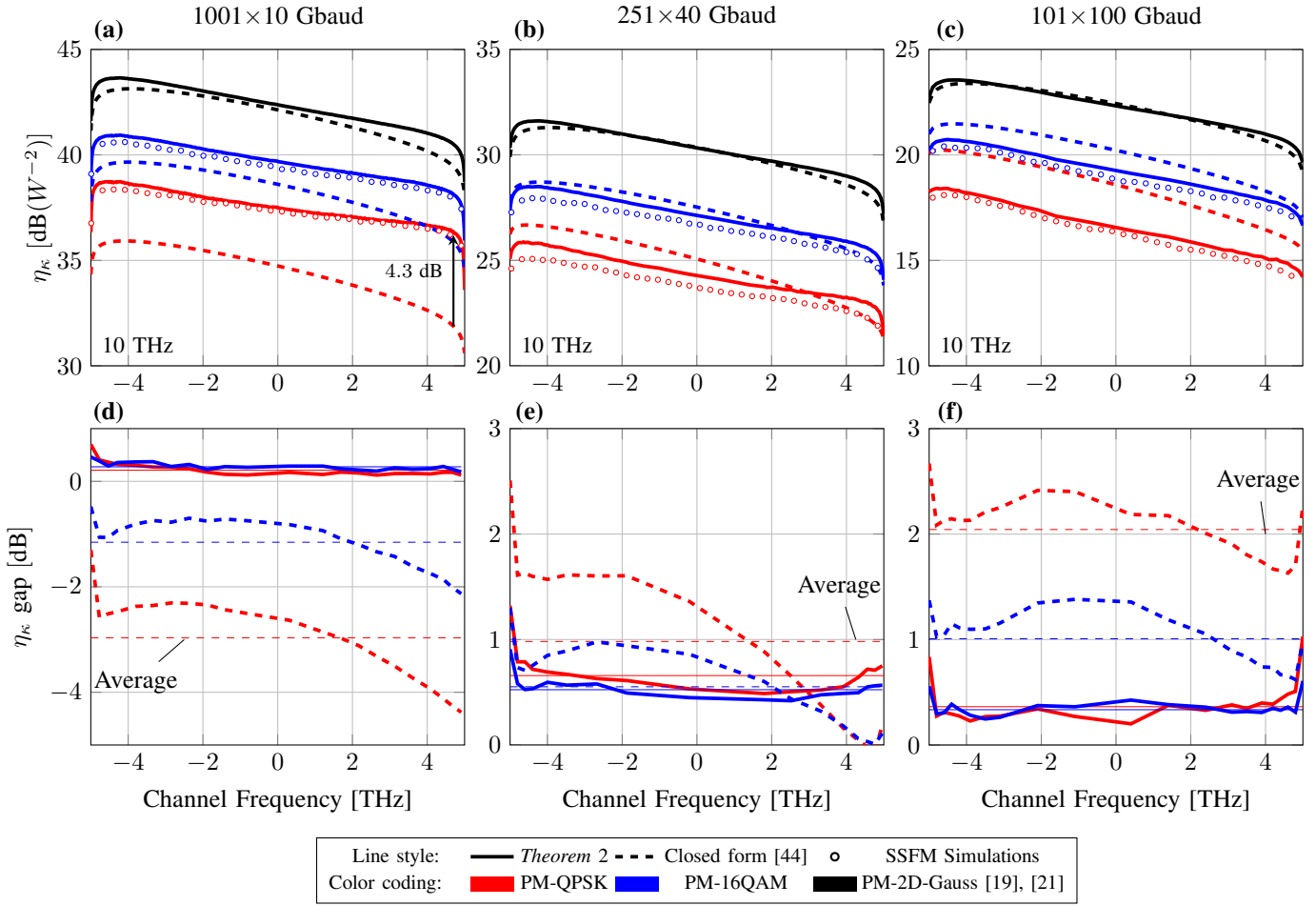
\begin{figure*}[!t]
\pgfplotsset{compat=1.3}
\begin{center}

\begin{tikzpicture}[scale=1]
\begin{groupplot}[group style={
group name=my plots, group size=3 by 2,horizontal sep=18pt,vertical sep=25pt},
width=6.8cm,height=6cm]

\def\ysh{5.325}
\def\xsh{5.875}
\def\xin{0.25}
\def\yin{4.675}
\node at (\xin,\yin) {\textbf{(a)}};
\node at (\xin+\xsh,\yin){\textbf{(b)}};
\node at (\xin+\xsh+\xsh,\yin){\textbf{(c)}};
\node at (\xin,\yin-\ysh) {\textbf{(d)}};
\node at (\xin+\xsh,\yin-\ysh){\textbf{(e)}};
\node at (\xin+\xsh+\xsh,\yin-\ysh){\textbf{(f)}};

\nextgroupplot 
[
legend columns=1,
ylabel={$\eta_\kappa \ [\text{dB}(W^{-2})]$},
ylabel shift=-5pt,
grid=major,
ymin=30,ymax=45, 
xmin=-5,xmax=5,
legend style={at={(0.002,0.215)},anchor=west,font=\footnotesize},
title={1001$\times$10 Gbaud},
]

\addplot [black,line width=0.5mm, x filter/.code={\pgfmathparse{#1/1000}\pgfmathresult}] table[x index=0,y index=1]
{10T-NI/10G/gauss-10G.dat};

\addplot [red, line width=0.5mm,x filter/.code={\pgfmathparse{#1/1000}\pgfmathresult}] table[x index=0,y index=1]
{10T-NI/10G/qpsk-10G.dat};

\addplot [red, line width=0.5mm,dashed,x filter/.code={\pgfmathparse{#1/1000}\pgfmathresult}] table[x index=0,y index=1]
{10T-NI/10G/qpsk-danielCF-10G.dat};

\addplot [red, only marks, mark=*, mark size=1pt, mark options={fill=white},mark repeat=30,x filter/.code={\pgfmathparse{#1/1000}\pgfmathresult}] table[x index=0,y index=1]{Dat/SSFM_New/ISRS_4QAM_Nch=1001x10GBaud_1x100km_Ptot=25dBm_PhiNL=0.05mmrad_Eta_vs_Ch.dat};

\addplot [blue, line width=0.5mm, x filter/.code={\pgfmathparse{#1/1000}\pgfmathresult}] table[x index=0,y index=1]
{10T-NI/10G/16qam-10G.dat};

\addplot [black, line width=0.5mm, dashed, x filter/.code={\pgfmathparse{#1/1000}\pgfmathresult}] table[x index=0,y index=1]
{10T-NI/10G/gauss-danielCF-10G.dat};




\addplot [blue, line width=0.5mm,dashed,x filter/.code={\pgfmathparse{#1/1000}\pgfmathresult}] table[x index=0,y index=1]
{10T-NI/10G/16qam-danielCF-10G.dat};

\addplot [blue, only marks, mark=*, mark size=1pt, mark options={fill=white},mark repeat=30,x filter/.code={\pgfmathparse{#1/1000}\pgfmathresult}] table[x index=0,y index=1]{Dat/SSFM_New/ISRS_16QAM_Nch=1001x10GBaud_1x100km_Ptot=25dBm_PhiNL=0.05mmrad_Eta_vs_Ch.dat};


\draw[->,>=stealth,thick] (axis cs:4.7,31.88) -- (axis cs:4.7,36.2);
\node at (axis cs:3.65,34.5) {\footnotesize{$4.3$ dB}};

\node at (axis cs:-3.75,31) {\small{10 THz}};

\nextgroupplot
[
legend columns=4,
legend entries={ISRS GN Model, PM-16QAM ,PM-QPSK ({\it{Theorem}}~\ref{main.result})},
grid=major,
ymin=20,ymax=35, 
xmin=-5,xmax=5,
legend style={at={(0.50,-1.7)},anchor=south,font=\footnotesize},
title={251$\times$40 Gbaud},
]


\addlegendimage{empty legend}

\addlegendimage{black, line width=0.5mm}
\addlegendimage{black, line width=0.5mm,dashed}
\addlegendimage{black,only marks,solid, mark=*,mark size = 1pt, line width = 0.8, mark options={fill=white,solid,scale=1.5}}
\addlegendimage{empty legend}
\addlegendimage{area legend,draw opacity=0, fill=red}
\addlegendimage{area legend,draw opacity=0, fill=blue}
\addlegendimage{area legend,draw opacity=0, fill=black}

\addlegendentry{Line style: }

\addlegendentry{ {\it{Theorem}}~\ref{main.result}}
\addlegendentry{ Closed form \cite{Semrau_Correction}}
\addlegendentry{SSFM Simulations}

\addlegendentry{Color coding: \ \ \   }

\addlegendentry{ PM-QPSK}
\addlegendentry{ PM-16QAM}
\addlegendentry{ PM-2D-Gauss \cite{cantono2018interplay,semrau2018gaussian}}

\addplot [black, line width=0.5mm,x filter/.code={\pgfmathparse{\pgfmathresult/1000}\pgfmathresult}] table[x index=0,y index=1]
{10T-NI/40G/gauss-40G.dat};

\addplot [blue, line width=0.5mm,x filter/.code={\pgfmathparse{\pgfmathresult/1000}\pgfmathresult}] table[x index=0,y index=1]
{10T-NI/40G/16qam-40G.dat};

\addplot [red, line width=0.5mm,x filter/.code={\pgfmathparse{\pgfmathresult/1000}\pgfmathresult}] table[x index=0,y index=1]
{10T-NI/40G/qpsk-40G.txt};




\addplot [red, line width=0.5mm,dashed,x filter/.code={\pgfmathparse{\pgfmathresult/1000}\pgfmathresult}] table[x index=0,y index=1]
{10T-NI/40G/qpsk-danielCF-40G.dat};

\addplot [red, only marks, mark=*, mark size=1pt, mark options={fill=white},mark repeat=7,x filter/.code={\pgfmathparse{\pgfmathresult/1000}\pgfmathresult}] table[x index=0,y index=1]
{Dat/SSFM_New/ISRS_4QAM_Nch=251x40GBaud_1x100km_Ptot=25dBm_PhiNL=0.01mmrad_Eta_vs_Ch.dat};

\addplot [black, line width=0.5mm,dashed,x filter/.code={\pgfmathparse{\pgfmathresult/1000}\pgfmathresult}] table[x index=0,y index=1]
{10T-NI/40G/gauss-danielCF-40G.dat};

\addplot [blue, line width=0.5mm,dashed,x filter/.code={\pgfmathparse{\pgfmathresult/1000}\pgfmathresult}] table[x index=0,y index=1]
{10T-NI/40G/16qam-danielCF-40G.dat};


 \addplot [blue, only marks, mark=*, mark size=1pt, mark options={fill=white},mark repeat=7,x filter/.code={\pgfmathparse{\pgfmathresult/1000}\pgfmathresult}] table[x index=0,y index=1]
{Dat/SSFM_New/ISRS_16QAM_Nch=251x40GBaud_1x100km_Ptot=25dBm_PhiNL=0.01mmrad_Eta_vs_Ch.dat};

\node at (axis cs:-3.75,21) {\small{10 THz}};

\nextgroupplot 
[
legend columns=1,
grid=major,
ymin=10,ymax=25, 
xmin=-5,xmax=5,
legend style={at={(0.00,-0.0)},anchor=south,font=\footnotesize},
title={101$\times$100 Gbaud}
]

\addplot [red, only marks, mark=*, mark size=1pt, mark options={fill=white},mark repeat=3,x filter/.code={\pgfmathparse{\pgfmathresult/1000}\pgfmathresult}] table[x index=0,y index=1]
{Dat/SSFM_New/ISRS_4QAM_Nch=101x100GBaud_1x100km_Ptot=25dBm_PhiNL=0.02mmrad_Eta_vs_Ch.dat};


\addplot [blue, line width=0.5mm,x filter/.code={\pgfmathparse{\pgfmathresult/1000}\pgfmathresult}] table[x index=0,y index=1]
{10T-NI/100G/16qam-100G.dat};

\addplot [red, line width=0.5mm,x filter/.code={\pgfmathparse{\pgfmathresult/1000}\pgfmathresult}] table[x index=0,y index=1]
{10T-NI/100G/qpsk-100G.dat};

\addplot [black,line width=0.5mm,solid,x filter/.code={\pgfmathparse{\pgfmathresult/1000}\pgfmathresult}] table[x index=0,y index=1]
{10T-NI/100G/gauss-100G.dat};

\addplot [black, line width=0.5mm,dashed,x filter/.code={\pgfmathparse{\pgfmathresult/1000}\pgfmathresult}] table[x index=0,y index=1]
{10T-NI/100G/gauss-danielCF-100G.dat};

\addplot [blue, line width=0.5mm,dashed,x filter/.code={\pgfmathparse{#1/1000}\pgfmathresult}] table[x index=0,y index=1]
{10T-NI/100G/16qam-danielCF-100G.dat};

\addplot [red,line width=0.5mm,dashed,x filter/.code={\pgfmathparse{\pgfmathresult/1000}\pgfmathresult}] table[x index=0,y index=1]
{10T-NI/100G/qpsk-danielCF-100G.dat};

\addplot [blue, only marks, mark=*, mark size=1pt, mark options={fill=white},mark repeat=3,x filter/.code={\pgfmathparse{\pgfmathresult/1000}\pgfmathresult}] table[x index=0,y index=1]
{Dat/SSFM_New/ISRS_16QAM_Nch=101x100GBaud_1x100km_Ptot=25dBm_PhiNL=0.03mmrad_Eta_vs_Ch.dat};

\node at (axis cs:-3.75,11) {\small{10 THz}};

\nextgroupplot 
[
legend columns=1,
ylabel={$\eta_\kappa$ gap $[\text{dB}]$},
grid=major,
ymin=-5,ymax=1, 
xmin=-5,xmax=5,
xlabel={Channel Frequency  [THz]}, 
legend style={at={(0.00,-0.0)},anchor=south,font=\footnotesize},
]

\addplot [red,line width=0.5mm, each nth point=12] table[x index=0,y index=1]
{Dat/SSFM_New/Gaps/EtaGap_Model_vs_SSFM_10THz_4QAM_10GBaud.dat};

\addplot [red,line width=0.1mm, each nth point=12] table[x index=0,y index=1]
{Dat/SSFM_New/Gaps/Average_EtaGap_Model_vs_SSFM_10THz_4QAM_10GBaud.dat};

\addplot [red, line width=0.5mm, dashed, each nth point=12] table[x index=0,y index=1]
{Dat/SSFM_New/Gaps/EtaGap_ClosedForm_vs_SSFM_10THz_4QAM_10GBaud.dat};

\addplot [red, line width=0.1mm, dashed, each nth point=12] table[x index=0,y index=1]
{Dat/SSFM_New/Gaps/Average_EtaGap_ClosedForm_vs_SSFM_10THz_4QAM_10GBaud.dat};

\addplot [blue,line width=0.5mm,each nth point=12] table[x index=0,y index=1]
{Dat/SSFM_New/Gaps/EtaGap_Model_vs_SSFM_10THz_16QAM_10GBaud.dat};

\addplot [blue,line width=0.1mm,each nth point=12] table[x index=0,y index=1]
{Dat/SSFM_New/Gaps/Average_EtaGap_Model_vs_SSFM_10THz_16QAM_10GBaud.dat};

\addplot [blue, line width=0.5mm, dashed,each nth point=12] table[x index=0,y index=1]
{Dat/SSFM_New/Gaps/EtaGap_ClosedForm_vs_SSFM_10THz_16QAM_10GBaud.dat};

\addplot [blue, line width=0.1mm, dashed,each nth point=12] table[x index=0,y index=1]
{Dat/SSFM_New/Gaps/Average_EtaGap_ClosedForm_vs_SSFM_10THz_16QAM_10GBaud.dat};

\node[align=center] at (axis cs:-3.8,-3.8) (av) {Average}; 
\draw[thin] (av)--(axis cs:-2.5,-3);

\nextgroupplot 
[
legend columns=1,
grid=major,
ymin=0,ymax=3, 
xmin=-5,xmax=5,
xlabel={Channel Frequency [THz]}, 
legend style={at={(0.00,-0.0)},anchor=south,font=\footnotesize},
]



\addplot [red,line width=0.5mm, each nth point=5] table[x index=0,y index=1]
{Dat/SSFM_New/Gaps/EtaGap_Model_vs_SSFM_10THz_4QAM_40GBaud.dat};

\addplot [red,line width=0.1mm] table[x index=0,y index=1]
{Dat/SSFM_New/Gaps/Average_EtaGap_Model_vs_SSFM_10THz_4QAM_40GBaud.dat};

\addplot [red, line width=0.5mm, dashed, each nth point=5] table[x index=0,y index=1]
{Dat/SSFM_New/Gaps/EtaGap_ClosedForm_vs_SSFM_10THz_4QAM_40GBaud.dat};

\addplot [red, line width=0.1mm, dashed, each nth point=5] table[x index=0,y index=1]
{Dat/SSFM_New/Gaps/Average_EtaGap_ClosedForm_vs_SSFM_10THz_4QAM_40GBaud.dat};

\addplot [blue,line width=0.5mm,each nth point=5,mark phase=2] table[x index=0,y index=1]
{Dat/SSFM_New/Gaps/EtaGap_Model_vs_SSFM_10THz_16QAM_40GBaud.dat};

\addplot [blue,line width=0.1mm,each nth point=5,mark phase=2] table[x index=0,y index=1]
{Dat/SSFM_New/Gaps/Average_EtaGap_Model_vs_SSFM_10THz_16QAM_40GBaud.dat};

\addplot [blue, line width=0.5mm, dashed,each nth point=5,mark phase=2] table[x index=0,y index=1]
{Dat/SSFM_New/Gaps/EtaGap_ClosedForm_vs_SSFM_10THz_16QAM_40GBaud.dat};

\addplot [blue, line width=0.1mm, dashed,each nth point=5,mark phase=2] table[x index=0,y index=1]
{Dat/SSFM_New/Gaps/Average_EtaGap_ClosedForm_vs_SSFM_10THz_16QAM_40GBaud.dat};

\node[align=center] at (axis cs:3.8,1.5) (av) {Average}; 
\draw[thin] (av)--(axis cs:4.25,1);

\nextgroupplot 
[
legend columns=1,
grid=major,
ymin=0,ymax=3, 
xmin=-5,xmax=5,
xlabel={Channel Frequency [THz]}, 
legend style={at={(0.00,-0.0)},anchor=south,font=\footnotesize},
]

\addplot [red,line width=0.5mm, each nth point=2] table[x index=0,y index=1]
{Dat/SSFM_New/Gaps/EtaGap_Model_vs_SSFM_10THz_4QAM_100GBaud.dat};

\addplot [red,line width=0.1mm] table[x index=0,y index=1]
{Dat/SSFM_New/Gaps/Average_EtaGap_Model_vs_SSFM_10THz_4QAM_100GBaud.dat};

\addplot [red, line width=0.5mm, dashed, each nth point=2] table[x index=0,y index=1]
{Dat/SSFM_New/Gaps/EtaGap_ClosedForm_vs_SSFM_10THz_4QAM_100GBaud.dat};

\addplot [red, line width=0.1mm, dashed, each nth point=2] table[x index=0,y index=1]
{Dat/SSFM_New/Gaps/Average_EtaGap_ClosedForm_vs_SSFM_10THz_4QAM_100GBaud.dat};

\addplot [blue,line width=0.5mm,each nth point=2,mark phase=2] table[x index=0,y index=1]
{Dat/SSFM_New/Gaps/EtaGap_Model_vs_SSFM_10THz_16QAM_100GBaud.dat};

\addplot [blue,line width=0.1mm,each nth point=2,mark phase=2] table[x index=0,y index=1]
{Dat/SSFM_New/Gaps/Average_EtaGap_Model_vs_SSFM_10THz_16QAM_100GBaud.dat};

\addplot [blue, line width=0.5mm, dashed,each nth point=2,mark phase=2] table[x index=0,y index=1]
{Dat/SSFM_New/Gaps/EtaGap_ClosedForm_vs_SSFM_10THz_16QAM_100GBaud.dat};

\addplot [blue, line width=0.1mm, dashed,each nth point=2,mark phase=2] table[x index=0,y index=1]
{Dat/SSFM_New/Gaps/Average_EtaGap_ClosedForm_vs_SSFM_10THz_16QAM_100GBaud.dat};

\node[align=center] at (axis cs:3.8,2.5) (av) {Average}; 
\draw[thin] (av)--(axis cs:4,2);

\end{groupplot}
\end{tikzpicture}

\vspace*{-15pt}
\end{center}
\caption{In the first row, $\eta_\kappa$ as a function of channel number $\kappa$ for 10 THz transmission and after a single $L=100$ km fiber span. PM-QPSK (red), PM-16QAM (blue), and PM-2D-Gauss constellation (black) performance are shown for (a) 10 Gbaud, (b) 40 Gbaud, and (c) 100 Gbaud. Circles refer to SSFM simulation results, whilst solid lines and dashed lines refer to MC integrations and closed-form expressions, resp. In the second row, $\eta_\kappa$ gap from SSFM estimates for {\it{Theorem}}~\ref{main.result} (solid lines) and the closed-form expression in \cite[eq.~(3)]{Semrau_Correction} (dashed lines) for (d) 10 Gbaud, (e) 40 Gbaud, and (f) 100 Gbaud. Horizontal lines indicate the average gap across the whole optical spectrum.} 
\label{fig:10T10G40G100G}
\end{figure*}




As expected, we observe the largest values of $\eta_{\kappa}$ for the PM-2D-Gauss constellation (black). The agreement between {\it{Theorem}}~\ref{main.result} and the closed-form approximation in \cite[eq.~(3)]{Semrau_Correction} is within 0.6 dB.  \HR{The average gap between \cite[integral model]{Semrau.2019.MD.jlt} and the black solid line in Fig.~\ref{nlc_fig} (a) is about 1.6 dB. In particular,  \cite[integral model]{Semrau.2019.MD.jlt} and \cite[eq.~(3)]{Semrau_Correction} systematically underestimate the NLI power for all modulation formats. The model in \cite[integral model]{Semrau.2019.MD.jlt} underestimates the NLI for PM-QPSK by about 3.8 dB at $f=-400$ GHz.} This is explained by the fact that this approximated expression only accounts for the SCI and XPM, and neglects other NLI contributions. As depicted in Fig.~\ref{nlc_fig}, changing the modulation format significantly impacts $\eta_\kappa$. For example, $\eta_\kappa$ for PM-2D-Gauss (black dashed line) is approximately $3.9$ dB higher than PM-QPSK (red circles) for the center channel frequency $f = 0$ GHz. 
The gap between PM-QPSK and PM-16QAM (blue circles) is approximately \HR{$1.7$ dB at $f=-200$ GHz}. 
This comes from the fact that PM-QPSK has the lowest excess kurtosis (given in Table~\ref{tab2}) among the exploited modulation formats.

The modulation format dependence of $\eta_\kappa$ in the presence of SRS is well predicted by the model presented in this paper. The SSFM results are practically coinciding with the curves obtained using {\it{Theorem}}~\ref{main.result}. Average gaps between our model and SSFM simulations are approximately $0.18$ dB for PM-QPSK in the presence of SRS. The same match is not observed for the results using the model in \HR{\cite[integral model]{Semrau.2019.MD.jlt}, \cite{serena2020sims,Lasagniecoc}, \cite[eq.~(3)]{Semrau_Correction}. Figs.~\ref{nlc_fig} (a), (b), and (c) show that the models proposed in \cite{serena2020sims,Lasagniecoc} cannot accurately predict the NLI for PM-16QAM and PM-2D-Gauss, whereas they follow the SSFM results for PM-QPSK very closely. The models in \cite{serena2020sims, Lasagniecoc} inaccurately underestimate the NLI of PM-2D-Gauss by around 3.2 dB. This deviation may be rooted in the simplifying assumptions made in \cite[eq.~(14)]{Semrau2019jlt}.} For PM-16QAM, the model in \cite[eq.~(3)]{Semrau_Correction} (blue dashed lines) predicts $\eta_\kappa$ $2.2$ dB lower than the SSFM simulation results at $f = -400$ GHz (blue circles). For PM-QPSK, this gap increases to $4.3$ dB. This remarkable discrepancy stems from the fact that \cite[eq.~(3)]{Semrau_Correction} only considers SCI (for Gaussian signal) and XPM (for non-Gaussian signal) nonlinear terms and discards the XCI and MCI terms whose contributions at low symbol rates are substantial.

\subsection{10 THz Results}
The results for a 10 THz optical transmission bandwidth are presented in Fig.~\ref{fig:10T10G40G100G}. In this section we compare the SSFM results with our proposed model and with \cite{semrau2018gaussian,cantono2018interplay,Semrau.2019.MD.jlt,Semrau_Correction}. The same modulation formats as in Sec.~\ref{subsec:1THz_results} are shown also for this scenario.
Figs.~\ref{fig:10T10G40G100G}~(a)-(c) show $\eta_{\kappa}$ as a function of the channel frequency, where the total optical bandwidth is partitioned in (a) 1001$\times$10 Gbaud, (b) 251$\times$40 Gbaud, and (c) 101$\times$100 Gbaud channels. The rest of the system parameters are listed in Table~\ref{system_parameters_table}. In Figs.~\ref{fig:10T10G40G100G}~(d)-(f), the gaps of the expressions presented in Figs.~\ref{fig:10T10G40G100G}~(a)-(c) to their corresponding SSFM estimates are shown for each investigated symbol rate.

 For the 10 Gbaud case (Fig.~\ref{fig:10T10G40G100G}~(a)), our model is in very good agreement with the SSFM results across the entire transmitted optical bandwidth for both PM-QPSK (red) and PM-16QAM (blue) modulation formats. The closed-form expression in \cite[eq.~(3)]{Semrau_Correction} results in a significant underestimation of the $\eta_{k}$, which is more pronounced for PM-QPSK. The level of accuracy of the compared analytical expressions is shown in more detail in Fig.~\ref{fig:10T10G40G100G}~(d), where gaps with SSFM estimates are illustrated as a function of the channel frequency for both our model and \cite[eq.~(3)]{Semrau_Correction}. The average gap across the entire optical bandwidth is also shown (horizontal lines). The model in {\it{Theorem}}~\ref{main.result} (solid lines with no markers) is on average approximately 0.2 dB and 0.3 dB above the SSFM estimates for PM-QPSK and PM-16QAM formats, resp. The closed-form formula in \cite[eq.~(3)]{Semrau_Correction} can be seen to underestimate on average the SSFM results by 1.2 dB and 3 dB, for PM-QPSK and PM-16QAM, resp. To explain the source of this inaccuracy, we observe that for the PM-2D-Gauss format the closed-form expression (black dashed line in Fig.~\ref{fig:10T10G40G100G}) follows closely the prediction given by the integral form (black solid line) for frequencies around the center of the optical spectrum. However, an increasing gap is observed as we move away from the central channel frequency $f=0$ (up to 1 dB at $f=4$ THz). We conclude that for the 10 Gbaud transmission scenario the inaccuracy of \cite[eq.~(3)]{Semrau_Correction} is due to the missing MCI terms in the modulation format correction term and, to a minor extent, to its Gaussian component \cite{Semrau2019jlt}. As discussed in Sec.~\ref{subsec:1THz_results} for the 1 THz transmission case, the MCI terms bring a significant contribution in relatively low symbol rate scenarios such as 10 Gbaud channels (see \cite[ Sec.~II]{poggiolini2016analytical}). As confirmed by results in Figs.~\ref{fig:10T10G40G100G}~(a) and (d), this  contribution is still very noticeable for 10 THz transmission.

The results on the 251$\times$40 Gbaud channel transmission case are shown in Figs.~\ref{fig:10T10G40G100G}~(b) and (d). In general, it can be noticed from Fig.~\ref{fig:10T10G40G100G}~(b) that all compared models are in good agreement with the SSFM results. This is due to the increased dominance of the SCI and XPM terms (see, e.g., \cite[Sec.~II]{poggiolini2016analytical}) over the MCI ones, which is confirmed by the fact that the GN closed-form expression \cite{Semrau2019jlt} agrees very well with its integral form  (black lines with squares) across the whole optical spectrum. Fig.~\ref{fig:10T10G40G100G}~(b) shows an average gap of the model in {\it{Theorem}}~\ref{main.result} from SSFM $\eta_{\kappa}$ estimates of 0.6 dB and 0.5 dB for PM-QPSK and PM-16QAM, resp. It can also be observed that the closed form in \cite[eq.~(3)]{Semrau_Correction} is fairly accurate (approx 0.5 dB away from SSFM estimates) for PM-16QAM, but still showing an average 1 dB gap from SSFM estimates for PM-QPSK, where the modulation-format correction term is more dominant. 

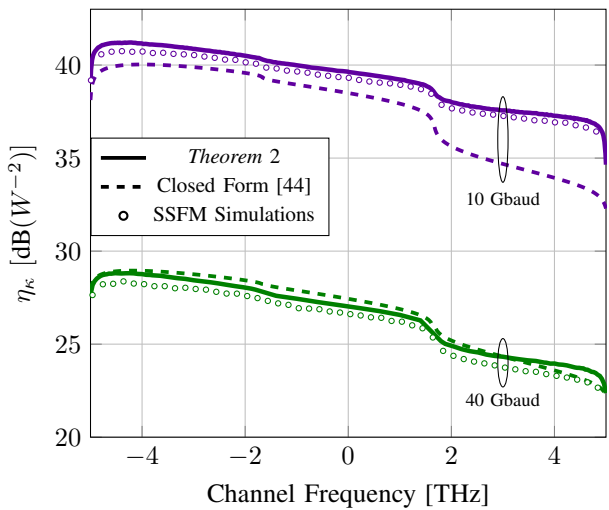
\begin{figure}[!t]
\pgfplotsset{compat=1.3}
\begin{center}

\begin{tikzpicture}[scale=1]
\begin{axis}
[
legend columns=1,
legend style={at={(0.0,0.585)},anchor=west,font=\footnotesize},
legend entries={10 Gbaud Hybrid ({\it{Theorem}}~\ref{main.result}), 40 GBd Hybrid ({\it{Theorem}}~\ref{main.result}), Closed Form \cite{Semrau.2019.MD.jlt}  ,SSFM Simulations },
ylabel={$\eta_\kappa \ [\text{dB}( W^{-2})]$},
grid=major,
ymin=20,ymax=43, 
xmin=-5,xmax=5,
xlabel={Channel Frequency [THz]}
]

\addlegendimage{black, line width=0.5mm}
\addlegendimage{black, line width=0.5mm,dashed}
\addlegendimage{black,only marks,solid, mark=*,mark size = 1pt, line width = 0.8, mark options={fill=white,solid,scale=1.5}}

\addlegendentry{{\it{Theorem}}~\ref{main.result}}
\addlegendentry{Closed Form \cite{Semrau_Correction}}
\addlegendentry{SSFM Simulations}

\addplot [purple!50!blue, line width=0.6mm,x filter/.code={\pgfmathparse{\pgfmathresult/1000}\pgfmathresult}] table[x index=0,y index=1]
{10T-NI/hybrid/64-16-qpsk-10G.txt};

\addplot [green!50!black, line width=0.6mm,x filter/.code={\pgfmathparse{\pgfmathresult/1000}\pgfmathresult}] table[x index=0,y index=1]
{10T-NI/hybrid/64-16-qpsk-40G.txt};

\addplot [purple!50!blue, line width=0.5mm,dashed,x filter/.code={\pgfmathparse{\pgfmathresult/1000}\pgfmathresult}] table[x index=0,y index=1]
{10T-NI/hybrid/64-16-qpsk-danielCF-10G.txt};

\addplot [purple!50!blue, only marks, mark=*, mark size=1pt, mark options={fill=white},mark repeat=20,x filter/.code={\pgfmathparse{\pgfmathresult/1000}\pgfmathresult}] table[x index=0,y index=1]
{Dat/SSFM_New/ISRS_HybridQAM_Nch=1001x10GBaud_1x100km_Ptot=25dBm_PhiNL=0.05mmrad_Eta_vs_Ch.dat};

\addplot [green!50!black, line width=0.5mm,dashed,x filter/.code={\pgfmathparse{\pgfmathresult/1000}\pgfmathresult}] table[x index=0,y index=1]
{10T-NI/hybrid/64-16-qpsk-danielCF-40G.txt};

\addplot [green!50!black, only marks, mark=*, mark size=1pt, mark options={fill=white},mark repeat=5,x filter/.code={\pgfmathparse{\pgfmathresult/1000}\pgfmathresult}] table[x index=0,y index=1]
{Dat/SSFM_New/ISRS_HybridQAM_Nch=251x40GBaud_1x100km_Ptot=25dBm_PhiNL=0.03mmrad_Eta_vs_Ch.dat};

\draw (axis cs:3,24) ellipse (10 and 13) {};
\node (tx1) at (axis cs:3,22) {\scriptsize{40 Gbaud}};

\draw (axis cs:3,36) ellipse (10 and 23) {};
\node (tx1) at (axis cs:3,32.8) {\scriptsize{10 Gbaud}};

\end{axis}
\end{tikzpicture}

\vspace*{-2pt}
\end{center}
\setlength{\belowcaptionskip}{-12pt}
\caption{Hybrid modulation format simulation using PM-64QAM, PM-16QAM, and PM-QPSK in a 10 Gbaud and 40 Gbaud channel transmission.} 
\label{fig:hybrid}
\end{figure}

Finally, the 101$\times$100 Gbaud transmission results are shown in Figs.~\ref{fig:10T10G40G100G}~(c) and (f). In Fig.~\ref{fig:10T10G40G100G}~(c), the model in {\it{Theorem}}~\ref{main.result} can be observed to be still in very good agreement with SSFM results for both PM-QPSK and PM-16QAM. Moreover, for the PM-2D-Gauss case \cite[eq.~(3)]{Semrau_Correction}  approximates very well the expression in {\it{Theorem}}~\ref{main.result}. However, for PM-QPSK and PM-16QAM formats  \cite[eq.~(3)]{Semrau_Correction} significantly overestimates the $\eta_{\kappa}$ across the entire optical bandwidth. This can be attributed to the increasingly dominant SCI terms as the symbol rate is increased for a fixed total optical bandwidth. In Fig.~\ref{fig:10T10G40G100G}~(f), it can be seen that this results in an average gap of the closed-form expression compared to SSFM results of approximately 2 dB and 1 dB for PM-QPSK and PM-16QAM, resp. Only an average 0.4 dB gap is instead observed for the model in {\it{Theorem}}~\ref{main.result}, in both PM-QPSK and PM-16QAM cases.

\HR{It is clear from Figs.~\ref{fig:10T10G40G100G} (a), (b), and (c) that the model given in \cite[eq. (3)]{Semrau.2019.MD.jlt,Semrau_Correction} yields different NLI tilt over the entire spectrum, compared to the SSFM results. It is very likely that this tilt discrepancy results from the simplifying assumptions made in \cite{Semrau.2019.MD.jlt}.}

\subsection{Different Modulation Formats}

To further confirm the validity of our model, in Fig.~\ref{fig:hybrid}, $\eta_k$ is shown as a function of the channel frequency for a scenario where different modulation formats are transmitted over different WDM channels. Both 10 Gbaud and 40 Gbaud transmission scenarios are analyzed for a total 10.011 THz and 10.041 THz bandwidth, resp. PM-64QAM channels are transmitted over the first third of the optical bandwidth, PM-16QAM channels are transmitted over the second third, and PM-QPSK channels are transmitted over the last third. This results in a channel distribution of for the 3 modulation formats PM-64QAM, PM-16QAM and PM-QPSK of (334,333,334) and (84,83,84) for the 10 Gbaud case, and 40 Gbaud case, resp. For the 10 Gbaud channel transmission (purple lines), it can be seen that our model (solid line) is matching quite well the SSFM results (circles), with deviations within 0.4 dB across the entire optical bandwidth. The closed-form expression in \cite[eq.~(3)]{Semrau_Correction} increasingly  understimates $\eta_{k}$ as we move towards the right side of the spectrum, where lower-order formats are transmitted. For the right-most part of the spectrum (PM-QPSK format transmitted) the gap between SSFM results and \cite[eq.~(3)]{Semrau_Correction} can be in excess of 3 dB. For 40 Gbaud channels (green lines), estimates, {\it{Theorem}}~\ref{main.result} and \cite[eq.~(3)]{Semrau_Correction} are in good agreement, with maximum deviations across the entire optical spectrum of about 0.7 dB.


\HR{
\subsection{Heterogeneous Spans}
Finally, the formula proposed in {\it{Theorem}}~\ref{main.result} was validated in a transmission transmission scenario comprising 3 fibre spans of length (80, 100, 120) km, for the first, second, and third span, respectively. The EDFA amplifiers' gain was fixed to 20 dB for all three spans, leading to a non-uniform power profile in each fiber span. The other selected transmission parameters were identical to the 40 Gbaud scenario studied previously in this section. The effect of SRS on the signal power profile was not compensated at the end of each span, as, for instance, in \cite{Semrau_Correction}, but only at the end of the transmission link. This also includes intra-channel compensation of the SRS gain tilt, which can result in an overestimation of $\eta_{\kappa}$ when such an estimation is performed in SSFM simulations via \eqref{eq:snr} and \eqref{eq:snr2}.

In a multi-span transmission scenario, both the numerical integration and the SSFM simulations are particularly challenging due to the increased accuracy and number of computations required for a multi-span system. To partly relax the computational burden, numerical integration was performed using importance sampling for the integration over distance $z'$ in the $\mu_s$ term in Table~\ref{terms.different}. The distribution chosen for the points in the distance domain was obtained from the power profile by normalizing the function $\text{exp}(-\alpha z)$ from 0 to $L_s$ to match a probability distribution. The number of simulated channels was also reduced to 44 to reduce simulation time. The chosen channels were densely located in the borders of the simulation bandwidth and more spaced for middle frequencies. 
For the SSFM simulations, the maximum nonlinear phase rotation was reduced to $8\cdot 10^{-6}$ [rad] to achieve convergence of the SNR results within 0.1 dB. All the other simulation parameters adopted were the ones described in Sec.~\ref{subsec:SSFMnumsims}.

Fig.~\ref{fig:het3spans} shows $\eta_{\kappa}$ estimates obtained using both the numerical integration of \eqref{EGN_fixed} and SSFM simulations, and for 2 different modulation formats: PM-QPSK and PM-16QAM. We observe that, in general, there is good agreement numerical integration the SSFM results with an average gap between the $\eta_{\kappa}$ estimates of approx.~0.8 dB for both modulation formats shown.          

\begin{figure}
    \centering
   \begin{tikzpicture}[scale=1]
\begin{axis}
[
legend columns=1,
legend style={at={(0.025,0.15)},anchor=west,font=\footnotesize},
legend entries={10 Gbaud Hybrid (\textit{Corollary }\ref{main.result.identical}), 40 GBd Hybrid (\textit{Corollary }\ref{main.result.identical}), 
,SSFM Simulations },
ylabel={$\eta_\kappa \ [\text{dB}( W^{-2})]$},
grid=major,
ymin=25,ymax=52, 
xmin=-5,xmax=5,
xlabel={Channel Frequency [THz]}
]

\addlegendimage{black, line width=0.5mm}
\addlegendimage{black,only marks,solid, mark=*,mark size = 1pt, line width = 0.8, mark options={fill=white,solid,scale=1.5}}

\addlegendentry{\textit{Theorem}~ \ref{main.result}}
\addlegendentry{SSFM Simulations}


\addplot [blue, line width=0.5mm] table[x index=0,y index=1]
{10T-NI/3Spans/H3S_16qam.txt};

\addplot [red, line width=0.5mm] table[x index=0,y index=1]
{10T-NI/3Spans/H3S_qpsk.txt};

\addplot [red, only marks, mark=o] table[x index=0,y index=1]
{10T-NI/3Spans/H3S_qpsk_SSFM.txt};

\addplot+[blue, only marks, mark=o,
    x filter/.code={\pgfmathparse{\pgfmathresult/1e12}\pgfmathresult},mark repeat=10]
table[x index=0,y index=1]
{10T-NI/3Spans/H3S_16qam_SSFM.txt};    
\end{axis}
\end{tikzpicture}

    \caption{Simulation results for 3 consecutive spans of $80$, $100$, and $120$ km. An amplifier of $20$ dB gain was added in the end of each span.}
    \label{fig:het3spans}
\end{figure}
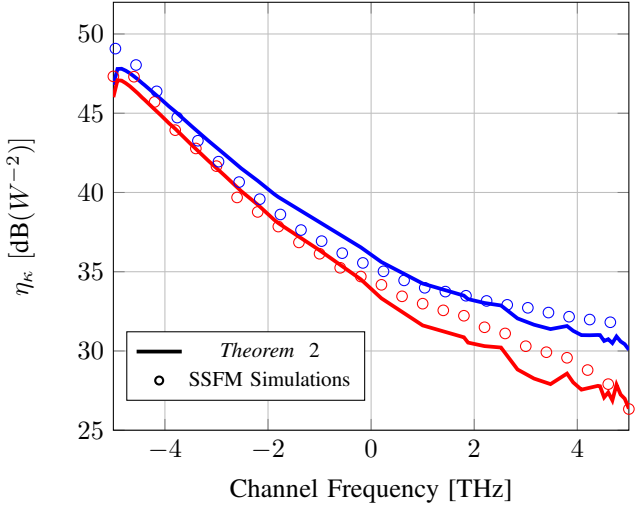}


\section{Conclusions}\label{conclusion.sec}
 
\HR{We proposed an analytical model in the presence of SRS, which has improved features, including the enhanced link function and a new signal power profile. To this new signal power profile was added the frequency-dependent fiber loss profile. Additionally, a new formulation for the NLI terms was proposed, thus enabling us to compute all these terms for partially loaded spectrum, populated with channels with variable bandwidths, more quickly and efficiently.} 
 
\HR{We conclude that the presented model is able to predict the NLI power with a good level of accuracy across all symbol rates and modulation formats. The agreement between analytical and numerical results was found to be within 0.8 dB for all the transmission scenarios investigated in this paper.} The model is particularly effective whenever relatively low or high symbol rates (e.g., 10 Gbaud and 100 Gbaud, resp.) are used in combination with low-order modulation formats. In these scenarios, available closed-form approximations can result in a marked underestimation or overestimation of the NLI power potentially exceeding 3 dB. 


We remark, however, that closed-form expressions, whenever achieving the required accuracy, are orders of magnitude faster to evaluate than integral forms such as the one in {\it{Theorem}}~\ref{main.result}. More work is, thus, needed to precisely characterize the accuracy/complexity trade-offs arising when using a closed-form as opposed to an integral form in different transmission regimes. Future works also include convex power optimization and physical layer impairment-aware optical networking using the proposed model across the C+L band. In addition, the derived model provides a powerful tool for efficient design of capacity-maximizing modulation formats over C+L band systems.

\section{Acknowledgments}

The authors would like to thank Daniel Semrau (University College London) for useful discussions on earlier versions of this manuscript. The work of H. Rabbani, G. Liga, and A. Alvarado has received funding from the European Research Council (ERC) under the European Union's Horizon 2020 research and innovation programme (grant agreement No 757791). The work of A. Alvarado and V. Oliari is supported by the Netherlands Organisation for Scientific Research (NWO) via the VIDI Grant ICONIC (project number 5685).
\appendices

\HR{\section{Signal power profile}\label{signal.power.profile}
In this appendix we solve \eqref{integro.differential.equations} perturbatively since the term proportional to $C_r$ is small relative to the other terms, and thus the solution to \eqref{integro.differential.equations} can be written as
\begin{align}
    \label{sol.power.prof}
    G(z,f)=G_0(z,f)G_1(z,f),
\end{align}
where $G_0(z,f)$ is the solution for no SRS ($C_r=0$), and $G_1(z,f)$ appears as the first order perturbation correction factor.
In the absence of SRS ($C_r=0$), a solution to \eqref{integro.differential.equations} can be written as
\begin{align}\label{solution.integro.woSRS}
    G_0(z,f)=G(0,f)\text{e}^{-\alpha(f)z}.
\end{align}
By dividing both sides of \eqref{integro.differential.equations} by $G(z,f)$, and taking the derivative with respect to $f$, we have
\begin{align}
    \label{int.dif.1}
    \frac{\text{d}}{\text{d}f}\frac{1}{G(z,f)}\frac{\text{d}G(z,f)}{\text{d}z}=-\frac{\text{d}}{\text{d}f}\alpha(f)-C_r\int\text{d}\omega G(z,\omega).
\end{align}
where $\int\text{d}\omega G(z,\omega)$ is the total launch power at distance $z$
\begin{align}
    \label{total.pow.z.f}
 P_{\text{tot}}(z)=\int\text{d}\omega G(z,\omega).
\end{align}
By substituting \eqref{solution.integro.woSRS} into \eqref{total.pow.z.f}, and then plugging it into \eqref{int.dif.1}, we can solve \eqref{int.dif.1} approximately, which is written as
\begin{align}
    \label{int.dif.2}
    \frac{\text{d}}{\text{d}f}\frac{1}{G(z,f)}\frac{\text{d}G(z,f)}{\text{d}z}=\!-\frac{\text{d}}{\text{d}f}\alpha(f)-C_r\!\!\int\!\!\text{d}\omega G(0,\omega)\text{e}^{-\alpha(\omega)z},
\end{align}
By integrating over $z$, and then over $f$, \eqref{int.dif.2} can be solved as
\begin{align}
    \label{G.f.z.1}
    G(z,f)=G(0,f)\text{e}^{-\alpha(f)z-C_rf\int\text{d}\omega G(0,\omega)L_{\text{eff}}(z,\omega)+A(f)+F(z)},
\end{align}
where $L_{\text{eff}}(\cdot)$ is given in \eqref{L_eff.gnrl}. Combining \eqref{sol.power.prof}, \eqref{solution.integro.woSRS}, and \eqref{G.f.z.1}, we have
\begin{align}
    \label{g.1.first.purt}
    G_1(z,f)=\text{e}^{-C_rf\int\text{d}\omega G(0,\omega)L_{\text{eff}}(z,\omega)+A(f)+F(z)}.
\end{align}
Using the fact that $G(z,f)|_{z=0}=G(0,f)$, we find that $A(f)=0$. Integrating over $f$ in \eqref{G.f.z.1} gives
\begin{align}
    \label{G.f.z.1.int}
    &\int\text{d}f G(z,f)=\\&\nonumber\int\text{d}fG(0,f)\text{e}^{-\alpha(f)z-C_rf\int\text{d}\omega G(0,\omega)L_{\text{eff}}(z,\omega)+F(z)}.
\end{align}
Substituting \eqref{solution.integro.woSRS} into the left hand side of \eqref{G.f.z.1.int}, results in
\begin{align}
    \label{f.z}
    F(z)=\ln{\left(\frac{\int \text{d}fG(0,f)\text{e}^{-\alpha(f)z}}{\int \text{d}fG(0,f)\text{e}^{-\alpha(f)z-C_r f\int\text{d}\omega G(0,\omega)L_{\text{eff}}(z,\omega)}}\right)}.
\end{align}
By inserting \eqref{f.z} into \eqref{G.f.z.1}, we can  write \eqref{G.f.z.1} as 
\begin{align}
    \label{g.final.srs}
    &G(z,f)=G(0,f)\\&\nonumber\cdot\frac{\int\text{d}\omega'G(0,\omega')\text{e}^{-\alpha(\omega')z}\cdot \text{e}^{-\alpha(f)z-C_rf\int\text{d}\omega G(0,\omega)L_{\text{eff}}(z,\omega)}}{\int\text{d}\omega'G(0,\omega')\text{e}^{-\alpha(\omega')z-C_r\omega'\int\text{d}\omega'' G(0,\omega'')L_{\text{eff}}(z,\omega'')}}.
\end{align}
Considering \eqref{rho.3}, the signal power profile is written as \eqref{general.link.function}.
}


\section{Proof of Theorem \ref{main.result}}\label{AppendixB}

This appendix contains two Lemmas. In \textit{Lemma}~\ref{multiple.span.extention}, we derive the nonlinear electrical field in \eqref{MAN_NLI_sol} at the end of a link with different spans.  \textit{Lemma}~2 uses this result to derive the NLI PSD in \eqref{G_x.noBias_Def}, which is then used to compute the nonlinear power in \eqref{EGN_fixed} via \eqref{NLI_Power_Channel_Kappa}. The proofs are relegated to the end of this appendix.

\newlength{\hsep}
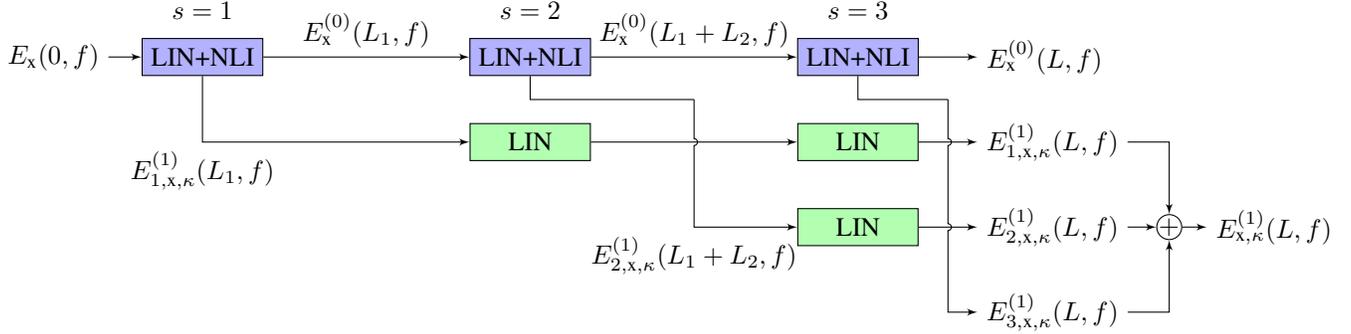
\begin{figure*}
\begin{tikzpicture}[>=latex', scale=0.92]

\setlength{\hsep}{0.7em}

\def\blockH{2} 
\def\blockW{6.5}
\def\radius{0.2} 

\def\listblock{1/1,2/1,3/1,2/2,3/2,3/3}
\def\listaux{1,2,4}
\def\listYEpos{1,2,3,4}
\def\endblocks{3,5,6}

\def\spaceX{19.25}
\def\spaceY{5}

\def\distnum{3.45}
\def\distnumini{1.95}

\def\cirR{1}
\def\distSum{2.45}

\def\distTot{1.45}

\def\distLabel{0.5}

\tikzstyle{blockBlue} = [draw,fill=blue!30,rectangle, 
    minimum height=\blockH\hsep, minimum width=\blockW\hsep] 
\tikzstyle{blockGreen} = [draw,fill=green!30,rectangle, 
    minimum height=\blockH\hsep, minimum width=\blockW\hsep]

 \pgfmathtruncatemacro{\index}{0};
\foreach \x / \y [evaluate=\x / \y as \index using {int((\y-1)*2+\x-max(0,\y-2))}] in \listblock{
    \ifnum \y>1
        \node[blockGreen] (block\index) at (\spaceX*\x\hsep,-\spaceY*\y\hsep) {LIN};
    \else    
        \node[blockBlue,label={[label distance=\distLabel\hsep] $s=\x$}] (block\index) at (\spaceX*\x\hsep,-\spaceY*\y\hsep)  {LIN+NLI};
    \fi
}

\foreach \x  [evaluate=\x as \index using {int(max(\x-3,1))}] in \endblocks{
    \coordinate (endall\x) at ($({block\x}.east)+(\distnum\hsep,0)$);
    \draw[->] ({block\x}.east) -- (endall\x);
}
\coordinate (extraendall) at ($(endall3)+(0,-\spaceY*3\hsep)$);

\node[anchor=east] (E0) at ($({block1}.west)+(-\distnumini\hsep,0)$) {$E_{\text{x}}(0,f)$};
\draw[->] (E0) -- (block1.west);

\node[anchor=west] (Elin) at (endall3) {$E_{\text{x}}^{(0)}(L,f)$};
\node[anchor=west] (Enli1) at (endall5) {$E_{1,\text{x},\kappa}^{(1)}(L,f)$};
\node[anchor=west] (Enli2) at (endall6) {$E_{2,\text{x},\kappa}^{(1)}(L,f)$};
\node[anchor=west] (Enli3) at (extraendall) {$E_{3,\text{x},\kappa}^{(1)}(L,f)$};

\draw[->] (block1.east) -- node[above] {$E_{\text{x}}^{(0)}(L_1,f)$} (block2.west);
\draw[->] (block2.east) -- node[above] {$E_{\text{x}}^{(0)}(L_1+L_2,f)$} (block3.west);
\draw[->] (block4.east) -- (block5.west);


\coordinate (semibranch1) at  (intersection of {block2--block5} and {block3--block4});
\coordinate (semibranch2) at  (intersection of {block3.east--endall5.west} and {block5.east--endall3.west});

\draw[->] (block1.south) |- node[below] {$E_{1,\text{x},\kappa}^{(1)}(L_1,f)$} (block4.west);
\draw (block2.south) |- (semibranch1);
\draw (block3.south) |- (semibranch2);

\tikzstyle{s}=[shift={(0mm,\radius\hsep)}]

\foreach \c in {1,2} {
\draw[->] (semibranch\c) {}{
            -- ([s]block4 -| semibranch\c) arc(90:-90:\radius\hsep)
        }
    \ifnum \c=1
         |- node[below] {$E_{2,\text{x},\kappa}^{(1)}(L_1+L_2,f)$} (block6);
    \else
        -- ([s]block6 -| semibranch\c) arc(90:-90:\radius\hsep)
        |- (extraendall);
    \fi
 }

\node (MinS) [minimum width=\cirR\hsep,circle,draw] at ($(Enli2.east)+(\distSum\hsep,0)$){};
\node at (MinS){$+$};

 \foreach \c in {1,3} {
    \draw[->] (Enli\c) -| (MinS);
 }
 \draw[->] (Enli2) -- (MinS);
 
 \node[anchor=west] (Tot) at ($(MinS.east)+(\distTot\hsep,0)$) {$E_{\text{x},\kappa}^{(1)}(L,f)$};

\draw[->] (MinS.east) -- (Tot.west);

\end{tikzpicture}
\caption{Illustration of the coherent accumulation of NLI along a multispan link resulting from the RP approach. The link is here composed of $N=3$ spans, i.e., $L=L_1+L_2+L_3$.}
\label{coherent.accumulation.fig}
\end{figure*}

In the following Lemma we derive, following the regular perturbation (RP) approach \cite{Armano_RP_VolteraSeris2002}, the total nonlinear electrical field at the end of a link with different spans for one of the 2 transmitted orthogonal polarizations (here referred to as x-polarization). The same result can also be used for the y-polarization field under the substitution x$\rightarrow$y, y$\rightarrow$x. 


 \begin{lemma}\label{multiple.span.extention}
 The total nonlinear electrical field at the end of a link with $N$ different spans can be written as
 \eqlab{NLI_elec_tot_detail}
{
&E_{\text{x},\kappa}^{(1)}(L,f)=\imath\frac{8f_0^{3/2}}{9}\sum_{\kappa_1,\kappa_2,\kappa_3 \in \mathcal{T}_{\kappa}}\sum_{i=-\infty}^{\infty}\delta(f-if_0-\Omega)\nonumber\\&\cdot\sum_{m,n,p\in \mathcal{S}_{i,\kappa}}\Big(\xi_{\text{x},\kappa_1,m}\xi_{\text{x},\kappa_3,n}^{*}\xi_{\text{x},\kappa_2,p}\nonumber\\&+\xi_{\text{y},\kappa_1,m}\xi_{\text{y},\kappa_3,n}^{*}\xi_{\text{x},\kappa_2,p}
\Big)\varsigma_{\kappa_1,\kappa_2,\kappa_3}(m,n,p),
} 
where $f_0=1/T_0$ and $T_0$ is the period of the transmitted signal. In \eqref{NLI_elec_tot_detail}, $L=\sum_{s=1}^{N}\!L_{s}$ is the total length of the link, $\mathcal{T}_\kappa$ is given in \eqref{tk},
and
\begin{align}\label{si}
\mathcal{S}_{i,\kappa}=&\Big\{(m,n,p)\in\mathbb{Z}^3: m-n+p=i,\nonumber\\& 
\,\,\,\,\nu_{\kappa} -\frac{B_\kappa}{2}\leq if_0+(\nu_{\kappa_1}-\nu_{\kappa_3}+\nu_{\kappa_2})\leq \nu_{\kappa} +\frac{B_\kappa}{2}\Big\}.
\end{align}
The coefficient $\varsigma_{\kappa_1,\kappa_2,\kappa_3}(m,n,p)$ in \eqref{NLI_elec_tot_detail} is given by
\eqlab{varsigma}{
\varsigma_{\kappa_1,\kappa_2,\kappa_3}(m,n,p)=\sum_{s=1}^{N}\gamma_s\Xi^{(s)}_{\kappa_1,\kappa_2,\kappa_3}(m,n,p)\cdot\zeta^{(s)}_{\kappa_1,\kappa_2,\kappa_3}(m,n,p),
}
in which
\begin{align}\label{Xi}
&\Xi^{(s)}_{\kappa_1,\kappa_2,\kappa_3}(m,n,p)=\nonumber\\&\prod_{s'=1}^{s-1}\sqrt{g_{s'}(mf_0+\nu_{\kappa_1})\rho_{s'}(L_{s'},mf_0+\nu_{\kappa_1})}\nonumber\\&\cdot \sqrt{g_{s'}(nf_0+\nu_{\kappa_3})}\nonumber\\&\cdot\sqrt{\rho_{s'}(L_{s'},nf_0+\nu_{\kappa_3})}\nonumber\\&
\cdot \sqrt{g_{s'}(pf_0+\nu_{\kappa_2})\rho_{s'}(L_{s'},pf_0+\nu_{\kappa_2})}\nonumber\\&\cdot\text{exp}\Big(\imath 4\pi^{2}\beta_{2,{s'}}L_{s'}\Big[
\Big((m-n)f_0+\nu_{\kappa_1}-\nu_{\kappa_3}\Big)\nonumber\\&\cdot\Big((p-n)f_0+\nu_{\kappa_2}-\nu_{\kappa_3}\Big)\Big]\Big)\nonumber\\&\cdot\text{exp}\Big(\imath 4\pi^3\beta_{3,{s'}}L_{s'}\Big[\Big((m-n)f_0+\nu_{\kappa_1}-\nu_{\kappa_3}\Big)\nonumber\\&\cdot\Big((p-n)f_0+\nu_{\kappa_2}-\nu_{\kappa_3}\Big)\nonumber\\&\cdot\Big((m+p)f_0+\nu_{\kappa_1}+\nu_{\kappa_2}\Big)\Big]\Big)\nonumber\\&\cdot
    \prod_{s'=s}^{N}\sqrt{g_{s'}(if_0+\Omega)\rho_{s'}(L_{s'},if_0+\Omega)} \nonumber\\&\cdot\prod_{s'=1}^{N}\text{exp}\Big([\imath 2\pi^{2}\beta_{2,{s'}} (if_0+\Omega)^{2}\nonumber\\&+\imath\frac{4}{3}\pi^3 \beta_{3,{s'}}(if_0+\Omega)^3]L_{s'}\Big),
\end{align}
and
\eqlab{zeta.s}{
&\zeta^{(s)}_{\kappa_1,\kappa_2,\kappa_3}(m,n,p)=\int_{0}^{L_{s}}\text{d}z'\sqrt{\rho_{s}(z',mf_0+\nu_{\kappa_1})}\nonumber\\&\cdot\frac{\sqrt{\rho_{s}(z',nf_0+\nu_{\kappa_3})}\sqrt{\rho_{s}(z',pf_0+\nu_{\kappa_2})}}{\sqrt{\rho_{s}(z',if_0+\Omega)}}\nonumber\\&\cdot
\text{exp}\Big(\imath 4\pi^{2}\beta_{2,{s}}z'\Big[ 
\Big((m-n)f_0+\nu_{\kappa_1}-\nu_{\kappa_3}\Big)\nonumber\\&\cdot\Big((p-n)f_0+\nu_{\kappa_2}-\nu_{\kappa_3}\Big)\Big]\Big)
\nonumber\\&\cdot\text{exp}\Big(\imath 4\pi^3\beta_{3,{s}}z'\Big[\Big((m-n)f_0+\nu_{\kappa_1}-\nu_{\kappa_3}\Big)\nonumber\\&\cdot\Big((p-n)f_0+\nu_{\kappa_2}-\nu_{\kappa_3}\Big)\nonumber\\&\cdot\Big((m+p)f_0+\nu_{\kappa_1}+\nu_{\kappa_2}\Big)\Big]\Big),
}
where $g_s$ is the gain of the amplifier located at the end of span $s$, and $\rho_{s}(z,f)$ is given in Table~\ref{terms.different}.
\end{lemma}
 \begin{table*}[t]
    \footnotesize
    \centering
    \caption{Integral expressions for the terms used in \textit{Lemma}~\ref{Lemma.psd}. The term $\Upsilon(\cdot)$ is given in Tables~\ref{terms.different} and \ref{terms.identical}.}
    \label{bar.DEFGH}
    \begin{tabular}{|c|l|}
    \hline
    
    \hline
    Term & Integral Expression \\
    \hline
    
    \hline
$\displaystyle \bar{D}$ &  $\frac{3}{8} B_{\kappa_1}B_{\kappa_2}B_{\kappa_3}\int_{-B_{\kappa_1}/2}^{B_{\kappa_1}/2}\text{d}f_1\int_{-B_{\kappa_2}/2}^{B_{\kappa_2}/2}\text{d}f_2|\Upsilon(f_1+\nu_{\kappa_1},f_2+\nu_{\kappa_2},f)|^2 |S_{\kappa_1}(f_1)|^{2}|S_{\kappa_3}(f_1-f+f_2+\Omega)|^{2}|S_{\kappa_2}(f_2)|^{2}$
\\
\hline
$\displaystyle \bar{E}$ &  $ B_{\kappa_1}B_{\kappa_2}\int_{-B_{\kappa_1}/2}^{B_{\kappa_1}/2}\text{d}f_1\int_{-B_{\kappa_2}/2}^{B_{\kappa_2}/2}\text{d}f_2\int_{-B_{\kappa_1}/2}^{B_{\kappa_1}/2}\text{d}f_1'
   \Upsilon(f_1+\nu_{\kappa_1},f_2+\nu_{\kappa_2},f)S_{\kappa_1}(f_1)|S_{\kappa_2}(f_2)|^{2}\Upsilon^*(f_1'+\nu_{\kappa_1},f_2+\nu_{\kappa_2},f)$\\&$\cdot S_{\kappa_3}^*(f_1-f+\Omega+f_2)S_{\kappa_1}^*(f_1') S_{\kappa_3}(f_1'-f+f_2+\Omega)\Phi_{\kappa_1}$
\\
\hline
$\displaystyle \bar{F}$ &
$ B_{\kappa_1}B_{\kappa_2}\int_{-B_{\kappa_1}/2}^{B_{\kappa_1}/2}\text{d}f_1\int_{-B_{\kappa_2}/2}^{B_{\kappa_2}/2}\text{d}f_2\int_{-B_{\kappa_2}/2}^{B_{\kappa_2}/2}\text{d}f_2'\Upsilon(f_1+\nu_{\kappa_1},f_2+\nu_{\kappa_2},f)|S_{\kappa_1}(f_1)|^{2} \Upsilon^*(f_1+\nu_{\kappa_1},f_2'+\nu_{\kappa_2},f)S_{\kappa_2}(f_2)$\\&$\cdot S_{\kappa_3}^{*}(f_1-f+\Omega+f_2)S_{\kappa_2}^{*}(f_2')S_{\kappa_3}(f_1-f+\Omega+f_2') \Phi_{\kappa_2}$
\\
\hline
$\displaystyle \bar{G}$ &
$ B_{\kappa_1}B_{\kappa_3}\int_{-B_{\kappa_1}/2}^{B_{\kappa_1}/2}\text{d}f_1\int_{-B_{\kappa_2}/2}^{B_{\kappa_2}/2}\text{d}f_2\int_{-B_{\kappa_1}/2}^{B_{\kappa_1}/2}\text{d}f_1'S_{\kappa_1}^*(f_1')S_{\kappa_2}^*(f_1-f_1'+f_2)S_{\kappa_2}(f_2)|S_{\kappa_3}(f_1-f+f_2+\Omega)|^{2}$\\&$\cdot S_{\kappa_1}(f_1)\Upsilon(f_1+\nu_{\kappa_1},f_2+\nu_{\kappa_2},f)\Upsilon^*(f_1'+\nu_{\kappa_1},f_1+f_2-f_1'+\nu_{\kappa_2},f)  \Phi_{\kappa_1}$
\\
\hline
$\displaystyle \bar{H}$ &
$\frac{1}{8} B_{\kappa_1}\int_{-B_{\kappa_1}/2}^{B_{\kappa_1}/2}\text{d}f_1\int_{-B_{\kappa_2}/2}^{B_{\kappa_2}/2}\text{d}f_2\int_{-B_{\kappa_1}/2}^{B_{\kappa_1}/2}\text{d}f_1'\int_{-B_{\kappa_2}/2}^{B_{\kappa_2}/2}\text{d}f_2'\Upsilon(f_1+\nu_{\kappa_1},f_2+\nu_{\kappa_2},f)\Upsilon^*(f_1'+\nu_{\kappa_1},f_2'+\nu_{\kappa_2},f)$\\&$\cdot S_{\kappa_1}(f_1)S_{\kappa_1}^*(f_1')S_{\kappa_2}(f_2)S_{\kappa_2}^*(f_2')S_{\kappa_3}(f_1-f+f_2+\Omega)S_{\kappa_3}^*(f_1'-f+\Omega+f_2')  \Psi_{\kappa_2}$
\\
\hline
    \end{tabular}
\end{table*}
 \begin{lemma}\label{Lemma.psd}
 The PSD of \eqref{NLI_elec_tot_detail} is
 \eqlab{G_x_d1}
{
G_{\text{\small{NLI}},\text{x},\kappa}(f)&=\frac{64}{81}\sum_{\kappa_1,\kappa_2,\kappa_3\in \mathcal{T}_{\kappa}}P_{\kappa_1}P_{\kappa_2}P_{\kappa_3}\nonumber\\&\cdot\Bigg(\bar{\delta}_{\kappa_2,\kappa_3}\bar{\delta}_{\kappa_1,\kappa_3} \bar{\delta}_{\kappa_1,\kappa_2}\bar{D}\nonumber\\&+
\delta_{\kappa_1,\kappa_3}\bar{\delta}_{\kappa_1,\kappa_2}\left(\bar{D}+\frac{3}{8}\bar{E}\right)\nonumber\\&+\delta_{\kappa_2,\kappa_3}\bar{\delta}_{\kappa_1,\kappa_2}\left(\bar{D}+\frac{2}{8}\bar{F}\right)\nonumber\\&+
\delta_{\kappa_1,\kappa_2}\bar{\delta}_{\kappa_2,\kappa_3}\left(\bar{D}+\frac{1}{8}\bar{G}\right)\nonumber\\&+\delta_{\kappa_1,\kappa_2}\delta_{\kappa_2,\kappa_3}\left(\!\bar{D}\!+\!\frac{5}{8}\bar{F}\!+\!\frac{1}{8} \bar{G}\!+\!\bar{H}\right)\!\!\!\Bigg),
} 
    where $\bar{D}$, $\bar{E}$, $\bar{F}$, $\bar{G}$, and $\bar{H}$ are given in Table~\ref{bar.DEFGH}, and $\delta_{i,j}$ and $\bar{\delta}_{i,j}$ are given in Sec.~\ref{notation.sec}.
\end{lemma}


The last step in the proof of Theorem~\ref{main.result} is to use \eqref{G_x_d1} in \eqref{G_x.noBias_Def} and \eqref{NLI_Power_Channel_Kappa}. To this end, we multiply the PSD in \eqref{G_x_d1} by two (see \eqref{G_x.noBias_Def}), we integrate it over the frequencies of channel $\kappa$ (see \eqref{NLI_Power_Channel_Kappa}), we change the variable $f$ into $f'+\nu_\kappa$,we group the delta functions in \eqref{G_x_d1}, and we use the fact that $\sum_{\kappa_1,\kappa_2,\kappa_3\in \tau_\kappa}P_{\kappa_1}P_{\kappa_2}P_{\kappa_3}\delta_{\kappa_1,\kappa_3}\bar{\delta}_{\kappa_1,\kappa_2}3/8\bar{E}+\sum_{\kappa_1,\kappa_2,\kappa_3\in \tau_\kappa}P_{\kappa_1}P_{\kappa_2}P_{\kappa_3}\delta_{\kappa_2,\kappa_3}\bar{\delta}_{\kappa_1,\kappa_2}2/8\bar{F}=\sum_{\kappa_1,\kappa_2,\kappa_3\in \tau_\kappa}P_{\kappa_1}P_{\kappa_2}P_{\kappa_3}\delta_{\kappa_1,\kappa_3}\bar{\delta}_{\kappa_1,\kappa_2}5/8\bar{E}$ since $\sum_{\kappa_1,\kappa_2,\kappa_3\in \tau_\kappa}P_{\kappa_1}P_{\kappa_2}P_{\kappa_3}\delta_{\kappa_1,\kappa_3}\bar{\delta}_{\kappa_1,\kappa_2}\bar{E}=\sum_{\kappa_1,\kappa_2,\kappa_3\in \tau_\kappa}P_{\kappa_1}P_{\kappa_2}P_{\kappa_3}\delta_{\kappa_2,\kappa_3}\bar{\delta}_{\kappa_1,\kappa_2}\bar{F}$. This process gives the total NLI power on channel $\kappa$, which is given by \eqref{EGN_fixed}.


\begin{ProofLemma}
Under the RP approach, we can write the nonlinear electrical field of channel $\kappa$ at the end of the link with $N$ different spans as (see e.g.,~\cite[eq.~(9)]{Armano_RP_VolteraSeris2002})
\eqlab{NLI_elec_tot}
{
E_{\text{x},\kappa}^{(1)}(L,f)=\sum_{s=1}^{N}E_{s,\text{x},\kappa}^{(1)}(L,f),
}
where $E_{s,\text{x},\kappa}^{(1)}(L,f)$ is the nonlinear electrical field of channel $\kappa$ generated in span $s$, which linearly propagates until the end of the link of length $L$, as schematically shown in Fig.~\ref{coherent.accumulation.fig}. To derive each of the terms in \eqref{NLI_elec_tot}, the linear electrical field is first needed.  

By solving \eqref{eq1} in the absence of the forcing NL term ${\boldsymbol{Q}}(z,f)$ we obtain that the linear electrical field at the input of span $s$ is
\eqlab{LIN_elec_span}{
E_{\text{x}}^{(0)}(L'_s,f)&=\prod_{s'=1}^{s-1}g_{s'}^{1/2}(f)\cdot\text{e}^{\tilde{\Gamma}_{s'}(L_{s'},f)}E_{\text{x}}(0,f),
}
where $L'_s={\scriptstyle\sum_{s'=1}^{s-1}L_{s'}}$, and
\eqlab{gamma.tilde.s}{
\tilde{\Gamma}_{s}(z,f)=\int_{0}^{z}\text{d}z'{\Gamma}_{s}(z',f), \qquad 0\leq z\leq L_s,
}
in which
\begin{equation}\label{gamma.s}
{\Gamma}_s(z,f)=\frac{g_s(z,f)}{2}+\imath 2\pi^{2}\beta_{2,s} f^{2}+\imath\frac{4}{3}\pi^3 \beta_{3,s}f^3, 
\end{equation}
where $g_s(z,f)$ is the generic frequency- and distance-dependent gain coefficient of span $s$, and $E_{\text{x}}(0,f)$ is given by
\eqref{E_input.final}.
Equations \eqref{LIN_elec_span}--\eqref{gamma.s} represent the electrical field passing through $s-1$ spans influenced only by chromatic dispersion, span losses, SRS gain/loss, and amplifier gains. The linear field in span $s$ can then be written as
\eqlab{LIN_elec_z}{
&E_{s,\text{x}}^{(0)}(z,f)=\text{e}^{\tilde{\Gamma}_s(z,f)}E_{\text{x}}^{(0)}(L'_s,f),\quad 0\leq z\leq L_s,
} where $E_{\text{x}}^{(0)}(L'_s,f)$ is given by \eqref{LIN_elec_span}.
In order to find the contribution at the $s$-th span to the nonlinear optical field in \eqrf{NLI_elec_tot} 
we define, similar to \eqref{Kerr-term1},
\eqlab{Kerr-term2}
{
&{\boldsymbol{Q}}_s(z,f)=\imath \gamma_s\frac{8}{9}\Big[E_{s,\text{x}}^{(0)}(z,f)*E_{s,\text{x}}^{(0)\scriptsize{*}}(z,-f)\nonumber\\&+E_{s,\text{y}}^{(0)}(z,f)*E_{s,\text{y}}^{(0)\scriptsize{*}}(z,-f)\Big]*{\boldsymbol{E}}^{(0)}_s(z,f),
}
whose first component (for the x polarization) can be written using \eqref{E_input.final} as
\eqlab{eq_6_1}{
&Q_{s,\text{x}}(z,f)=\imath\gamma_s\frac{8}{9}\Big[\nonumber\\
&f_0\Big(\text{e}^{\tilde{\Gamma}_s(z,f)}\prod_{s'=1}^{s-1}g_{s'}^{1/2}(f)\cdot\text{e}^{\tilde{\Gamma}_{s'}(L_{s'},f)}\sum_{\kappa_1=1}^{M}\sum_{m=-\infty}^{\infty}\xi_{\text{x},\kappa_1,m}\nonumber\\&\cdot\delta(f-mf_0-\nu_{\kappa_1})\Big)\nonumber\\
&*\Big(\text{e}^{\tilde{\Gamma}_{s}^*(z,f)}\prod_{s'=1}^{s-1}g_{s'}^{1/2}(f)\cdot\text{e}^{\tilde{\Gamma}_{s'}^{*}(L_{s'},f)}\sum_{\kappa_3=1}^{M}\sum_{n=-\infty}^{\infty}\xi^*_{\text{x},\kappa_3,n}\nonumber\\
&\cdot\delta(-f-nf_0-\nu_{\kappa_3})\Big)\nonumber\\
&+f_0\Big(\text{e}^{\tilde{\Gamma}_{s}(z,f)}\prod_{s'=1}^{s-1}g_{s'}^{1/2}(f)\cdot\text{e}^{\tilde{\Gamma}_{s'}(L_{s'},f)}\sum_{\kappa_1=1}^{M}\sum_{m=-\infty}^{\infty}\xi_{\text{y},\kappa_1,m}\nonumber\\
&\cdot\delta(f-mf_0-\nu_{\kappa_1})\Big)\nonumber\\
&*\Big(\text{e}^{\tilde{\Gamma}_{s}^*(z,f)}\prod_{s'=1}^{s-1}g_{s'}^{1/2}(f)\cdot\text{e}^{\tilde{\Gamma}_{s'}^{*}(L_{s'},f)}\sum_{\kappa_3=1}^{M}\sum_{m=-\infty}^{\infty}\xi^*_{\text{y},\kappa_3,n}\nonumber\\
&\cdot\delta(-f-nf_0-\nu_{\kappa_3})\Big)\Big]\nonumber\\
&*{f_0}^{1/2}\Big(\text{e}^{\tilde{\Gamma}_{s}(z,f)}\prod_{s'=1}^{s-1}g_{s'}^{1/2}(f)\cdot\text{e}^{\tilde{\Gamma}_{s'}(L_{s'},f)}\sum_{\kappa_2=1}^{M}\sum_{p=-\infty}^{\infty}\xi_{\text{x},\kappa_2,p}\nonumber\\
&\cdot\delta(f-pf_0-\nu_{\kappa_2})\Big).
}
We now use the property
\eqlab{delta.property}{
&a(f)\delta(f-f_a)*b(f)\delta(-f-f_b)*c(f)\delta(f-f_c)=\nonumber\\&a(f_a)b(f_b)c(f_c)\delta(f-(f_a-f_b+f_c)),
}
to express \eqref{eq_6_1} as
\begin{equation}\label{eq_6_1_1}
\begin{split}
&Q_{s,\text{x}}(z,f)=\imath\gamma_s\frac{8f_0^{3/2}}{9}\sum_{\kappa_1=1}^{M}\sum_{\kappa_3=1}^{M}\sum_{\kappa_2=1}^{M}\sum_{ m=-\infty}^{\infty}\sum_{n=-\infty}^{\infty}\\&\sum_{ p=-\infty}^{\infty}\delta\big(f-(m-n+p)f_0-(\nu_{\kappa_1}-\nu_{\kappa_3}+\nu_{\kappa_2})\big)\\&\cdot\Big(\xi_{\text{x},\kappa_1,m}\xi_{\text{x},\kappa_3,n}^{*}\xi_{\text{x},\kappa_2,p}+\xi_{\text{y},\kappa_1,m}\xi_{\text{y},\kappa_3,n}^{*}\xi_{\text{x},\kappa_2,p}\Big)\\&\cdot \text{e}^{\tilde{\Gamma}_{s}(z,mf_0+\nu_{\kappa_1})+\tilde{\Gamma}_{s}^*(z,nf_0+\nu_{\kappa_3})+\tilde{\Gamma}_{s}(z,pf_0+\nu_{\kappa_2} R)}\prod_{s'=1}^{s-1} \\&g_{s'}^{1/2}(mf_0+\nu_{\kappa_1})g_{s'}^{1/2}(nf_0+\nu_{\kappa_3})g_{s'}^{1/2}(pf_0+\nu{\kappa_2})\\&\cdot \text{e}^{\tilde{\Gamma}_{s'}(L_{s'},mf_0+\nu_{\kappa_1})+\tilde{\Gamma}_{s'}^*(L_{s'},nf_0+\nu_{\kappa_3})+\tilde{\Gamma}_{s'}(L_{s'},pf_0+\nu_{\kappa_2})}].
\end{split}
\end{equation}
%
%
Using \eqref{si}, we now restrict the frequency components in \eqref{eq_6_1_1} to be only those in the $\kappa$-th channel, 
which gives rise to
\begin{equation}\label{eq_6_si}
\begin{split}
&Q_{s,\text{x},\kappa}(z,f)=
\imath\gamma_s\frac{8f_0^{3/2}}{9}\sum_{\kappa_1=1}^{M}\sum_{\kappa_3=1}^{M}\sum_{\kappa_2=1}^{M}\sum_{i=-\infty}^{\infty}\\
&\delta(f-if_0-(\nu_{\kappa_1}-\nu_{\kappa_3}+\nu_{\kappa_2}))\\
&\cdot\sum_{m,n,p\in {\mathcal{S}}_{i,\kappa}}
\Big(
\xi_{\text{x},\kappa_1,m}\xi_{\text{x},\kappa_3,n}^{*}\xi_{\text{x},\kappa_2,p}+\xi_{\text{y},\kappa_1,m}\xi_{\text{y},\kappa_3,n}^{*}\xi_{\text{x},\kappa_2,p}
\Big)\\
&\cdot \text{e}^{\tilde{\Gamma}_{s}(z,mf_0+\nu_{\kappa_1})+\tilde{\Gamma}_{s}^*(z,nf_0+\nu_{\kappa_3})+\tilde{\Gamma}_{s}(z,pf_0+\nu_{\kappa_2})}\prod_{s'=1}^{s-1} \\&g_{s'}^{1/2}(mf_0+\nu_{\kappa_1})g_{s'}^{1/2}(nf_0+\nu_{\kappa_3})g_{s'}^{1/2}(pf_0+\nu_{\kappa_2})\\&\cdot \text{e}^{\tilde{\Gamma}_{s'}(L_{s'},mf_0+\nu_{\kappa_1})+\tilde{\Gamma}_{s'}^*(L_{s'},nf_0+\nu_{\kappa_3})+\tilde{\Gamma}_{s'}(L_{s'},pf_0+\nu_{\kappa_2})},
\end{split}
\end{equation}
where we use the notation $Q_{s,\text{x},\kappa}(z,f)$ to show the Kerr term in channel $\kappa$.
The rectangular spectral shape $S_{\kappa_3}(nf_0)$  with center frequency $f=0$ in \eqref{FourierSerisCoef1} implies that    $nf_0$ should satisfy $-\frac{B_{\kappa_3}}{2}\leq nf_0 \leq\frac{B_{\kappa_3}}{2}$. A similar interpretation can be used on $mf_0$ and $pf_0$, and thus,
\begin{align*}
&-\frac{B_{\kappa_1}}{2}\leq mf_0 \leq \frac{B_{\kappa_1}}{2},\quad
-\frac{B_{\kappa_3}}{2}\leq nf_0 \leq\frac{B_{\kappa_3}}{2},\\&\nonumber
-\frac{B_{\kappa_2}}{2}\leq pf_0 \leq\frac{B_{\kappa_2}}{2},
\end{align*}
which gives 
\begin{equation}
-\frac{B_{\kappa_1}+B_{\kappa_2}+B_{\kappa_3}}{2}\leq(m-n+p)f_0 \leq\frac{B_{\kappa_1}+B_{\kappa_2}+B_{\kappa_3}}{2}. 
\label{mnp_ineq}
\end{equation}


Combining the inequalities in the definition of the set ${\mathcal{S}}_{i,\kappa}$ in \eqref{si} with \eqref{mnp_ineq}, we obtain\footnote{To obtain \eqref{boundary.kappa'}, we use the fact that for any $a,b,c,d,x
\in\mathbb{R}$ the two inequalities $a\leq x \leq b$ and $c\leq x \leq d$, imply that if $|c-d|\leq|b-a|$ we have $d\leq b$ and $a\leq c$. On the other hand, if $|b-a|\leq|c-d|$ we have $b\leq d$ and $c\leq a$.}
\eqlab{boundary.kappa'}{
&-\frac{|\hat{B}-B_{\kappa}|}{2}\leq\Omega-\nu_{\kappa}\leq\frac{|\hat{B}-B_{\kappa}|}{2},
}
where $\Omega$ and $\hat{B}$ are defined in {\it{Theorem}}~\ref{main.result}.
Using \eqref{boundary.kappa'}, we express \eqref{eq_6_si} as
\begin{equation}\label{qx_2}
\begin{split}
&Q_{s,\text{x},\kappa}(z,f)=\imath\gamma_s\frac{8f_0^{3/2}}{9}\sum_{\kappa_1,\kappa_2,\kappa_3\in \mathcal{T}_\kappa}\sum_{i=-\infty}^{\infty}\\&\delta\big(f-if_0-\Omega\big)\sum_{m,n,p\in \mathcal{S}_{i,\kappa}}\Big(\xi_{\text{x},\kappa_1,m}\xi_{\text{x},\kappa_3,n}^{*}\\&\cdot\xi_{\text{x},\kappa_2,p}+\xi_{\text{y},\kappa_1,m}\xi_{\text{y},\kappa_3,n}^{*}\xi_{\text{x},\kappa_2,p}\Big)\\&\cdot \text{e}^{\tilde{\Gamma}_s(z,mf_0+\nu_{\kappa_1})+\tilde{\Gamma}_s^*(z,nf_0+\nu_{\kappa_3})+\tilde{\Gamma}_s(z,pf_0+\nu_{\kappa_2})}\\&\cdot\prod_{s'=1}^{s-1} [g_{s'}^{1/2}(mf_0+\nu_{\kappa_1})g_{s'}^{1/2}(nf_0+\nu_{\kappa_3})\\&\cdot g_{s'}^{1/2}(pf_0+\nu_{\kappa_2}) \text{e}^{\tilde{\Gamma}_{s'}(L_{s'},mf_0+\nu_{\kappa_1})}\\&\cdot\text{e}^{\tilde{\Gamma}_{s'}^*(L_{s'},nf_0+\nu_{\kappa_3})+\tilde{\Gamma}_{s'}(L_{s'},pf_0+\nu_{\kappa_2})}]
\end{split}
\end{equation}
where the set $\mathcal{S}_{i,\kappa}$  and $\tau_\kappa$ are defined in \eqref{si} and \eqref{tk}, resp.
In what follows, we will use \eqref{qx_2} to calculate the nonlinear electrical field on this channel, i.e., $E_{s,\text{x},\kappa}^{(1)}(z,f)$.
The nonlinear electrical field in the $s$-th span is obtained by combining \eqref{qx_2} and \eqref{MAN_NLI_sol}, which gives
\begin{equation}\label{NLIN_expression_x_1.delta}
\begin{split}
&E_{s,\text{x},\kappa}^{(1)}(z,f)=\imath\gamma_s\frac{8f_0^{3/2}}{9}\sum_{\kappa_1,\kappa_2,\kappa_3\in \mathcal{T}_\kappa}\sum_{i=-\infty}^{\infty}\\&\delta(f-if_0-\Omega)\sum_{m,n,p\in \mathcal{S}_{i,\kappa}}\\&\!\!\!\Big(\xi_{\text{x},\kappa_1,m}\xi_{\text{x},\kappa_3,n}^{*}\xi_{\text{x},\kappa_2,p}+\xi_{\text{y},\kappa_1,m}\xi_{\text{y},\kappa_3,n}^{*}\xi_{\text{x},\kappa_2,p}
\Big)\\&
\cdot\prod_{s'=1}^{s-1} [g_{s'}^{1/2}(mf_0+\nu_{\kappa_1})g_{s'}^{1/2}(nf_0+\nu_{\kappa_3})\\&\!\!\!\cdot g_{s'}^{1/2}(pf_0+\nu_{\kappa_2}) \text{e}^{\tilde{\Gamma}_{s'}(L_{s'},mf_0+\nu_{\kappa_1})+\tilde{\Gamma}_{s'}^*(L_{s'},nf_0+\nu_{\kappa_3})}\\
&\cdot\text{e}^{\tilde{\Gamma}_{s'}(L_{s'},pf_0+\nu_{\kappa_2})}]\cdot\text{e}^{\tilde{\Gamma}_{s}(z,if_0+\Omega)}\\
&\cdot\int_{0}^{z}\text{d}z' \text{e}^{\tilde{\Gamma}_s(z',mf_0+\nu_{\kappa_1})}\cdot\text{e}^{\tilde{\Gamma}_s^*(z',nf_0+\nu_{\kappa_3})}\\&\cdot\text{e}^{\tilde{\Gamma}_s(z',pf_0+\nu_{\kappa_2})}\cdot\text{e}^{-\tilde{\Gamma}_s(z',if_0+\Omega)},
\end{split}
\end{equation}
in which the delta function property $A(f)\delta(f-f_a)=A(f_a)\delta(f-f_a)$ is used.
Considering the amplifier gain at the end of span $s$, we have
\eqlab{NLI_elec_kappa_end_ns}{
&E_{s,\text{x},\kappa}^{(1)}(L'_s+L_s,f)=g_{s}^{1/2}(f)\cdot E_{s,\text{x},\kappa}^{(1)}(L_s,f),
}
which using \eqref{NLIN_expression_x_1.delta} is equal to
\begin{equation}\label{NLIN_expression_x_1.end.span.s}
\begin{split}
&E_{s,\text{x},\kappa}^{(1)}(L'_s+L_s,f)=\imath\gamma_s\frac{8f_0^{3/2}}{9}\sum_{\kappa_1,\kappa_2,\kappa_3\in \mathcal{T}_\kappa}\sum_{i=-\infty}^{\infty}\\&\delta(f-if_0-\Omega)\sum_{m,n,p\in \mathcal{S}_{i,\kappa}}\\&\Big(\xi_{\text{x},\kappa_1,m}\xi_{\text{x},\kappa_3,n}^{*}\xi_{\text{x},\kappa_2,p}+\xi_{\text{y},\kappa_1,m}\xi_{\text{y},\kappa_3,n}^{*}\xi_{\text{x},\kappa_2,p}
\Big)\\&
\cdot g_{s}^{1/2}(if_0+\Omega)\prod_{s'=1}^{s-1} [g_{s'}^{1/2}(mf_0+\nu_{\kappa_1})\\&\cdot g_{s'}^{1/2}(nf_0+\nu_{\kappa_3}) g_{s'}^{1/2}(pf_0+\nu_{\kappa_2})\\&\cdot\text{e}^{\tilde{\Gamma}_{s'}(L_{s'},mf_0+\nu_{\kappa_1})+\tilde{\Gamma}_{s'}^*(L_{s'},nf_0+\nu_{\kappa_3})}\\&\cdot\text{e}^{\tilde{\Gamma}_{s'}(L_{s'},pf_0+\nu_{\kappa_3})}]\cdot \text{e}^{\tilde{\Gamma}_{s}(L_s,if_0+\Omega)}\\
&\cdot\int_{0}^{L_{s}}\text{d}z' \text{e}^{\tilde{\Gamma_s}(z',mf_0+\nu_{\kappa_1})}\cdot\text{e}^{\tilde{\Gamma}_{s}^*(z',nf_0+\nu_{\kappa_3}))}\\&\cdot\text{e}^{\tilde{\Gamma_s}(z',pf_0+\nu_{\kappa_2})}\cdot\text{e}^{-\tilde{\Gamma_s}(z',if_0+\Omega)}.
\end{split}
\end{equation}

From the RP approach, $E_{s,\text{x},\kappa}^{(1)}(L'_s+L_s,f)$ in \eqref{NLIN_expression_x_1.end.span.s} propagates linearly over $N-s$ spans (see Fig.~\ref{coherent.accumulation.fig}), i.e.,
\eqlab{}{
E_{s,\text{x},\kappa}^{(1)}(L,f)=&\prod_{s'=s+1}^{N}g_{s'}^{1/2}(f)\cdot\text{e}^{\tilde{\Gamma}_{s'}(L_{s'},f)} E_{s,\text{x},\kappa}^{(1)}(L'_s+L_s,f),
}
which using \eqref{NLIN_expression_x_1.end.span.s} gives
\begin{equation}\label{NLIN_expression_x_1.s.end.link}
\begin{split}
&E_{s,\text{x},\kappa}^{(1)}(L
,f)=\imath\gamma_s\frac{8f_0^{3/2}}{9}\sum_{\kappa_1,\kappa_2,\kappa_3\in \mathcal{T}_\kappa}\sum_{i=-\infty}^{\infty}\\&\delta(f-if_0-\Omega)\sum_{m,n,p\in \mathcal{S}_{i,\kappa}}\\&\Big(\xi_{\text{x},\kappa_1,m}\xi_{\text{x},\kappa_3,n}^{*}\xi_{\text{x},\kappa_2,p}+\xi_{\text{y},\kappa_1,m}\xi_{\text{y},\kappa_3,n}^{*}\xi_{\text{x},\kappa_2,p}
\Big)\\&\cdot \prod_{s'=1}^{s-1}g_{s'}^{1/2}(mf_0+\nu_{\kappa_1})g_{s'}^{1/2}(nf_0+\nu_{\kappa_3})\\&\!\!\!\cdot g_{s'}^{1/2}(pf_0+\nu_{\kappa_2}) \text{e}^{\tilde{\Gamma}_{s'}(L_{s'},mf_0+\nu_{\kappa_1})+\tilde{\Gamma}_{s'}^*(L_{s'},nf_0+\nu_{\kappa_3})}\\&\cdot\text{e}^{\tilde{\Gamma}_{s'}(L_{s'},pf_0+\nu_{\kappa_2})}\cdot g_{s}^{1/2}(if_0+\Omega) \text{e}^{\tilde{\Gamma}_{s}(L_s,if_0+\Omega)}\\&\cdot\prod_{s'=s+1}^{N}g_{s'}^{1/2}(if_0+\Omega)\cdot\text{e}^{\tilde{\Gamma}_{s'}(L_{s'},if_0+\Omega)}\\&\cdot\int_{0}^{L_{s}}\text{d}z' \text{e}^{\tilde{\Gamma}(z',mf_0+\nu_{\kappa_1})}\cdot\text{e}^{\tilde{\Gamma}^*(z',nf_0+\nu_{\kappa_3})}\\&\cdot\text{e}^{\tilde{\Gamma}(z',pf_0+\nu_{\kappa_2})}\cdot\text{e}^{-\tilde{\Gamma}(z',if_0+\Omega)}.
\end{split}
\end{equation}
Using (\ref{NLI_elec_tot}) and \eqref{NLIN_expression_x_1.s.end.link} gives 
 \begin{equation}\label{NLIN_expression_x_1.end.link.total}
\begin{split}
&E_{\text{x},\kappa}^{(1)}(L,f)=\imath\frac{8f_0^{3/2}}{9}\sum_{\kappa_1,\kappa_2,\kappa_3\in \mathcal{T}_\kappa}\sum_{i=-\infty}^{\infty}\\&\delta(f-if_0-\Omega)\sum_{m,n,p\in \mathcal{S}_{i,\kappa}}\\&\!\!\!\!\Big(\xi_{\text{x},\kappa_1,m}\xi_{\text{x},\kappa_3,n}^{*}\xi_{\text{x},\kappa_2,p}+\xi_{\text{y},\kappa_1,m}\xi_{\text{y},\kappa_3,n}^{*}\xi_{\text{x},\kappa_2,p}
\Big)\\&\cdot \sum_{s=1}^{N}\gamma_s\Big[ \prod_{s'=1}^{s-1} [g_{s'}^{1/2}(mf_0+\nu_{\kappa_1})\\&\cdot g_{s'}^{1/2}(nf_0+\nu_{\kappa_3}) g_{s'}^{1/2}(pf_0+\nu_{\kappa_2})\\&\!\!\!\cdot \text{e}^{\tilde{\Gamma}_{s'}(L_{s'},mf_0+\nu_{\kappa_1})+\tilde{\Gamma}_{s'}^*(L_{s'},nf_0+\nu_{\kappa_3})+\tilde{\Gamma}_{s'}(L_{s'},pf_0+\nu_{\kappa_2})}]\\&\cdot g_{s}^{1/2}(if_0+\Omega)\text{e}^{\tilde{\Gamma}_{s}(L_s,if_0+\Omega)}\\&\cdot\prod_{s'=s+1}^{N}g_{s'}^{1/2}(if_0+\Omega)\cdot\text{e}^{\tilde{\Gamma}_{s'}(L_{s'},if_0+\Omega)}\Big]\\&\cdot\int_{0}^{L_s}\text{d}z' \text{e}^{\tilde{\Gamma}_s(z',mf_0+\nu_{\kappa_1})}\cdot\text{e}^{\tilde{\Gamma}_s^*(z',nf_0+\nu_{\kappa_3})}\\&\cdot\text{e}^{\tilde{\Gamma}_s(z',pf_0+\nu_{\kappa_2})}\cdot\text{e}^{-\tilde{\Gamma}_s(z',if_0+\Omega)}.
\end{split}
\end{equation}

We will now show that \eqref{Xi} and \eqref{zeta.s} stem from the product terms and integral term in \eqref{NLIN_expression_x_1.end.link.total}, respectively.
The product terms in \eqref{NLIN_expression_x_1.end.link.total}, namely
\begin{align}\label{product.terms.NLIN_expression_x_1.end.link.total}
    &g_{s}^{1/2}(if_0+\Omega)\text{e}^{\tilde{\Gamma}_{s}(L_s,if_0+\Omega)} \prod_{s'=1}^{s-1}g_{s'}^{1/2}(mf_0+\nu_{\kappa_1})\nonumber\\&\cdot g_{s'}^{1/2}(nf_0+\nu_{\kappa_3}) g_{s'}^{1/2}(pf_0+\nu_{\kappa_2})\nonumber\\&\!\!\!\cdot \text{e}^{\tilde{\Gamma}_{s'}(L_{s'},mf_0+\nu_{\kappa_1})}\text{e}^{\tilde{\Gamma}_{s'}^*(L_{s'},nf_0+\nu_{\kappa_3})}\text{e}^{\tilde{\Gamma}_{s'}(L_{s'},pf_0+\nu_{\kappa_2})}\nonumber\\
    &\cdot\prod_{s'=s+1}^{N}g_{s'}^{1/2}(if_0+\Omega)\text{e}^{\tilde{\Gamma}_{s'}(L_{s'},if_0+\Omega)},
\end{align}
in which
\begin{align}
   & g_{s}^{1/2}(if_0+\Omega)\text{e}^{\tilde{\Gamma}_{s}(L_s,if_0+\Omega)}\\&\cdot\prod_{s'=s+1}^{N}g_{s'}^{1/2}(if_0+\Omega)\text{e}^{\tilde{\Gamma}_{s'}(L_{s'},if_0+\Omega)}=\nonumber\\&
    \prod_{s'=s}^{N}g_{s'}^{1/2}(if_0+\Omega)\text{e}^{\tilde{\Gamma}_{s'}(L_{s'},if_0+\Omega)},
\end{align}
can be expressed as
\begin{align}\label{product.terms.2.NLIN_expression_x_1.end.link.total}
    &\prod_{s'=1}^{s-1}g_{s'}^{1/2}(mf_0+\nu_{\kappa_1})g_{s'}^{1/2}(nf_0+\nu_{\kappa_3})\nonumber\\&\cdot  g_{s'}^{1/2}(pf_0+\nu_{\kappa_2}) \text{e}^{\tilde{\Gamma}_{s'}(L_{s'},mf_0+\nu_{\kappa_1})}\text{e}^{\tilde{\Gamma}_{s'}^*(L_{s'},nf_0+\nu_{\kappa_3})}\nonumber\\&\!\!\!\cdot\text{e}^{\tilde{\Gamma}_{s'}(L_{s'},pf_0+\nu_{\kappa_2})}\prod_{s'=s}^{N}g_{s'}^{1/2}(if_0+\Omega)\nonumber\\&\cdot\text{e}^{\tilde{\Gamma}_{s'}(L_{s'},if_0+\Omega)}.
\end{align}
By substituting \eqref{gamma.s} into \eqref{gamma.tilde.s} with $s=s'$, and taking exponentials on both sides of the resulting equation gives
\begin{align}
\label{exp.gamma.tilde.expansion}
\text{e}^{\tilde{\Gamma}_{s'}(z,f)}&=\text{e}^{\frac{1}{2}\int_{0}^{z}\text{d}z'g_{s'}(z',f)}\cdot\text{e}^{\int_{0}^{z}\text{d}z'[\imath 2\pi^{2}\beta_{2,{s'}} f^{2}+\imath\frac{4}{3}\pi^3 \beta_{3,{s'}}f^3]}\\
\label{exp.gamma.tilde.expansion.2}
&=\sqrt{\rho_{s'}(z,f)} \cdot\text{e}^{[\imath 2\pi^{2}\beta_{2,{s'}} f^{2}+\imath\frac{4}{3}\pi^3 \beta_{3,{s'}}f^3]z}
\end{align}
where $\rho_{s'}(\cdot)$ is given in Table~\ref{terms.different}. The first term in \eqref{exp.gamma.tilde.expansion.2} follows from the definition of the normalized signal power profile definition in \eqref{rho.1} and the second from the fact that $\beta_{2,{s'}}$ and $\beta_{3,{s'}}$ are $z$-independent.

Using \eqref{exp.gamma.tilde.expansion.2}, the first product term in \eqref{product.terms.2.NLIN_expression_x_1.end.link.total} can be written as
\begin{align}\label{first.product.NLIN_expression_x_1.end.link.total.2}
&\prod_{s'=1}^{s-1} g_{s'}^{1/2}(mf_0+\nu_{\kappa_1}) g_{s'}^{1/2}(nf_0+\nu_{\kappa_3})\nonumber\\&\cdot g_{s'}^{1/2}(pf_0+\nu_{\kappa_2})\nonumber\\&\cdot \text{e}^{\tilde{\Gamma}_{s'}(L_{s'},mf_0+\nu_{\kappa_1})+\tilde{\Gamma}_{s'}^*(L_{s'},nf_0+\nu_{\kappa_3})}\nonumber\\&\cdot\text{e}^{\tilde{\Gamma}_{s'}(L_{s'},pf_0+\nu_{\kappa_2})}=\prod_{s'=1}^{s-1}g_{s'}^{1/2}(mf_0+\nu_{\kappa_1}) \nonumber\\&\cdot g_{s'}^{1/2}(nf_0+\nu_{\kappa_3}) g_{s'}^{1/2}(pf_0+\nu_{\kappa_2})\nonumber\\&\cdot \sqrt{\rho_{s'}(L_{s'},mf_0+\nu_{\kappa_1})}\nonumber\\&\cdot \sqrt{\rho_{s'}(L_{s'},nf_0+\nu_{\kappa_3})}\nonumber\\&
\cdot \sqrt{\rho_{s'}(L_{s'},pf_0+\nu_{\kappa_2})}\text{exp}\Big(\imath 2\pi^2\beta_{2,{s'}}L_{s'}\big[(mf_0+\nu_{\kappa_1})^2\nonumber\\&-(nf_0+\nu_{\kappa_3})^2+(pf_0+\nu_{\kappa_2})^2\big]\Big)\nonumber\\&\cdot\text{exp}\Big(\imath\frac{4}{3}\pi^3\beta_{3,{s'}}L_{s'}\big[(mf_0+\nu_{\kappa_1})^3\nonumber\\&-(nf_0+\nu_{\kappa_3})^3+(pf_0+\nu_{\kappa_2})^3\big]\Big)
\end{align}
We now use the equality
\begin{align}\label{equality.quadratic}
& x^2-y^2+z^2=(x-y+z)^2+
2(x-y)(z-y),
\end{align}
to express the first exponential term in the right hand side of \eqref{first.product.NLIN_expression_x_1.end.link.total.2} as
\eqlab{first.exponential.NLIN_expression_x_1.end.link.total.2}{
&\text{exp}\Big(\imath 2\pi^2\beta_{2,{s'}}L_{s'}\Big[\Big((m-n+p)f_0+\Omega\Big)^2\Big]\nonumber\\&+\imath 4\pi^{2}\beta_{2,{s'}}L_{s'}\Big[ 
\Big((m-n)f_0+\nu_{\kappa_1}-\nu_{\kappa_3}\Big)\nonumber\\&\cdot\Big((p-n)f_0+\nu_{\kappa_2}-\nu_{\kappa_3}\Big)\Big]\Big).
}
We also use
\begin{align}\label{equality.cube}
x^3-y^3+z^3&=(x-y+z)^3+3(x-y)(z-y)(x+z),
\end{align}
to express the second exponential term in the right hand side of \eqref{first.product.NLIN_expression_x_1.end.link.total.2} as
\eqlab{second.exponential.NLIN_expression_x_1.end.link.total.2}{
&\text{exp}\Big(\imath\frac{4}{3}\pi^3\beta_{3,{s'}}L_{s'}\Big[\Big((m-n+p)f_0+\Omega\Big)^3\Big]\nonumber\\&+\imath 4\pi^3\beta_{3,{s'}}L_{s'}\Big[\Big((m-n)f_0+\nu_{\kappa_1}-\nu_{\kappa_3}\Big)\nonumber\\&\cdot\Big((p-n)f_0+\nu_{\kappa_2}-\nu_{\kappa_3}\Big)\nonumber\\&\cdot\Big((m+p)f_0+\nu_{\kappa_1}+\nu_{\kappa_2}\Big)\Big]\Big).
}
Using \eqref{first.exponential.NLIN_expression_x_1.end.link.total.2} and \eqref{second.exponential.NLIN_expression_x_1.end.link.total.2}, and observing from \eqref{si} that $m-n+p = i$,  \eqref{first.product.NLIN_expression_x_1.end.link.total.2} can be written as
\begin{align}\label{first.product.NLIN_expression_x_1.end.link.total.2.alternative}
&\prod_{s'=1}^{s-1}g_{s'}^{1/2}(mf_0+\nu_{\kappa_1}) g_{s'}^{1/2}(nf_0+\nu_{\kappa_3})\nonumber\\&\cdot g_{s'}^{1/2}(pf_0+\nu_{\kappa_2}) \sqrt{\rho_{s'}(L_{s'},mf_0+\nu_{\kappa_1})}\nonumber\\&\cdot \sqrt{\rho_{s'}(L_{s'},nf_0+\nu_{\kappa_3})}\nonumber\\&
\cdot \sqrt{\rho_{s'}(L_{s'},pf_0+\nu_{\kappa_2})}\nonumber\\&\cdot\text{exp}\Big(\imath 2\pi^2\beta_{2,{s'}}L_{s'}\Big[\Big(if_0+\Omega\Big)^2\Big]\nonumber\\&+\imath 4\pi^{2}\beta_{2,{s'}}L_{s'}\Big[
\Big((m-n)f_0+\nu_{\kappa_1}-\nu_{\kappa_3}\Big)\nonumber\\&\cdot\Big((p-n)f_0+\nu_{\kappa_2}-\nu_{\kappa_3}\Big)\Big]\Big)\nonumber\\&\cdot\text{exp}\Big(\imath\frac{4}{3}\pi^3\beta_{3,{s'}}L_{s'}\Big[\Big(if_0+\Omega\Big)^3\Big]\nonumber\\&+\imath 4\pi^3\beta_{3,{s'}}L_{s'}\Big[\Big((m-n)f_0+\nu_{\kappa_1}-\nu_{\kappa_3}\Big)\nonumber\\&\cdot\Big((p-n)f_0+\nu_{\kappa_2}-\nu_{\kappa_3}\Big)\nonumber\\&\cdot\Big((m+p)f_0+\nu_{\kappa_1}+\nu_{\kappa_2}\Big)\Big]\Big).
\end{align}

According to \eqref{exp.gamma.tilde.expansion.2}, the last product term in \eqref{product.terms.2.NLIN_expression_x_1.end.link.total} can also be written as
\begin{align}\label{last.product.term.NLIN_expression_x_1.end.link.total.2}
    &\prod_{s'=s}^{N}g_{s'}^{1/2}(if_0+\Omega)\text{e}^{\tilde{\Gamma}_{s'}(L_{s'},if_0+\Omega)}=\nonumber\\&
    \prod_{s'=s}^{N}g_{s'}^{1/2}(if_0+\Omega)\sqrt{\rho_{s'}(L_{s'},if_0+\Omega)} \nonumber\\&\cdot\text{exp}\Big([\imath 2\pi^{2}\beta_{2,{s'}} (if_0+\Omega)^{2}\nonumber\\&+\imath\frac{4}{3}\pi^3 \beta_{3,{s'}}(if_0+\Omega)^3]L_{s'}\Big).
\end{align}
Using \eqref{first.product.NLIN_expression_x_1.end.link.total.2.alternative} and \eqref{last.product.term.NLIN_expression_x_1.end.link.total.2}, \eqref{product.terms.2.NLIN_expression_x_1.end.link.total} is equal to
\begin{align}\label{product.terms.2.NLIN_expression_x_1.end.link.total.alternative}
&\prod_{s'=1}^{s-1}g_{s'}^{1/2}(mf_0+\nu_{\kappa_1}) g_{s'}^{1/2}(nf_0+\nu_{\kappa_3})\nonumber\\&\cdot g_{s'}^{1/2}(pf_0+\nu_{\kappa_2}) \sqrt{\rho_{s'}(L_{s'},mf_0+\nu_{\kappa_1})}\nonumber\\&\cdot \sqrt{\rho_{s'}(L_{s'},nf_0+\nu_{\kappa_3})}
\nonumber\\&\cdot \sqrt{\rho_{s'}(L_{s'},pf_0+\nu_{\kappa_2})}\nonumber\\&\cdot\text{exp}\Big(\imath 2\pi^2\beta_{2,{s'}}L_{s'}(if_0+\Omega)^2+\imath 4\pi^{2}\beta_{2,{s'}}L_{s'}\nonumber\\&\cdot
((m-n)f_0+\nu_{\kappa_1}-\nu_{\kappa_3})\cdot((p-n)f_0\nonumber\\&+\nu_{\kappa_2}-\nu_{\kappa_3})\Big)\cdot\text{exp}\Big(\imath\frac{4}{3}\pi^3\beta_{3,{s'}}L_{s'}(if_0+\Omega)^3\nonumber\\&+\imath 4\pi^3\beta_{3,{s'}}L_{s'}((m-n)f_0+\nu_{\kappa_1}-\nu_{\kappa_3})\nonumber\\&\cdot((p-n)f_0+\nu_{\kappa_2}-\nu_{\kappa_3})\cdot((m+p)f_0+\nu_{\kappa_1}+\nu_{\kappa_2})\Big)\nonumber\\&\cdot
    \prod_{s'=s}^{N}g_{s'}^{1/2}(if_0+\Omega)\sqrt{\rho_{s'}(L_{s'},if_0+\Omega)} \nonumber\\&\cdot\text{exp}\Big([\imath 2\pi^{2}\beta_{2,{s'}} (if_0+\Omega)^{2}\nonumber\\&+\imath\frac{4}{3}\pi^3 \beta_{3,{s'}}(if_0+\Omega)^3]L_{s'}\Big),
\end{align}
which can be written as \eqref{Xi}.

We now use \eqref{exp.gamma.tilde.expansion.2} to express the integral in \eqref{NLIN_expression_x_1.end.link.total} as
\eqlab{last.four.terms.exp.gamma}{
&\int_{0}^{L_{s}}\text{d}z'\text{e}^{\tilde{\Gamma}_s(z',mf_0+\nu_{\kappa_1})}\text{e}^{\tilde{\Gamma}_s^*(z',nf_0+\nu_{\kappa_3})}\nonumber\\&\cdot\text{e}^{\tilde{\Gamma}_s(z',pf_0+\nu_{\kappa_2})}\text{e}^{-\tilde{\Gamma}_s(z',if_0+\Omega)}=\nonumber\\&
\int_{0}^{L_{s}}\text{d}z'\sqrt{\rho_{s}(z',mf_0+\nu_{\kappa_1})}\nonumber\\&\cdot\frac{\sqrt{\rho_{s}(z',nf_0+\nu_{\kappa_3})}\sqrt{\rho_{s}(z',pf_0+\nu_{\kappa_2})}}{\sqrt{\rho_{s}(z',if_0+\Omega)}}\nonumber\\&\cdot
\text{e}^{[\imath 2\pi^{2}\beta_{2,{s}} (mf_0+\nu_{\kappa_1})^{2}+\imath\frac{4}{3}\pi^3 \beta_{3,{s}}(mf_0+\nu_{\kappa_1})^3]z'}
\nonumber\\&\cdot
\text{e}^{[-\imath 2\pi^{2}\beta_{2,{s}} (nf_0+\nu_{\kappa_3})^{2}-\imath\frac{4}{3}\pi^3 \beta_{3,{s}}(nf_0+\nu_{\kappa_3})^3]z'}
\nonumber\\&\cdot
\text{e}^{[\imath 2\pi^{2}\beta_{2,{s}} (pf_0+\nu_{\kappa_2})^{2}+\imath\frac{4}{3}\pi^3 \beta_{3,{s}}(pf_0+\nu_{\kappa_2})^3]z'}
\nonumber\\&\cdot
\text{e}^{[-\imath 2\pi^{2}\beta_{2,{s}} (if_0+\Omega)^{2}-\imath\frac{4}{3}\pi^3 \beta_{3,{s}}(if_0+\Omega)^3]z'}.
}
The last step in the proof is therefore to show that
the arguments of the four exponentials in \eqref{last.four.terms.exp.gamma} corresponds to the arguments of the two exponentials in \eqref{zeta.s}. We do this by first grouping the quadratic and cubic terms in the exponentials in \eqref{last.four.terms.exp.gamma} as
\eqlab{beta2.beta3.expon}{
&\text{exp}\Big(\imath 2\pi^{2}\beta_{2,{s}}z'\nonumber\\&\cdot\Big[ (mf_0+\nu_{\kappa_1})^{2}- (nf_0+\nu_{\kappa_3})^{2}\nonumber\\&+ (pf_0+\nu_{\kappa_2})^{2}- (if_0+\Omega)^{2}\Big]\Big)\nonumber\\&\cdot
\text{exp}\Big(\imath\frac{4}{3}\pi^3 \beta_{3,{s}}z'\Big[(mf_0+\nu_{\kappa_1})^3\nonumber\\&-(nf_0+\nu_{\kappa_3})^3+(pf_0+\nu_{\kappa_2})^3\nonumber\\&-if_0+\Omega)^3\Big]\Big).
}
We now use the equality in \eqref{equality.quadratic}
to express the first exponential term in \eqref{beta2.beta3.expon} as
\eqlab{first.exponential.beta2}{
&\text{exp}\Big(\imath 4\pi^{2}\beta_{2,{s}}z'\Big[ 
\Big((m-n)f_0+\nu_{\kappa_1}-\nu_{\kappa_3}\Big)\nonumber\\&\cdot\Big((p-n)f_0+\nu_{\kappa_2}-\nu_{\kappa_3}\Big)\Big]\Big).
}
We also use \eqref{equality.cube}
to express the second exponential term in \eqref{beta2.beta3.expon} as
\eqlab{second.exponential.beta2}{
&\text{exp}\Big(\imath 4\pi^3\beta_{3,{s}}z'\Big[\Big((m-n)f_0+\nu_{\kappa_1}-\nu_{\kappa_3}\Big)\nonumber\\&\cdot\Big((p-n)f_0+\nu_{\kappa_2}-\nu_{\kappa_3}\Big)\nonumber\\&\cdot\Big((m+p)f_0+\nu_{\kappa_1}+\nu_{\kappa_2}\Big)\Big]\Big).
}
By replacing the exponential terms in \eqref{last.four.terms.exp.gamma} by the multiplication of \eqref{first.exponential.beta2} and  \eqref{second.exponential.beta2}, we can rewrite the right hand side of \eqref{last.four.terms.exp.gamma} as \eqref{zeta.s}.
This completes the proof. 

\QEDA
\end{ProofLemma}

 \begin{ProofLemma}
 To evaluate the PSD of the nonlinear electrical field given in \textit{Lemma}~\ref{multiple.span.extention}, we ignore the triplets $(m,n,p)$ in which $m=n$ or $p=n$, as these terms create a constant phase shift and can be interpreted as bias or non-fluctuating terms \cite[Sec.~VIII]{Mecozzi2012}, \cite[Sec.~IV-B]{Pontus_JLT_modeling_2013}, \cite[Sec.~III]{dar2013properties}, and thus, irrelevant for the noise variance we would like to compute. The set $\mathcal{S}_{i,\kappa}$ in \eqref{si} can therefore be written as 
 \begin{align}\label{sil}
\mathcal{S}_{i,\kappa}=&\Big\{(m,n,p)\in\mathbb{Z}^3: m-n+p=i,\; m\neq n,\;  p \neq n, \nonumber\\& 
\nu_{\kappa} -\frac{B_\kappa}{2}\leq if_0+(\nu_{\kappa_1}-\nu_{\kappa_3}+\nu_{\kappa_2})\leq \nu_{\kappa} +\frac{B_\kappa}{2} \Big\}.
\end{align}
The total nonlinear electrical field given in \eqref{NLI_elec_tot_detail} has therefore the form
\eqlab{tot.nli.fild.periodic}{
E_{\text{x},\kappa}^{(1)}(L,f)=\sum_{i=-\infty}^{\infty}I_{i}\delta(f-if_0-\Omega)
}
where
\eqlab{i.second}{
&I_{i}=\imath \frac{8f_0^{3/2}}{9}\sum_{\kappa_1,\kappa_2,\kappa_3 \in \mathcal{T}_{\kappa}}\sum_{m,n,p\in \mathcal{S}_{i,\kappa}}\Big(\xi_{\text{x},\kappa_1,m}\xi_{\text{x},\kappa_3,n}^{*}\nonumber\\&\cdot\xi_{\text{x},\kappa_2,p}+\xi_{\text{y},\kappa_1,m}\xi_{\text{y},\kappa_3,n}^{*}\xi_{\text{x},\kappa_2,p}
\Big)\varsigma_{\kappa_1,\kappa_2,\kappa_3}(m,n,p).
} 
Considering \cite[Eqs.~(60), (61), (62)]{poggiolini2012detailed}, we find that the power spectral density of \eqref{tot.nli.fild.periodic} can be expressed as 
\eqlab{psd.tot.nli.fild.periodic.f.final2}{
&G_{\text{\small{NLI}},\text{x},\kappa}(f)=
\sum_{i=-\infty}^{\infty}\mathbb{E}\bigl\{|I_{i}|^2\bigr\}\delta(f-if_0-\Omega).
}
Using the fact that the symbols in the y polarization are independent from those in the x polarization (see Sec.~\ref{key_results}), we obtain from \eqref{i.second} 
\eqlab{G_x}
{
&\mathbb{E}\bigl\{|I_{i}|^2\bigr\}=\frac{64}{81}f_0^{3}\sum_{\kappa_1,\kappa_2,\kappa_3 \in \mathcal{T}_{\kappa}}\sum_{\kappa_1',\kappa_2',\kappa_3 \in \mathcal{T}_{\kappa}}\nonumber\\
&\sum_{m,n,p\in \mathcal{S}_{i,\kappa}}\sum_{m',n',p'\in \mathcal{S}_{i,\kappa}}\varsigma_{\kappa_1,\kappa_2,\kappa_3}(m,n,p)\varsigma^{*}_{\kappa_1',\kappa_2',\kappa_3'}(m',n',p')\nonumber\\
&\cdot\Big(C_{\text{sp}}+C_{\text{xp}}+\mathbb{E}\big\{\xi_{\text{x},\kappa_1,m}\xi^{*}_{\text{x},\kappa_3,n}\xi_{\text{x},\kappa_2,p}\xi^{*}_{\text{x},\kappa_2',p'}\big\}\nonumber\\
&\cdot\mathbb{E}\big\{\xi^{*}_{y,\kappa_1',m'}\xi_{y,\kappa_3',n'}\big\}\nonumber+\mathbb{E}\big\{\xi_{y,\kappa_1,m}\xi^{*}_{y,\kappa_3,n}\big\}\\
&\cdot\mathbb{E}\big\{\xi^{*}_{\text{x},\kappa'_1,m'}\xi_{\text{x},\kappa_3',n'}\xi^{*}_{\text{x},\kappa'_2,p'}\xi_{\text{x},\kappa_2,p}\big\}\Big),
} 
where
 \eqlab{Csp}
{
&C_{\text{sp}}=\nonumber\\
&\mathbb{E}\big\{\xi_{\text{x},\small{\kappa_1},m}\xi^{*}_{\text{x},\small{\kappa_3},n}\xi_{\text{x},\small{\kappa_2},p}\xi^{*}_{\text{x},\small{\kappa_1'},m'}\xi_{\text{x},\small{\kappa_3'},n'}\nonumber\\&\cdot\xi^{*}_{\text{x},\small{\kappa_2'},p'}\big\},
}
and
\eqlab{Cxp}
{
&C_{\text{xp}}=\mathbb{E}\big\{\xi_{y,\kappa_1,m}\xi^{*}_{y,\kappa_3,n}\xi^{*}_{y,\kappa_1',m'}\xi_{y,\kappa_3',n'}\big\}\nonumber\\
&\cdot
\mathbb{E}\big\{\xi_{\text{x},\kappa_2,p}\xi^{*}_{\text{x},\kappa_2',p'}\big\}.
}
We will now show that for any given $\kappa\in\{-M,\ldots,M\}$, and $i \in \mathbb{Z}$, 
$\mathbb{E}\big\{\xi^{*}_{y,\kappa_1',m'}\xi_{y,\kappa_3',n'}\big\}=0$ $\forall \kappa_{1}',\kappa_{3}' \in\mathcal{T}_{\kappa}$ and $\forall m',n',p' \in\mathcal{S}_{i,\kappa}$. Similarly, for any given $\kappa\in\{-M,\ldots,M\}$, and $i \in \mathbb{Z}$, 
we will also prove that 
$\mathbb{E}\big\{\xi_{y,\kappa_1,m}\xi^{*}_{y,\kappa_3,n}\big\}=0$, $\forall \kappa_{1},\kappa_{3} \in\mathcal{T}_{\kappa}$ and $\forall m,n,p \in\mathcal{S}_{i,\kappa}$. These two cases will prove that only $C_{\text{sp}}$ and $C_{\text{xp}}$ contribute to \eqref{G_x}.

We start by using \eqref{FourierSerisCoef1}, which gives
\begin{align}\label{independent.xi}
  & \mathbb{E}\big\{\xi^{*}_{y,\kappa_1',m'}\xi_{y,\kappa_3',n'}\big\}=f_0S^*(m' f_0)S(n' f_0)\\&\cdot\nonumber \sum_{w_1=1}^{W}\sum_{w_2=1}^{W}\mathbb{E}\big\{{{b}}^{*}_{\text{y},{\kappa_1'},w_1}{{b}}_{\text{y},{\kappa_3'},w_2}\}\text{e}^{\imath\frac{2\pi}{W}(m'w_1-n'w_2)}.
\end{align}
To show that \eqref{independent.xi} is indeed equal to zero, two cases should be taken into consideration: $\kappa_1'\neq \kappa_3'$ and $\kappa_1'= \kappa_3'$.

In the $\kappa_1'\neq \kappa_3'$ case, the expectation term $\mathbb{E}\big\{{{b}}^{*}_{\text{y},{\kappa_1'},w_1}{{b}}_{\text{y},{\kappa_3'},w_2}\}$ in \eqref{independent.xi} can be written as
    \begin{align}\label{ee}
    &\mathbb{E}\big\{{{b}}^{*}_{\text{y},{\kappa_1'},w_1}\}\mathbb{E}\big\{{{b}}_{\text{y},{\kappa_3'},w_2}\},
    \end{align}
    because symbols in different WDM channels are independent. The expectation in \eqref{ee} is zero because the constellations have zero mean.

In the $\kappa_1'= \kappa_3'$ case, \eqref{independent.xi} is 
\begin{align}\label{independent.xi.k1equalk2}
  &f_0S^*(m' f_0)S(n' f_0)\sum_{w_1=1}^{W}\sum_{w_2=1}^{W}\mathbb{E}\big\{{{b}}^{*}_{\text{y},{\kappa_1'},w_1}{{b}}_{\text{y},{\kappa_1'},w_2}\}\\&\nonumber\cdot\text{e}^{\imath\frac{2\pi}{W}(m'w_1-n'w_2)}.
\end{align}
When $w_1\neq w_2$ we have 
\begin{align}\label{rr}
  \mathbb{E}\big\{{{b}}^{*}_{\text{y},{\kappa_1'},w_1}{{b}}_{\text{y},{\kappa_1'},w_2}\}=\mathbb{E}\big\{b^{*}_{\text{y},{\kappa_1'},w_1}\}\mathbb{E}\big\{b_{\text{y},{\kappa_1'},w_2}\}=0
\end{align}
which follows from the zero-mean and independence assumption on the symbols across WDM channels. Using \eqref{rr}, \eqref{independent.xi.k1equalk2} becomes
    \begin{align}\label{independent.xi.k1equalk2.w1equalw2}
  &f_0S^*(m' f_0)S(n' f_0)\mathbb{E}\big\{|{{b}}_{\text{y},{\kappa_1'}}|^2\}\sum_{w_1=1}^{W}\text{e}^{\imath\frac{2\pi}{W}(m'-n')w_1},
\end{align}
where we have used the stationary $\mathbb{E}\big\{|{{b}}_{\text{y},{\kappa_1'}}|^2\}=\mathbb{E}\big\{|{{b}}_{\text{y},{\kappa_1'},w_i}|^2\}$ (see \cite[Appendix~E]{carena2014accuracy}). Furthermore, the sum in \eqref{independent.xi.k1equalk2.w1equalw2} is
\begin{align}\label{pr}
    \sum_{w_1=1}^{W}\text{e}^{\imath\frac{2\pi}{W}(m'-n')w_1}=
\left\{ \begin{array}{l}
    0 \quad m'-n'\neq pW\\
    W \quad m'-n'=pW
  \end{array}, \right.
\end{align}
where $p\in\mathbb{Z}\setminus \{0\}$ ($m'=n'$ is not included in $\mathcal{S}_{i,\kappa}$) and $W=B_{\kappa_1}/f_0$ (see Sec.~\ref{key_results}). When $m'-n'=pW$ (i.e., the second case in \eqref{pr}), the coefficients $S^*(m' f_0)S(n' f_0)$ in \eqref{independent.xi.k1equalk2.w1equalw2} are
\begin{align*}
    S^*(m' f_0)S((m'-pW) f_0)=S^*(m' f_0)S(m'f_0-pB_{\kappa_1})=0
\end{align*}
where the second equality is due to the fact that $S(f)=0$ for $|f|>B_{\kappa_1}/{2}$. 

The procedure above shows that indeed $\mathbb{E}\big\{\xi^{*}_{y,\kappa_1',m'}\xi_{y,\kappa_3',n'}\big\}=0$. The same procedure can be followed to prove that $\mathbb{E}\big\{\xi_{y,\kappa_1,m}\xi^{*}_{y,\kappa_3,n}\big\}=0$, which makes \eqref{psd.tot.nli.fild.periodic.f.final2} equal to 
\eqlab{psd.tot.nli.fild.periodic.f.final}{
&G_{\text{\small{NLI}},\text{x},\kappa}(f)=
\sum_{i=-\infty}^{\infty}\delta(f-if_0-\Omega)\nonumber\\&\cdot\frac{64}{81}f_0^{3}\sum_{\kappa_1,\kappa_2,\kappa_3 \in \mathcal{T}_{\kappa}}\sum_{\kappa_1',\kappa_2',\kappa_3' \in \mathcal{T}_{\kappa}}\sum_{m,n,p\in \mathcal{S}_{i,\kappa}}\sum_{m',n',p'\in \mathcal{S}_{i,\kappa}}\nonumber\\&\varsigma_{\kappa_1,\kappa_2,\kappa_3}(m,n,p)\varsigma^{*}_{\kappa_1',\kappa_2',\kappa_3'}(m',n',p')(C_{\text{sp}}+C_{\text{xp}}).
}

The rest of the proof contains the evaluation of $C_{\text{sp}}$ and $C_{\text{xp}}$ given in \eqref{Csp} and \eqref{Cxp}, resp.
To evaluate \eqref{Csp}, we study all the possible values of $\kappa_1, \kappa_2,\kappa_3\in \mathcal{T}_{\kappa}$ and $\kappa_1', \kappa_2',\kappa_3'\in \mathcal{T}_{\kappa}$ using eight different cases, namely 
\begin{enumerate}
    \item {$\kappa_1=\kappa_2=\kappa_3=\kappa_1'=\kappa_2'=\kappa_3'$}
     \item {$\kappa_1=\kappa_3=\kappa_2'=\kappa_3'\neq\kappa_2=\kappa_1'$}
      \item {$\kappa_1=\kappa_1'=\kappa_3=\kappa_3'\neq\kappa_2=\kappa_2'$}
       \item {$\kappa_1=\kappa_1'=\kappa_2=\kappa_2'\neq\kappa_3=\kappa_3'$}
        \item {$\kappa_2=\kappa_2'=\kappa_3=\kappa_3'\neq\kappa_1=\kappa_1'$}
         \item {$\kappa_2=\kappa_3=\kappa_1'=\kappa_3'\neq\kappa_1=\kappa_2'$}
          \item {$\kappa_1=\kappa_1'\neq\kappa_2'=\kappa_2'\neq\kappa_3=\kappa_3'$}
           \item {$\kappa_1=\kappa_2'\neq\kappa_2=\kappa_1'\neq\kappa_3=\kappa_3'$},
\end{enumerate}
and hence, we can write \eqref{Csp} as
\begin{align}\label{momentum.expansion}
   &C_{\text{sp}}=\\&\nonumber
   \delta_{\kappa_1,\kappa_1'}\delta_{\kappa_2,\kappa_2'}\delta_{\kappa_3,\kappa_3'}\delta_{\kappa_1,\kappa_2}\delta_{\kappa_1,\kappa_3}\\&\nonumber\cdot
   \mathbb{E}\big\{\xi_{\text{x},\small{\kappa_1},m}\xi^{*}_{\text{x},\small{\kappa_1},n}\xi_{\text{x},\small{\kappa_1},p}\xi^{*}_{\text{x},\small{\kappa_1},m'}\xi_{\text{x},\small{\kappa_1},n'}\xi^{*}_{\text{x},\small{\kappa_1},p'}\big\}\\&\nonumber+
\delta_{\kappa_1,\kappa_2'}\delta_{\kappa_2,\kappa_1'}\delta_{\kappa_3,\kappa_3'}\delta_{\kappa_1,\kappa_3}\bar{\delta}_{\kappa_1,\kappa_2}\\&\nonumber\cdot\mathbb{E}\big\{\xi_{\text{x},\small{\kappa_1},m}\xi^{*}_{\text{x},\small{\kappa_1},n}\xi_{\text{x},\small{\kappa_1},n'}\xi^{*}_{\text{x},\small{\kappa_1},p'}\big\}\mathbb{E}\big\{\xi_{\text{x},\small{\kappa_2},p}\xi^{*}_{\text{x},\small{\kappa_2},m'}\big\}\\&\nonumber+
\delta_{\kappa_1,\kappa_1'}\delta_{\kappa_2,\kappa_2'}\delta_{\kappa_3,\kappa_3'}\delta_{\kappa_1,\kappa_3}\bar{\delta}_{\kappa_1,\kappa_2}\\&\nonumber\cdot\mathbb{E}\big\{\xi_{\text{x},\small{\kappa_1},m}\xi^{*}_{\text{x},\small{\kappa_1},n}\xi_{\text{x},\small{\kappa_1},n'}\xi^{*}_{\text{x},\small{\kappa_1},m'}\big\}\mathbb{E}\big\{\xi_{\text{x},\small{\kappa_2},p}\xi^{*}_{\text{x},\small{\kappa_2},p'}\big\}\\&\nonumber+
\delta_{\kappa_1,\kappa_1'}\delta_{\kappa_2,\kappa_2'}\delta_{\kappa_3,\kappa_3'}\delta_{\kappa_1,\kappa_2}\bar{\delta}_{\kappa_1,\kappa_3}\\&\nonumber\cdot\mathbb{E}\big\{\xi_{\text{x},\small{\kappa_1},m}\xi_{\text{x},\small{\kappa_1},p}\xi^{*}_{\text{x},\small{\kappa_1},m'}\xi^{*}_{\text{x},\small{\kappa_1},p'}\big\}\\&\nonumber\cdot\mathbb{E}\nonumber\big\{\xi^{*}_{\text{x},\small{\kappa_3},n}\xi_{\text{x},\small{\kappa_3},n'}\big\}\\&\nonumber+
\delta_{\kappa_1,\kappa_1'}\delta_{\kappa_2,\kappa_2'}\delta_{\kappa_3,\kappa_3'}\delta_{\kappa_2,\kappa_3}\bar{\delta}_{\kappa_1,\kappa_2}\\&\nonumber\cdot\mathbb{E}\big\{\xi_{\text{x},\small{\kappa_2},p}\xi^{*}_{\text{x},\small{\kappa_2},n}\xi_{\text{x},\small{\kappa_2},n'}\xi^{*}_{\text{x},\small{\kappa_2},p'}\big\}\mathbb{E}\big\{\xi_{\text{x},\small{\kappa_1},m}\xi^{*}_{\text{x},\small{\kappa_1},m'}\big\}\\&\nonumber+
\delta_{\kappa_1,\kappa_2'}\delta_{\kappa_2,\kappa_1'}\delta_{\kappa_3,\kappa_3'}\delta_{\kappa_2,\kappa_3}\bar{\delta}_{\kappa_1,\kappa_2}\\&\nonumber\cdot\mathbb{E}\big\{\xi_{\text{x},\small{\kappa_2},p}\xi^{*}_{\text{x},\small{\kappa_2},n}\xi^{*}_{\text{x},\small{\kappa_2},m'}\xi_{\text{x},\small{\kappa_2},n'}\big\}\mathbb{E}\big\{\xi^{*}_{\text{x},\small{\kappa_1},p'}\xi_{\text{x},\small{\kappa_1},m}\big\}\\&\nonumber+
\delta_{\kappa_1,\kappa_1'}\delta_{\kappa_2,\kappa_2'}\delta_{\kappa_3,\kappa_3'}\bar{\delta}_{\kappa_1,\kappa_2}\bar{\delta}_{\kappa_2,\kappa_3}\bar{\delta}_{\kappa_1,\kappa_3}\\&\nonumber\cdot
\mathbb{E}\big\{\xi_{\text{x},\small{\kappa_1},m}\xi^{*}_{\text{x},\small{\kappa_1},m'}\big\}\mathbb{E}\big\{\xi^{*}_{\text{x},\small{\kappa_3},n}\xi_{\text{x},\small{\kappa_3},n'}\big\}\\&\cdot\nonumber\mathbb{E}\big\{\xi_{\text{x},\small{\kappa_2},p}\xi^{*}_{\text{x},\small{\kappa_2},p'}\big\}\\&\nonumber+
\delta_{\kappa_1,\kappa_2'}\delta_{\kappa_2,\kappa_1'}\delta_{\kappa_3,\kappa_3'}\bar{\delta}_{\kappa_1,\kappa_2}\bar{\delta}_{\kappa_2,\kappa_3}\bar{\delta}_{\kappa_1,\kappa_3}\\&\nonumber\cdot\mathbb{E}\big\{\xi_{\text{x},\small{\kappa_1},m}\xi^{*}_{\text{x},\small{\kappa_1},p'}\big\}\mathbb{E}\big\{\xi^{*}_{\text{x},\small{\kappa_3},n}\xi_{\text{x},\small{\kappa_3},n'}\big\}\\&\nonumber\cdot\mathbb{E}\big\{\xi_{\text{x},\small{\kappa_2},p}\xi^{*}_{\text{x},\small{\kappa_2},m'}\big\}.
\end{align}
In \eqref{momentum.expansion}, we recognize the sixth order moment (the first term in \eqref{momentum.expansion}), a mix of second and fourth order moments (second to sixth terms in \eqref{momentum.expansion}), and a mix of second order moments (seventh and eighth terms in \eqref{momentum.expansion}).

Using the sixth, fourth, and second order moments given in \cite[eq.~(105)]{carena2014accuracy}, \cite[eq.~(100)]{carena2014accuracy}, and \cite[eq.~(95)]{carena2014accuracy}, resp., and removing the bias terms from them assuming $p=0$, we can write \eqref{momentum.expansion} as
\begin{align}\label{momentum.expansion.2}
   C_{\text{sp}}&=
   \delta_{\kappa_1,\kappa_1'}\delta_{\kappa_2,\kappa_2'}\delta_{\kappa_3,\kappa_3'}\delta_{\kappa_1,\kappa_2}\delta_{\kappa_1,\kappa_3}\cdot A_6
   \\&\nonumber+
\delta_{\kappa_1,\kappa_2'}\delta_{\kappa_2,\kappa_1'}\delta_{\kappa_3,\kappa_3'}\delta_{\kappa_1,\kappa_3}\bar{\delta}_{\kappa_1,\kappa_2}\cdot B_{4,2}\\&\nonumber+
\delta_{\kappa_1,\kappa_1'}\delta_{\kappa_2,\kappa_2'}\delta_{\kappa_3,\kappa_3'}\delta_{\kappa_1,\kappa_3}\bar{\delta}_{\kappa_1,\kappa_2}\cdot C_{4,2}\\&\nonumber+
\delta_{\kappa_1,\kappa_1'}\delta_{\kappa_2,\kappa_2'}\delta_{\kappa_3,\kappa_3'}\delta_{\kappa_1,\kappa_2}\bar{\delta}_{\kappa_1,\kappa_3}\cdot  D_{4,2}\\&\nonumber+
\delta_{\kappa_1,\kappa_1'}\delta_{\kappa_2,\kappa_2'}\delta_{\kappa_3,\kappa_3'}\delta_{\kappa_2,\kappa_3}\bar{\delta}_{\kappa_1,\kappa_2}\cdot E_{4,2}\\&\nonumber+
\delta_{\kappa_1,\kappa_1'}\delta_{\kappa_2,\kappa_2'}\delta_{\kappa_3,\kappa_3'}\bar{\delta}_{\kappa_1,\kappa_2}\bar{\delta}_{\kappa_2,\kappa_3}\bar{\delta}_{\kappa_1,\kappa_3}\cdot F_{4,2}\\&\nonumber+
\delta_{\kappa_1,\kappa_1'}\delta_{\kappa_2,\kappa_2'}\bar{\delta}_{\kappa_1,\kappa_2}\bar{\delta}_{\kappa_2,\kappa_3}\bar{\delta}_{\kappa_1,\kappa_3}\cdot
G_{2,2,2}\\&\nonumber+
\delta_{\kappa_1,\kappa_2'}\delta_{\kappa_2,\kappa_1'}\delta_{\kappa_3,\kappa_3'}\bar{\delta}_{\kappa_1,\kappa_2}\bar{\delta}_{\kappa_2,\kappa_3}\bar{\delta}_{\kappa_1,\kappa_3}\cdot H_{2,2,2},
\end{align}
in which $A_6$, $B_{4,2}$, $C_{4,2}$, $D_{4,2}$, $E_{4,2}$, $F_{4,2}$, $G_{2,2,2}$, and $H_{2,2,2}$ are given in Table~\ref{A6}. To be more specific, $A_6$ stems from the sixth order moment given in \cite[eq.~(105)]{carena2014accuracy}. The terms $B_{4,2}$, $C_{4,2}$, $D_{4,2}$, $E_{4,2}$, and $F_{4,2}$ are obtained using the fourth and second order moments given in \cite[eq.~(100)]{carena2014accuracy}, and \cite[eq.~(95)]{carena2014accuracy}, resp. The second order moment given in \cite[eq.~(95)]{carena2014accuracy} is used for computing $G_{2,2,2}$ and $H_{2,2,2}$. 
 \begin{table*}[t]
    \footnotesize
    \centering
    \caption{The expressions for the terms used in Eqs.~\eqref{momentum.expansion.2} and \eqref{cross-polarization NLI.2}. The $U$ functions are given by Eqs.~\eqref{Ut}--\eqref{U6}.}
    \label{A6}
    \begin{tabular}{|c|l|}
    \hline
    
    \hline
    Term & Expression \\
    \hline
    
    \hline
$\displaystyle A_6$ &  $2B_{\kappa_1}^{3}\mathbb{E}^{3}\{|b_{\text{x},\kappa_1}|^{2}\}U_{mnpm'n'p'}\delta_{m,m'}\delta_{n,n'}\delta_{p,p'}+B_{\kappa_1}^{2}f_0\mathbb{E}^{2}\{|b_{\text{x},\kappa_1}|^{2}\}[\mathbb{E}\{|b_{\text{x},\kappa_1}|^{4}\}-2\mathbb{E}\{|b_{\text{x},\kappa_1}|^{2}\}]U_{mnpm'n'p'}(4\delta_{m,m'}$\\&$\cdot\delta_{p-n+n'-p',0}+\delta_{n,n'}\delta_{m+p-m'-p',0})+B_{\kappa_1}f_0[\mathbb{E}\{|b_{\text{x},\kappa_1}|^{6}\}-9\mathbb{E}\{|b_{\text{x},\kappa_1}|^{4}\}\mathbb{E}^{2}\{|b_{\text{x},\kappa_1}|^{2}\}+12\mathbb{E}^{3}\{|b_{\text{x},\kappa_1}|^{2}\}] U_{mnpm'n'p'}$\\&$\cdot
\delta_{m-n+p-m'+n'-p',0}$
\\
\hline
$\displaystyle B_{4,2}$ &  $\!\!\!(B_{\kappa_1}^{2}\mathbb{E}^{2}\{|b_{\text{x},\kappa_1}|^{2}\}U_{mn'p'n}\delta_{m,p'}\delta_{n,n'}+B_{\kappa_1}f_0U_{mn'p'n}\delta_{m-n-p'+n',0}[\mathbb{E}\{|b_{\text{x},\kappa_1}|^{4}\}-2\mathbb{E}^{2}\{|b_{\text{x},\kappa_1}|^{2}\}])B_{\kappa_2}|S_{\kappa_2}(pf_0)|^{2}\mathbb{E}\{|b_{\text{x},\kappa_2}|^{2}\}\delta_{p,m'}$
\\
\hline
$\displaystyle C_{4,2}$ &
$\!\!\!(B_{\kappa_1}^{2}\mathbb{E}^{2}\{|b_{\text{x},\kappa_1}|^{2}\}U_{mn'm'n}\delta_{m,m'}\delta_{n,n'}+B_{\kappa_1}f_0U_{mn'm'n}\delta_{m-n-m'+n',0}[\mathbb{E}\{|b_{\text{x},\kappa_1}|^{4}\}-2\mathbb{E}^{2}\{|b_{\text{x},\kappa_1}|^{2}\}])B_{\kappa_2}|S_{\kappa_2}(pf_0)|^{2}\mathbb{E}\{|b_{\text{x},\kappa_2}|^{2}\}\delta_{p,p'}$
\\
\hline
$\displaystyle D_{4,2}$ &
$\!\!\!(2B_{\kappa_1}^{2}\mathbb{E}^{2}\{|b_{\text{x},\kappa_1}|^{2}\}U_{mm'p'p}\delta_{m,m'}\delta_{p,p'}+B_{\kappa_1}f_0U_{mm'p'p}\delta_{m-m'-p'+p,0}[\mathbb{E}\{|b_{\text{x},\kappa_1}|^{4}\}-2\mathbb{E}^{2}\{|b_{\text{x},\kappa_1}|^{2}\}])$\\&$\cdot B_{\kappa_3}|S_{\kappa_3}(nf_0)|^{2}\mathbb{E}\{|b_{\text{x},\kappa_3}|^{2}\}\delta_{n,n'}$
\\
\hline
$\displaystyle E_{4,2}$ &
$\!\!\!(B_{\kappa_2}^{2}\mathbb{E}^{2}\{|b_{\text{x},\kappa_2}|^{2}\}U_{pn'p'n}\delta_{p,p'}\delta_{n,n'}+B_{\kappa_2}f_0U_{pn'p'n}\delta_{p-n-p'+n',0}[\mathbb{E}\{|b_{\text{x},\kappa_2}|^{4}\}-2\mathbb{E}^{2}\{|b_{\text{x},\kappa_2}|^{2}\}])B_{\kappa_1}|S_{\kappa_1}(mf_0)|^{2}\mathbb{E}\{|b_{\text{x},\kappa_1}|^{2}\}\delta_{m,m'}$
\\
\hline
$\displaystyle F_{4,2}$ &$\!\!\!(B_{\kappa_2}^{2}\mathbb{E}^{2}\{|b_{\text{x},\kappa_2}|^{2}\}U_{npm'n'}\delta_{p,m'}\delta_{n,n'}+B_{\kappa_2}f_0U_{npm'n'}\delta_{p-n-m'+n',0}[\mathbb{E}\{|b_{\text{x},\kappa_2}|^{4}\}-2\mathbb{E}^{2}\{|b_{\text{x},\kappa_2}|^{2}\}])B_{\kappa_1}|S_{\kappa_1}(mf_0)|^{2}\mathbb{E}\{|b_{\text{x},\kappa_1}|^{2}\}\delta_{m,p'}$
\\
\hline
$\displaystyle G_{2,2,2}$ &$B_{\kappa_1}|S_{\kappa_1}(mf_0)|^{2}\mathbb{E}\{|b_{\text{x},\kappa_1}|^{2}\}\delta_{m,m'}B_{\kappa_3}|S_{\kappa_3}(nf_0)|^{2}\mathbb{E}\{|b_{\text{x},\kappa_3}|^{2}\}\delta_{n,n'}B_{\kappa_2}|S_{\kappa_2}(pf_0)|^{2}\mathbb{E}\{|b_{\text{x},\kappa_2}|^{2}\}\delta_{p,p'}$
\\
\hline
$\displaystyle H_{2,2,2}$ &$B_{\kappa_1}|S_{\kappa_1}(mf_0)|^{2}\mathbb{E}\{|b_{\text{x},\kappa_1}|^{2}\}\delta_{m,p'}B_{\kappa_3}|S_{\kappa_3}(nf_0)|^{2}\mathbb{E}\{|b_{\text{x},\kappa_3}|^{2}\}\delta_{n,n'}B_{\kappa_2}|S_{\kappa_2}(pf_0)|^{2}\mathbb{E}\{|b_{\text{x},\kappa_2}|^{2}\}\delta_{p,m'}$
\\
\hline
\hline
$\displaystyle {X}_{1}$ &$(B_{\kappa_1}^2 \mathbb{E}^{2}\{|b_{\text{y},\kappa_1}|^{2}\}|S_{\kappa_1}(mf_0)|^2|S_{\kappa_3}(nf_0)|^2\delta_{m,m'}\delta_{n,n'}+B_{\kappa_1}f_0U_{mnm'n'}\delta_{m'-m-n'+n,0}[\mathbb{E}\{|b_{\text{y},\kappa_1}|^{4}\}-2\mathbb{E}^{2}\{|b_{\text{y},\kappa_1}|^{2}\}])$\\&$\cdot B_{\kappa_2}|S_{\kappa_2}(pf_0)|^{2}\mathbb{E}\{|b_{\text{x},\kappa_2}|^{2}\}\delta_{p,p'}$
\\
\hline
$\displaystyle {X}_{2}$ &$B_{\kappa_1} |S_{\kappa_1}(mf_0)|^{2}\mathbb{E}\{|b_{\text{y},\kappa_1}|^{2}\}\delta_{m,m'}B_{\kappa_3}|S_{\kappa_3}(nf_0)|^{2}\mathbb{E}\{|b_{\text{y},\kappa_3}|^{2}\}\delta_{n,n'}B_{\kappa_2}|S_{\kappa_2}(pf_0)|^{2}\mathbb{E}\{|b_{\text{x},\kappa_2}|^{2}\}\delta_{p,p'}$
\\
\hline
    \end{tabular}
\end{table*}
Using the same procedure in \eqref{momentum.expansion} and \eqref{momentum.expansion.2}, $C_{\text{xp}}$ in \eqref{Cxp} is expressed as
\begin{align}\label{cross-polarization NLI.2}
C_{\text{xp}}& =\delta_{\kappa_1,\kappa_1'}\delta_{\kappa_2,\kappa_2'}\delta_{\kappa_3,\kappa_3'}\delta_{\kappa_1,\kappa_2}\delta_{\kappa_1,\kappa_3}\cdot X_1
   \\&\nonumber+
\delta_{\kappa_1,\kappa_1'}\delta_{\kappa_2,\kappa_2'}\delta_{\kappa_3,\kappa_3'}\delta_{\kappa_1,\kappa_3}\bar{\delta}_{\kappa_1,\kappa_2}\cdot X_1\\&\nonumber+
\delta_{\kappa_1,\kappa_1'}\delta_{\kappa_2,\kappa_2'}\delta_{\kappa_3,\kappa_3'}\delta_{\kappa_1,\kappa_2}\bar{\delta}_{\kappa_2,\kappa_3}\cdot  X_2\\&\nonumber+
\delta_{\kappa_1,\kappa_1'}\delta_{\kappa_2,\kappa_2'}\delta_{\kappa_3,\kappa_3'}\delta_{\kappa_2,\kappa_3}\bar{\delta}_{\kappa_1,\kappa_2}\cdot X_2\\&\nonumber+
\delta_{\kappa_1,\kappa_1'}\bar{\delta}_{\kappa_2,\kappa_2'}\delta_{\kappa_3,\kappa_3'}\delta_{\kappa_1,\kappa_2}\bar{\delta}_{\kappa_2,\kappa_3}\bar{\delta}_{\kappa_1,\kappa_3}\cdot
X_2,
\end{align}
where $X_1$ and $X_2$ are given in Table~\ref{A6}, and where the $U$ functions in Table~\ref{A6} are
\begin{align}
&U_{mnp'm'n'p}=\nonumber\\&\label{Ut}S_{\kappa_1}(mf_0)S_{\kappa_3}^*(nf_0)S_{\kappa_2}(pf_0)S_{\kappa_1}^*(m'f_0)S_{\kappa_3}(n'f_0)S_{\kappa_2}^*(p'f_0),\\&\label{U2}
U_{mn'p'n}=S_{\kappa_1}(mf_0)S_{\kappa_3}^*(nf_0)S_{\kappa_2}^*(p'f_0)S_{\kappa_3}(n'f_0),\\&\label{U3}
U_{mnm'n'}=S_{\kappa_1}(mf_0)S_{\kappa_3}^*(nf_0)S_{\kappa_1}^*(m'f_0)S_{\kappa_3}(n'f_0),\\&\label{U4}
U_{mm'p'p}=S_{\kappa_1}(mf_0)S_{\kappa_1}^*(m'f_0)S_{\kappa_2}^*(p'f_0)S_{\kappa_2}(pf_0),\\&\label{U5}
U_{npn'p'}=S_{\kappa_3}^*(nf_0)S_{\kappa_2}(pf_0)S_{\kappa_3}(n'f_0)S_{\kappa_2}^*(p'f_0),
\end{align}
and 
\begin{align}\label{U6}
U_{npn'm'}=S_{\kappa_3}^*(nf_0)S_{\kappa_2}(pf_0)S_{\kappa_3}(n'f_0)S_{\kappa_1}^*(m'f_0).
\end{align}

By substituting \eqref{momentum.expansion.2} and \eqref{cross-polarization NLI.2} into \eqref{psd.tot.nli.fild.periodic.f.final}, the total nonlinear PSD can be written as
\eqlab{G_x_d}
{
&G_{\text{\small{NLI}},\text{x},\kappa}(f)=\frac{64}{81}\sum_{\kappa_1,\kappa_2,\kappa_3\in \mathcal{T}_{\kappa}}\nonumber\\&\Big(\bar{\delta}_{\kappa_2,\kappa_3}\bar{\delta}_{\kappa_1,\kappa_3} \bar{\delta}_{\kappa_1,\kappa_2}\hat{D}\\&\nonumber+
\delta_{\kappa_1,\kappa_3}\bar{\delta}_{\kappa_1,\kappa_2}\hat{E}+\delta_{\kappa_2,\kappa_3}\bar{\delta}_{\kappa_1,\kappa_2}\hat{F}\\&\nonumber+
\delta_{\kappa_1,\kappa_2}\bar{\delta}_{\kappa_2,\kappa_3}\hat{G}+\delta_{\kappa_1,\kappa_2}\delta_{\kappa_2,\kappa_3}\hat{H}\Big),
}
where $\hat{D}$, $\hat{E}$, $\hat{F}$, $\hat{G}$, and $\hat{H}$ are given in Table~\ref{hat.DEFGH}. Finally, by using Table~\ref{A6} into Table~\ref{hat.DEFGH}, we get Table~\ref{hat.DEFGH.second}.

 \begin{table*}[t]
    \footnotesize
    \centering
    \caption{The expressions for the terms given in \eqref{G_x_d}. To obtain these expressions, the terms given in Table~\ref{A6} are used. }
    \label{hat.DEFGH}
    \begin{tabular}{|c|l|}
    \hline
    
    \hline
    Term & Expression \\
    \hline
    
    \hline
$\displaystyle \hat{D}$ &  $f_0^{3}\sum_{i=-\infty}^{\infty}\delta\big(f-if_0-\Omega\big)[\sum_{m,n,p\in \mathcal{S}_{i,\kappa}}\sum_{m',n',p'\in \mathcal{S}_{i,\kappa}}\varsigma_{\kappa_1,\kappa_2,\kappa_3}(m,n,p)\varsigma^{*}_{\kappa_1,\kappa_2,\kappa_3}(m',n',p')\cdot{G_{2,2,2}}+\sum_{m,n,p\in \mathcal{S}_{i,\kappa}}$\\&$\sum_{m',n',p'\in \mathcal{S}_{i,\kappa}}\varsigma_{\kappa_1,\kappa_2,\kappa_3}(m,n,p)\varsigma^{*}_{\kappa_2,\kappa_1,\kappa}(m',n',p')\cdot{H_{2,2,2}}+
\sum_{m,n,p\in \mathcal{S}_{i,\kappa}}\sum_{m',n',p'\in \mathcal{S}_{i,\kappa}}\varsigma_{\kappa_1,\kappa_2,\kappa_3}(m,n,p)\varsigma^{*}_{\kappa_1,\kappa_2,\kappa_3}(m',n',p')\cdot{X}_{2}]$
\\
\hline
$\displaystyle \hat{E}$ &  $f_0^{3}\sum_{i=-\infty}^{\infty}\delta(f-if_0-\Omega)[\sum_{m,n,p\in \mathcal{S}_{i,\kappa}}\sum_{m',n',p'\in \mathcal{S}_{i,\kappa}}\varsigma_{\kappa_1,\kappa_2,\kappa_3}(m,n,p)\varsigma^{*}_{\kappa_2,\kappa_1,\kappa}(m',n',p'){B_{4,2}}+\sum_{m,n,p\in \mathcal{S}_{i,\kappa}}$\\&$\sum_{m',n',p'\in \mathcal{S}_{i,\kappa}}\varsigma_{\kappa_1,\kappa_2,\kappa_3}(m,n,p)\varsigma^{*}_{\kappa_1,\kappa_2,\kappa_3}(m',n',p'){C_{4,2}}+
\sum_{m,n,p\in \mathcal{S}_{i,\kappa}}\sum_{m',n',p'\in \mathcal{S}_{i,\kappa}}\varsigma_{\kappa_1,\kappa_2,\kappa_3}(m,n,p)\varsigma^{*}_{\kappa_1,\kappa_2,\kappa_3}(m',n',p'){X}_{1}]$
\\
\hline
$\displaystyle \hat{F}$ &
$f_0^{3}\sum_{i=-\infty}^{\infty}\delta(f-if_0-\Omega)[\sum_{m,n,p\in \mathcal{S}_{i,\kappa}}\sum_{m',n',p'\in \mathcal{S}_{i,\kappa}}\varsigma_{\kappa_1,\kappa_2,\kappa_3}(m,n,p)\varsigma^{*}_{\kappa_1,\kappa_2,\kappa_3}(m',n',p'){E_{4,2}}+\sum_{m,n,p\in \mathcal{S}_{i,\kappa}}\sum_{m',n',p'\in \mathcal{S}_{i,\kappa}}$\\&$\varsigma_{\kappa_1,\kappa_2,\kappa_3}(m,n,p)\varsigma^{*}_{\kappa_2,\kappa_1,\kappa}(m',n',p'){F_{4,2}}+
\sum_{m,n,p\in \mathcal{S}_{i,\kappa}}\sum_{m',n',p'\in \mathcal{S}_{i,\kappa}}\varsigma_{\kappa_1,\kappa_2,\kappa_3}(m,n,p)\varsigma^{*}_{\kappa_1,\kappa_2,\kappa_3}(m',n',p'){X}_{2}]
$
\\
\hline
$\displaystyle \hat{G}$ &
$f_0^{3}\sum_{i=-\infty}^{\infty}\delta\big(f-if_0-\Omega\big)[\sum_{m,n,p\in \mathcal{S}_{i,\kappa}}\sum_{m',n',p'\in \mathcal{S}_{i,\kappa}}\varsigma_{\kappa_1,\kappa_2,\kappa_3}(m,n,p)\varsigma^{*}_{\kappa_1,\kappa_2,\kappa_3}(m',n',p')\cdot{D_{4,2}}$\\&$+\sum_{m,n,p\in \mathcal{S}_{i,\kappa}}\sum_{m',n',p'\in \mathcal{S}_{i,\kappa}}\varsigma_{\kappa_1,\kappa_2,\kappa_3}(m,n,p)\varsigma^{*}_{\kappa_1,\kappa_2,\kappa_3}(m',n',p')\cdot{X}_{2}]$
\\
\hline
$\displaystyle \hat{H}$ &
$f_0^{3}\sum_{i=-\infty}^{\infty}\delta\big(f-if_0-\Omega\big)[\sum_{m,n,p\in \mathcal{S}_{i,\kappa}}\sum_{m',n',p'\in \mathcal{S}_{i,\kappa}}\varsigma_{\kappa_1,\kappa_2,\kappa_3}(m,n,p)\varsigma^{*}_{\kappa_1,\kappa_2,\kappa_3}(m',n',p')\cdot{A_{6}}$\\&$+
\sum_{m,n,p\in \mathcal{S}_{i,\kappa}}\sum_{m',n',p'\in \mathcal{S}_{i,\kappa}}\varsigma_{\kappa_1,\kappa_2,\kappa_3}(m,n,p)\varsigma^{*}_{\kappa_1,\kappa_2,\kappa_3}(m',n',p')\cdot{X}_{1}]$
\\
\hline
    \end{tabular}
\end{table*}
 \begin{table*}[t]
    \footnotesize
    \centering
    \caption{The expansion of the expressions given in Table~\ref{hat.DEFGH} using the terms given in Tables~\ref{A6}.}
    \label{hat.DEFGH.second}
    \begin{tabular}{|c|l|}
    \hline
    
    \hline
    Term & Expression \\
    \hline
    
    \hline
$\displaystyle \hat{D}$ &  $f_0^3\sum_{i=-\infty}^{\infty}\delta\big(f-if_0-\Omega\big)\sum_{m,n,p\in \mathcal{S}_{i,\kappa}}\varsigma_{\kappa_1,\kappa_2,\kappa_3}(m,n,p)\varsigma^{*}_{\kappa_1,\kappa_2,\kappa_3}(m,n,p) B_{\kappa_1}|S_{\kappa_1}(mf_0)|^{2}\mathbb{E}\{|b_{\text{x},\kappa_1}|^{2}\}B_{\kappa_3}|S_{\kappa_3}(nf_0)|^{2}$\\&$\cdot\mathbb{E}\{|b_{\text{x},\kappa_3}|^{2}\}B_{\kappa_2}|S_{\kappa_2}(pf_0)|^{2}\mathbb{E}\{|b_{\text{x},\kappa_2}|^{2}\}+
f_0^3\sum_{i=-\infty}^{\infty}\delta\big(f-if_0-\Omega\big)\sum_{m,n,p\in \mathcal{S}_{i,\kappa}}\varsigma_{\kappa_1,\kappa_2,\kappa_3}(m,n,p)\varsigma^{*}_{\kappa_2,\kappa_1,\kappa}(p,n,m)$\\&$\cdot 
B_{\kappa_1}|S_{\kappa_1}(mf_0)|^{2}\mathbb{E}\{|b_{\text{x},\kappa_1}|^{2}\}B_{\kappa_3}|S_{\kappa_3}(nf_0)|^{2}\mathbb{E}\{|b_{\text{x},\kappa_3}|^{2}\}B_{\kappa_2}|S_{\kappa_2}(pf_0)|^{2}\mathbb{E}\{|b_{\text{x},\kappa_2}|^{2}\}+
f_0^3\sum_{i=-\infty}^{\infty}\delta\big(f-if_0-\Omega\big)\sum_{m,n,p\in \mathcal{S}_{i,\kappa}}$\\&$\varsigma_{\kappa_1,\kappa_2,\kappa_3}(m,n,p)\varsigma^{*}_{\kappa_1,\kappa_2,\kappa_3}(m,n,p)
B_{\kappa_1}|S_{\kappa_1}(mf_0)|^{2}\mathbb{E}\{|b_{\text{y},\kappa_1}|^{2}\}B_{\kappa_3}|S_{\kappa_3}(nf_0)|^{2}\mathbb{E}\{|b_{\text{y},\kappa_3}|^{2}\}B_{\kappa_2}|S_{\kappa_2}(pf_0)|^{2}\mathbb{E}\{|b_{\text{x},\kappa_2}|^{2}\}$
\\
\hline
$\displaystyle \hat{E}$ &  $f_0^3\sum_{i=-\infty}^{\infty}\delta\big(f-if_0-\Omega\big) \sum_{m,n,p\in \mathcal{S}_{i,\kappa}}\varsigma_{\kappa_1,\kappa_2,\kappa_3}(m,n,p)\varsigma^{*}_{\kappa_2,\kappa_1,\kappa}(p,n,m)\cdot B_{\kappa_1}^2B_{\kappa_2}|S_{\kappa_1}(mf_0)|^{2}|S_{\kappa_3}(nf_0)|^{2}|S_{\kappa_2}(pf_0)|^{2}\mathbb{E}^2\{|b_{\text{x},\kappa_1}|^{2}\}$\\&$\cdot\mathbb{E}\{|b_{\text{x},\kappa_2}|^{2}\}
+f_0^3\sum_{i=-\infty}^{\infty}\delta\big(f-if_0-\Omega\big)\sum_{m,n,p\in \mathcal{S}_{i,\kappa}}\sum_{m,n',p'\in \mathcal{S}_{i,\kappa}}\varsigma_{\kappa_1,\kappa_2,\kappa_3}(m,n,p)\varsigma^{*}_{\kappa_2,\kappa_1,\kappa}(p,n',p')\cdot B_{\kappa_1}B_{\kappa_2}f_0U_{mnp'n'}$\\&$\cdot|S_{\kappa_2}(pf_0)|^{2}\delta_{m-n-p'+n',0}\Big[\mathbb{E}\{|b_{\text{x},\kappa_1}|^{4}\}-2\mathbb{E}^{2}\{|b_{\text{x},\kappa_1}|^{2}\}\Big]\cdot\mathbb{E}\{|b_{\text{x},\kappa_2}|^{2}\}$\\
&$+
 f_0^3\sum_{i=-\infty}^{\infty}\delta\big(f-if_0-\Omega\big) \sum_{m,n,p\in \mathcal{S}_{i,\kappa}}\varsigma_{\kappa_1,\kappa_2,\kappa_3}(m,n,p)\varsigma^{*}_{\kappa_1,\kappa_2,\kappa_3}(m,n,p)\cdot B_{\kappa_1}^2B_{\kappa_2}|S_{\kappa_1}(mf_0)|^{2}|S_{\kappa_3}(nf_0)|^{2}|S_{\kappa_2}(pf_0)|^{2}\mathbb{E}^2\{|b_{\text{x},\kappa_1}|^{2}\}$\\&$\cdot\mathbb{E}\{|b_{\text{x},\kappa_2}|^{2}\}+f_0^3\sum_{i=-\infty}^{\infty}\delta\big(f-if_0-\Omega\big)\sum_{m,n,p\in \mathcal{S}_{i,\kappa}}\sum_{m',n',p\in \mathcal{S}_{i,\kappa}}\varsigma_{\kappa_1,\kappa_2,\kappa_3}(m,n,p)\varsigma^{*}_{\kappa_1,\kappa_2,\kappa_3}(m',n',p)\cdot B_{\kappa_1}B_{\kappa_2}f_0U_{mn'm'n}$\\&$\cdot|S_{\kappa_2}(pf_0)|^{2}\delta_{m-n-m'+n',0}\Big[\mathbb{E}\{|b_{\text{x},\kappa_1}|^{4}\}-2\mathbb{E}^{2}\{|b_{\text{x},\kappa_1}|^{2}\}\Big]\cdot\mathbb{E}\{|b_{\text{x},\kappa_2}|^{2}\}
$\\
&$+
f_0^3\sum_{i=-\infty}^{\infty}\delta\big(f-if_0-\Omega\big) \sum_{m,n,p\in \mathcal{S}_{i,\kappa}}\varsigma_{\kappa_1,\kappa_2,\kappa_3}(m,n,p)\varsigma^{*}_{\kappa_1,\kappa_2,\kappa_3}(m,n,p)\cdot B_{\kappa_1}^2B_{\kappa_2}|S_{\kappa_1}(mf_0)|^{2}|S_{\kappa_3}(nf_0)|^{2}|S_{\kappa_2}(pf_0)|^{2}\mathbb{E}^2\{|b_{\text{y},\kappa_1}|^{2}\}$\\&$\cdot\mathbb{E}\{|b_{\text{x},\kappa_2}|^{2}\}
+f_0^3\sum_{i=-\infty}^{\infty}\delta\big(f-if_0-\Omega\big) \sum_{m,n,p\in \mathcal{S}_{i,\kappa}}\varsigma_{\kappa_1,\kappa_2,\kappa_3}(m,n,p)\varsigma^{*}_{\kappa_1,\kappa_2,\kappa_3}(m',n',p)B_{\kappa_1}B_{\kappa_2}f_0U_{mnm'n'}|S_{\kappa_2}(pf_0)|^{2}$\\&$\cdot\delta_{m-n-m'+n',0}\Big[\mathbb{E}\{|b_{\text{y},\kappa_1}|^{4}\}-2\mathbb{E}^{2}\{|b_{\text{y},\kappa_1}|^{2}\}\Big]\mathbb{E}\{|b_{\text{x},\kappa_2}|^{2}\}$
\\
\hline
$\displaystyle \hat{F}$ &
$f_0^3\sum_{i=-\infty}^{\infty}\delta\big(f-if_0-\Omega\big) \sum_{m}\sum_{p}\varsigma_{\kappa_1,\kappa_2,\kappa_3}(m,n,p)\varsigma^{*}_{\kappa_1,\kappa_2,\kappa_3}(m,n,p)\cdot B_{\kappa_1}B_{\kappa_2}^2|S_{\kappa_1}(mf_0)|^{2}|S_{\kappa_3}(nf_0)|^{2}|S_{\kappa_2}(pf_0)|^{2}\mathbb{E}\{|b_{\text{x},\kappa_1}|^{2}\}$\\&$\cdot\mathbb{E}^2\{|b_{\text{x},\kappa_2}|^{2}\}+f_0^3\sum_{i=-\infty}^{\infty}\delta\big(f-if_0-\Omega\big)\sum_{m}\sum_{p}\sum_{p'}\varsigma_{\kappa_1,\kappa_2,\kappa_3}(m,n,p)\varsigma^{*}_{\kappa_1,\kappa_2,\kappa_3}(m,n',p')\cdot B_{\kappa_1}B_{\kappa_2}f_0U_{pn'p'n}|S_{\kappa_1}(mf_0)|^{2}$\\&$\cdot\delta_{p-n-p'+n',0}\mathbb{E}\{|b_{x,\kappa_1}|^{2}\}\cdot\Big[\mathbb{E}\{|b_{x,\kappa_2}|^{4}\}-2\mathbb{E}^{2}\{|b_{x,\kappa_2}|^{2}\}\Big]
$\\
&$+
f_0^3\sum_{i=-\infty}^{\infty}\delta\big(f-if_0-\Omega\big) \sum_{m}\sum_{p}\varsigma_{\kappa_1,\kappa_2,\kappa_3}(m,n,p)\varsigma^{*}_{\kappa_2,\kappa_1,\kappa}(p,n,m)\cdot B_{\kappa_1}B_{\kappa_2}^2|S_{\kappa_1}(mf_0)|^{2}|S_{\kappa_3}(nf_0)|^{2}|S_{\kappa_2}(pf_0)|^{2}\mathbb{E}\{|b_{\text{x},\kappa_1}|^{2}\}$\\&$\cdot\mathbb{E}^2\{|b_{\text{x},\kappa_2}|^{2}\}+f_0^3\sum_{i=-\infty}^{\infty}\delta\big(f-if_0-\Omega\big)\sum_{m}\sum_{p}\sum_{m'}\varsigma_{\kappa_1,\kappa_2,\kappa_3}(m,n,p)\varsigma^{*}_{\kappa_2,\kappa_1,\kappa}(m',n',m)\cdot B_{\kappa_1}B_{\kappa_2}f_0U_{pnm'n'}|S_{\kappa_1}(mf_0)|^{2}$\\&$\cdot\delta_{p-n-m'+n',0}\Big[\mathbb{E}\{|b_{\text{x},\kappa_2}|^{4}\}-2\mathbb{E}^{2}\{|b_{\text{x},\kappa_2}|^{2}\}\Big]\cdot \mathbb{E}\{|b_{\text{x},\kappa_1}|^{2}\}+
f_0^3\sum_{i=-\infty}^{\infty}\delta\big(f-if_0-\Omega\big)\sum_{m}\sum_{p}\varsigma_{\kappa_1,\kappa_2,\kappa_3}(m,n,p)$\\&$\cdot\varsigma^{*}_{\kappa_1,\kappa_2,\kappa_3}(m,n,p)\cdot B_{\kappa_1}B_{\kappa_2}^1|S_{\kappa_1}(mf_0)|^{2}|S_{\kappa_3}(nf_0)|^{2}|S_{\kappa_2}(pf_0)|^{2}\mathbb{E}\{|b_{\text{y},\kappa_1}|^{2}\}\mathbb{E}\{|b_{\text{y},\kappa_2}|^{2}\}\mathbb{E}\{|b_{\text{x},\kappa_2}|^{2}\}
$
\\
\hline
$\displaystyle \hat{G}$ &
$f_0^{3}\sum_{i=-\infty}^{\infty}\delta\big(f-if_0-\Omega\big)\sum_{m}\sum_{p}|\varsigma_{\kappa_1,\kappa_2,\kappa_3}(m,n,p)|^2\cdot
2B_{\kappa_1}^2B_{\kappa_3}\mathbb{E}^{2}\{|b_{x,\kappa_1}|^{2}\}\cdot\mathbb{E}\{|b_{\text{x},\kappa_3}|^{2}\}|S_{\kappa_1}(mf_0)|^2|S_{\kappa_2}(pf_0)|^2|S_{\kappa_3}(nf_0)|^{2}$\\&$
+f_0^{4}\sum_{i=-\infty}^{\infty}\delta\big(f-if_0-\Omega\big)\sum_{m}\sum_{p}\sum_{m'}\varsigma_{\kappa_1,\kappa_2,\kappa_3}(m,n,p)\varsigma^{*}_{\kappa_1,\kappa_2,\kappa_3}(m',n,p')\cdot B_{\kappa_1}B_{\kappa_3} S_{\kappa_1}(mf_0)S_{\kappa_1}(m'f_0)^*S_{\kappa_2}(p'f_0)^*S_{\kappa_2}(pf_0)$\\&$\cdot|S_{\kappa_3}(nf_0)|^{2}\delta_{m-m'-p'+p,0}\Big[\mathbb{E}\{|b_{\text{x},\kappa_1}|^{4}\}-2Rf_0\mathbb{E}^{2}\{|b_{\text{x},\kappa_1}|^{2}\}\Big]\cdot\mathbb{E}\{|b_{\text{x},\kappa_3}|^{2}\}
$\\
&$+
f_0^{3}\sum_{i=-\infty}^{\infty}\delta\big(f-if_0-\Omega\big)\sum_{m}\sum_{p}|\varsigma_{\kappa_1,\kappa_2,\kappa_3}(m,n,p)|^2\cdot B_{\kappa_1}^2B_{\kappa_3}|S_{\kappa_1}(mf_0)|^{2}\cdot |S_{\kappa_3}(nf_0)|^{2}|S_{\kappa_2}(pf_0)|^{2}\mathbb{E}\{|b_{\text{y},\kappa_1}|^{2}\}\mathbb{E}\{|b_{\text{y},\kappa_3}|^{2}\}$\\&$\cdot\mathbb{E}\{|b_{\text{x},\kappa_2}|^{2}\}$
\\
\hline
$\displaystyle \hat{H}$ &
$f_0^{3}\sum_{i=-\infty}^{\infty}\delta\big(f-if_0-\Omega\big)\sum_{m}\sum_{p}|\varsigma_{\kappa_1,\kappa_2,\kappa_3}(m,n,p)|^2\cdot
2B_{\kappa_1}^{3}\mathbb{E}^{3}\{|b_{\text{x},\kappa_1}|^{2}\}|S_{\kappa_1}(mf_0)|^2|S_{\kappa_3}(nf_0)|^2|S_{\kappa_2}(pf_0)|^2$\\&+$f_0^{3}\sum_{i=-\infty}^{\infty}\delta\big(f-if_0-\Omega\big)\sum_{m}\sum_{p}\sum_{p'}\varsigma_{\kappa_1,\kappa_2,\kappa_3}(m,n,p)\varsigma^{*}_{\kappa_1,\kappa_2,\kappa_3}(m,n',p')\cdot B_{\kappa_1}^{2}f_0\mathbb{E}^{2}\{|b_{\text{x},\kappa_1}|^{2}\}\Big[\mathbb{E}\{|b_{\text{x},\kappa_1}|^{4}\}-2\mathbb{E}\{|b_{\text{x},\kappa_1}|^{2}\}\Big]$\\&$\cdot U_{mnpm'n'p'}
\cdot4\delta_{m,m'}\delta_{p-n+n'-p',0}+f_0^{3}\sum_{i=-\infty}^{\infty}\delta\big(f-if_0-\Omega\big)\sum_{m}\sum_{p}\sum_{m'}\varsigma_{\kappa_1,\kappa_2,\kappa_3}(m,n,p)\varsigma^{*}_{\kappa_1,\kappa_2,\kappa_3}(m',n,p')$\\&$\cdot B_{\kappa_1}^{2}f_0\mathbb{E}^{2}\{|b_{\text{x},\kappa_1}|^{2}\}\Big[\mathbb{E}\{|b_{\text{x},\kappa_1}|^{4}\}-2\mathbb{E}\{|b_{\text{x},\kappa_1}|^{2}\}\Big]U_{mnpm'n'p'}\cdot\delta_{n,n'}\delta_{m+p-m'-p',0})+f_0^{3}\sum_{i=-\infty}^{\infty}\delta\big(f-if_0-\Omega\big)$\\&$\cdot\sum_{m}\sum_{p}\sum_{m'}\sum_{p'}\varsigma_{\kappa_1,\kappa_2,\kappa_3}(m,n,p)\varsigma^{*}_{\kappa_1,\kappa_2,\kappa_3}(m',n',p') B_{\kappa_1}f_0\cdot\Big[\mathbb{E}\{|b_{\text{x},\kappa_1}|^{6}\}-9\mathbb{E}\{|b_{\text{x},\kappa_1}|^{4}\}\mathbb{E}^{2}\{|b_{\text{x},\kappa_1}|^{2}\}+12\mathbb{E}^{3}\{|b_{\text{x},\kappa_1}|^{2}\}\Big]U_{mnpm'n'p'}$\\&$\cdot\delta_{m-n+p-m'+n'-p',0}
$\\&$+
f_0^{3}\sum_{i=-\infty}^{\infty}\delta\big(f-if_0-\Omega\big)\sum_{m}\sum_{p}|\varsigma_{\kappa_1,\kappa_2,\kappa_3}(m,n,p)|^2\cdot
B_{\kappa_1}^3 \mathbb{E}^{2}\{|b_{\text{y},\kappa_1}|^{2}\}\mathbb{E}\{|b_{\text{x},\kappa_2}|^{2}\}|S_{\kappa_1}(mf_0)|^2|S_{\kappa_3}(nf_0)|^2|S_{\kappa_2}(pf_0)|^{2}$\\&$+f_0^{3}\sum_{i=-\infty}^{\infty}\delta\big(f-if_0-\Omega\big)\sum_{m}\sum_{p}\sum_{m'}\varsigma_{\kappa_1,\kappa_2,\kappa_3}(m,n,p)\varsigma^{*}_{\kappa_1,\kappa_2,\kappa_3}(m',n',p)\cdot B_{\kappa_1}f_0U_{mnm'n'}\delta_{m-n-m'+n',0}$\\&$\cdot\Big[\mathbb{E}\{|b_{\text{y},\kappa_1}|^{4}\}-2\mathbb{E}^{2}\{|b_{\text{y},\kappa_1}|^{2}\}\Big]B_{\kappa_1}|S_{\kappa_2}(pf_0)|^{2}\mathbb{E}\{|b_{\text{x},\kappa_2}|^{2}\}\delta_{p,p'}$
\\
\hline
    \end{tabular}
\end{table*}
Since the single period of channel $h$ signal ${\boldsymbol{p}}_h(t)$ introduced in \eqref{PeriodicSignal_x} is chosen large enough, i.e., $f_0\rightarrow 0$, one may replace the discrete summation by a continuous integral, namely
$
\lim_{f_0\rightarrow0}f_0\sum_{m}\Lambda(mf_0)=\int\text{d}f \Lambda(f)
$.
Using \eqref{PowerX/Y}, considering the terms given in Table~\ref{hat.DEFGH.second}, and also canceling the delta function in Table~\ref{hat.DEFGH.second} via the integral over $f$, we can rewrite \eqref{G_x_d} in the continuous domain as \eqref{G_x_d1}. This completes the proof.

\QEDA
\end{ProofLemma}





\section{Proof of Corollary \ref{main.result.identical}}\label{AppendixG}

If the loss of each frequency is exactly compensated for at the end of the corresponding span, we have 
\eqlab{GL}{
&\prod_{s'=1}^{s-1}\sqrt{g_{s'}(f_1)\rho_{s'}(L_{s'},f_1)}\\&\cdot
     \sqrt{g_{s'}(f_1+f_2-f)\rho_{s'}(L_{s'},f_1-f+f_2)}\nonumber\\&\cdot\sqrt{g_{s'}(f_2)\rho_{s'}(L_{s'},f_2)}=1,
}
and
\eqlab{GLraman}{
\prod_{s'=s}^{N}\sqrt{g_{s'}(f)\rho_{s'}(L_{s'},f)}=1,
}
and, hence, $\Upsilon(\cdot)$ in Table~\ref{terms.different} can be written as
\eqlab{Upsilon_mltp_idtcl}{
&\Upsilon(f_1,f_2,f)=
\sum_{s=1}^{N}\gamma_{s}\mu_s(f_1,f_2,f)\nonumber\\
&\cdot{\text{e}^{\imath 4\pi^{2}(f_1-f)(f_2-f)\sum_{s'=1}^{s-1}(\beta_{2,s'}L_{s'}+\pi(f_1+f_2)\beta_{3,s'}L_{s'})}}.
}
For multiple identical spans of homogeneous fiber ($\alpha_1=\ldots =\alpha_N=\alpha$, $L_{s}=\ldots =L_N= L$, $\gamma_1=\ldots =\gamma_N=\gamma$, $\beta_{2,{s'}}=\ldots =\beta_{2,N}=\beta_2$, $\beta_{3,{s'}}=\ldots =\beta_{3,N}=\beta_3$), \eqref{Upsilon_mltp_idtcl} is equal to
\eqlab{Upsilon_mltp_idtcl_2}{
&\Upsilon(f_1,f_2,f)=
\gamma\mu(f_1,f_2,f)\nonumber\\
&\cdot[1+\text{e}^{\imath 4\pi^{2}(f_1-f)(f_2-f)(\beta_{2}L+\pi(f_1+f_2)\beta_{3})L}\nonumber\\&+\text{e}^{\imath 4\pi^{2}(f_1-f)(f_2-f)(\beta_{2}L+\pi(f_1+f_2)\beta_{3})2L}+\ldots\nonumber\\&+\text{e}^{\imath 4\pi^{2}(f_1-f)(f_2-f)(\beta_{2}L+\pi(f_1+f_2)\beta_{3})(N-1)L}],
}
where $\mu(\cdot)$ is given in Table~\ref{terms.identical}, and using the fact that
\eqlab{Ts}{
1+\text{e}^{\imath x}+\text{e}^{2\imath x}+\cdots+\text{e}^{\imath (N_s-1)x}
=\frac{1-\text{e}^{\imath N_sx}}{1-\text{e}^{\imath x}}
}
\eqref{Upsilon_mltp_idtcl_2} can be written as
\eqlab{Upsilon_mltp_idtcl_final}{
&\Upsilon(f_1,f_2,f)=\gamma\mu(f_1,f_2,f)\nonumber\\&
\cdot\frac{1-\text{e}^{4\pi^{2}(f_1-f)(f_2-f)NL[\beta_2+\pi\beta_3(f_1+f_2)]}}{1-\text{e}^{4\pi^{2}(f_1-f)(f_2-f)L[\beta_2+\pi\beta_3(f_1+f_2)]}}
}
which is equivalent to $\Upsilon(\cdot)$ given in Table~\ref{terms.identical}. 

\bibliographystyle{IEEEtran}
\bibliography{ref_beygi}
\end{document}